\documentclass[10pt]{llncs}

\usepackage{etex}
\usepackage[utf8]{inputenc}
\usepackage{epic}

\usepackage{amsmath,amssymb,wasysym}
\usepackage{algorithm}
\usepackage{eucal}
\usepackage[table,xcdraw]{xcolor}
\usepackage{booktabs}
\usepackage{multirow}
\usepackage{graphicx}
\usepackage{xspace}
\usepackage{array}
\usepackage{subcaption}

\usepackage{hyperref}

\usepackage{multicol}
\usepackage{wrapfig}
\usepackage{cite}
\usepackage{algorithm}
\usepackage{algorithmic}
\usepackage{pgfplots}

\pgfplotsset{compat=1.12}
\usetikzlibrary{backgrounds}
\usetikzlibrary{calc}
\usetikzlibrary{positioning}

\pgfplotsset{ 
    discard if not/.style 2 args={
        x filter/.code={
            \edef\tempa{\thisrow{#1}}
            \edef\tempb{#2}
            \ifx\tempa\tempb
            \else
                
            \fi
        }
    }, 
}

\renewcommand{\arraystretch}{1.1}
\newcommand*\pct{\scalebox{.7}{\%}} 

\newcolumntype{C}[1]{>{\centering\let\newline\\\arraybackslash\hspace{0pt}}m{#1}}

\usepackage{mathtools}

\usepackage{bbm}
\usepackage{url}

\newcommand\adj{\text{adj}}
\newcommand\id{\text{id}}
\newcommand\cp{\circ}
\newcommand\ext{\gamma}
\newcommand\cext[2]{\cp_{\!#1}^{#2}}

\newcommand{\adjedges}[1]{E_{\adj}^{#1}}
\newcommand{\selfedges}[1]{E_{\id}^{#1}}

\newcommand\FFR{F\!F\!R}
\newcommand\FFRC{\FFR_{\circ}}

\newcommand\CM{\widetilde{M}}
\newcommand\AB{A\!B}

\newcommand\BB{B\!B}
\renewcommand\AA{A\!A}

\newcommand\ma{\mathcal A}
\newcommand\maa{\ma_{\star}}
\newcommand\mb{\mathcal B}
\newcommand\mbb{\mb_{\star}}
\newcommand\mg{\mathcal G}

\newcommand\T[1]{{\tt #1}}
\newcommand\z{\mspace{2mu}}
\newcommand\tl[1]{#1^{\z\!t}}
\newcommand\hd[1]{#1^{\z\!h}}
\newcommand\rev[1]{\overline{#1}}

\newcommand{\A}{\protect\mathcal{A}} 
\newcommand{\B}{\protect\mathcal{B}} 
\newcommand{\G}{\protect\mathcal{G}} 

\newcommand{\wcost}{\textup{wd}_\textsc{dcj}}
\newcommand{\wcostid}{\textup{wd}_\textsc{dcj}^\textsc{id}}

\newcommand{\ddcj}{\textup{d}_\textup{\textsc{dcj}}}

\newcommand{\ddcjid}{\textup{d}_\textup{\textsc{dcj}}^\textup{\textsc{id}}}
\newcommand{\dffdcjid}{\textup{ffd}_\textup{\textsc{dcj}}^\textup{\textsc{id}}}
\newcommand{\dunwffdcjid}{\textup{unwffd}_\textup{\textsc{dcj}}^\textup{\textsc{id}}}

\newcommand\GS{\mathcal S}
\newcommand\GSx[1]{\GS_{#1}}
\newcommand\WR{W\!R}

\usepackage{color}

\begin{document}

\title{Natural family-free genomic distance}

\author{
Diego P. Rubert\inst{1}
\and F\'abio V. Martinez\inst{1}
\and Mar\'ilia D. V. Braga\inst{2}
}
\institute{Faculdade de Computação, Universidade Federal de Mato Grosso do Sul, Brazil\\
\{diego,fhvm\}@facom.ufms.br
\and Faculty of Technology and Center for Biotechnology (CeBiTec), Bielefeld University, Germany\\
mbraga@cebitec.uni-bielefeld.de
}

\maketitle




\begin{abstract}
A classical problem in comparative genomics is to compute the rearrangement
distance, that is the minimum number of large-scale rearrangements required to transform a given genome into another given genome.
While the most traditional approaches in this area are \emph{family-based}, i.e., require the classification of DNA fragments of both genomes into \emph{families}, more recently an alternative model was proposed, which, instead of family classification, simply uses the \emph{pairwise similarities} 
between DNA fragments of both genomes to compute their rearrangement distance. This model represents structural rearrangements by the generic \emph{double cut and join} (DCJ) operation and is then called \emph{family-free DCJ distance}. It computes the DCJ distance between the two genomes by searching for a matching of their genes based on the given pairwise similarities, therefore helping to find gene homologies. The drawback is that its computation is NP-hard. Another point is that the family-free DCJ distance must correspond to a maximal matching of the genes, due to the fact that unmatched genes are just ignored: maximizing the matching  prevents the \emph{free lunch} artifact of having empty or almost empty matchings giving the smaller distances. 

\medskip

In this paper, besides DCJ operations, we allow content-modifying operations of \emph{insertions} and \emph{deletions} of DNA segments and propose a 
new and more general family-free genomic distance.
In our model we use the pairwise similarities to assign weights to both matched and unmatched genes, so that 
an optimal solution does not necessarily maximize the matching. Our model then results in a \emph{natural family-free genomic distance}, that takes into consideration all given genes and has a search space composed of  matchings of any size. 
We provide an efficient ILP formulation to solve it, by extending the previous formulations for computing family-based genomic distances from Shao {\it et al.} ({\it J. Comput.\ Biol.}, 2015) and Bohnenk\"amper {\it et al.} ({\it Proc.\ of RECOMB}, 2020).
Our experiments show that the ILP can handle not only bacterial genomes, but also fungi and insects, or sets of chromosomes of mammals and plants. In a comparison study of six fruit fly genomes, we obtained accurate results.
\keywords{Comparative genomics, Genome rearrangement, DCJ-indel distance.}
\end{abstract}

\section{Introduction}

Genomes are subject to mutations or \emph{rearrangements} in the
course of evolution. A classical problem in comparative genomics is to compute the rearrangement
\emph{distance}, that is the minimum number of large-scale rearrangements required to transform a
given genome into another given genome~\cite{San92}. 
Typical large-scale rearrangements change the
number of chromosomes, and/or the positions and orientations of
DNA segments. Examples of such \emph{structural} rearrangements are inversions, translocations,
fusions and fissions.
One might also need to consider rearrangements that modify the content of a
genome, such as insertions and deletions (collectively called \emph{indels}) of DNA
segments.

In order to study the rearrangement distance, one usually adopts a high-level view
of genomes, in which only ``relevant'' fragments of the DNA (e.g., genes) are
taken into consideration. Furthermore, a pre-processing of the
data is required, so that we can compare the content of the genomes.
One popular method, adopted for more than 20 years, is to group the fragments in
both genomes into \emph{families}, so that two fragments in the same family are said to
be equivalent. This setting is said to be \emph{family-based}. Without
duplications, that is, with the additional restriction that each family occurs
at most once in each genome, many polynomial models have been proposed to
compute the genomic distance~\cite{BER-MIX-STO-2006,HAN-PEV-1995,YAN-ATT-FRI-2005,YAN-FRI-2009,BRA-WIL-STO-2011}.
However, when duplications are allowed the problem is more intricate and
all approaches proposed so far are NP-hard, see for
instance~\cite{San99,Bry00,BJ12b,AFRTV09,RUB-FEI-BRA-STO-MAR-2017,SHA-LIN-MOR-2015}.

The required pre-classification of DNA fragments into families is a drawback of the family-based approaches. Moreover, even with a careful pre-processing, it is not always possible to classify each fragment unambiguously into a single 
family. Due to these facts, an alternative to the family-based setting was
proposed and consists in studying the rearrangement distance without
prior family assignment.
Instead of families, the \emph{pairwise similarities} 
between fragments is directly used~\cite{DTS12,BCDJSTW13}. By letting structural rearrangements be represented by the generic \emph{double cut and join} (DCJ) operation~\cite{YAN-ATT-FRI-2005}, a first family-free genomic distance, called family-free DCJ distance, was already proposed~\cite{MAR-FEI-BRA-STO-15}. Its computation helps to match occurrences of duplicated genes and find homologies, but unmatched genes are simply ignored. 

In the family-based setting, the mentioned approaches that handle duplications either require the compared genomes to be \emph{balanced} (that is, have the same number of occurrences of each family)~\cite{RUB-FEI-BRA-STO-MAR-2017,SHA-LIN-MOR-2015} or adopt some approach to match genes, ignoring unmatched genes~\cite{San99,BJ12b}. 
Recently, a new family-based approach was proposed, allowing each family to occur any number of times in each genome and 
integrating DCJ operations 
and indels in a \emph{DCJ-indel} distance formula~\cite{BOH-BRA-DOE-STO-2020}. For its computation, that is NP-hard, an efficient ILP was proposed.  

Here we adapt the approach mentioned above and give an ILP formulation to compute 
a new family-free DCJ-indel distance. In the family-based approach from~\cite{BOH-BRA-DOE-STO-2020} as well as in the family-free DCJ distance proposed in~\cite{MAR-FEI-BRA-STO-15}, the search space needs to be restricted
to candidates that maximize the number of matched genes, in order to avoid the \emph{free lunch} artifact of having empty or almost empty matchings giving the smaller distances~\cite{YAN-FRI-2009}. In our formulation we use the pairwise similarities to assign weights to matched and unmatched genes, so that, for the first time, 
an optimal solution does not necessarily maximize the number of matched genes. 
For the fact that our model takes into consideration all given genes and has a search space composed of matchings of any size, we call it \emph{natural family-free genomic distance}.
Our simulated experiments show that our ILP can handle not only bacterial genomes, but also complete genomes of fungi and insects, or sets of chromosomes of mammals and plants. We use our implementation to generate pairwise distances and reconstruct the phylogeny of six species of fruit flies from the genus \emph{Drosophila}, obtaining accurate results.

This paper is organized as follows. In Section~\ref{sec:preliminaries} we give some basic definitions and previous results that are essential for the approach presented here. In Section~\ref{sec:ff} we define the new \emph{natural} family-free DCJ-indel distance and show its NP-hardness. In Section~\ref{sec:ff-diagram} we describe the optimization approach for computing the family-free DCJ-indel distance with the help of the \emph{family-free relational diagram}. In Section~\ref{sec:ilp} we present the ILP formulation and the experimental results. Finally, Section~\ref{sec:conclusion} concludes the text.

\section{Preliminaries}\label{sec:preliminaries}

We call \emph{marker} an oriented DNA fragment. A \emph{chromosome} is composed of markers and can be linear or circular. A marker~$m$ in a chromosome can be represented by the symbol~$m$ itself, if it is read in direct orientation, or the symbol~$\rev{m}$, if it is read in reverse orientation. We concatenate all markers of a chromosome $Z$ in a string $z$, which can be read in any of the two directions. If $Z$
is circular, we can start to read it at any marker and the string $z$ is
flanked by parentheses. A set of chromosomes comprises a \emph{genome}. As an example, 
let $A=\{\rev{\T{6}}\z\T{1}\z\T{7}\z\T{8}\z\rev{\T{4}},  \T{3}\z\rev{\T{5}}\z\T{2} \}$ be a genome composed of two linear chromosomes.
A genome can be transformed or \emph{sorted} into another genome
with the following types of mutations.

\begin{enumerate}
\item {\bf DCJ operations modify the organization of a genome:} 
A {\em cut} performed on a genome $A$ separates two adjacent markers of $A$.
A \emph{double-cut and join} or \emph{DCJ} applied on a genome $A$ is the
operation that performs cuts in two different positions of $A$, creating four
open ends, and joins these open ends in a different way~\cite{YAN-ATT-FRI-2005,BER-MIX-STO-2006}. 
For example, 
let $A=\{\rev{\T{6}}\z\T{1}\z\T{7}\z\T{8}\z\rev{\T{4}},  \T{3}\z\rev{\T{5}}\z\T{2} \}$, and consider a DCJ that cuts between markers $\T{1}$ and $\T{7}$ of its first chromosome and between markers $\T{5}$ and $\T{2}$ of its second chromosome, creating fragments $\rev{\T{6}}\z\T{1}\bullet$, $\bullet\T{7}\z\T{8}\z\rev{\T{4}}$,  $\T{3}\z\rev{\T{5}}\bullet$ and $\bullet\T{2}$ (where the symbols $\bullet$ represent the open ends). If we join the first with the third and the second with the fourth open end, we get $A'=\{\rev{\T{6}}\z\T{1}\z\T{2},  \T{3}\z\rev{\T{5}}\z\T{7}\z\T{8}\z\rev{\T{4}} \}$, that is, the described DCJ operation is a translocation transforming $A$ into $A'$. 
Indeed, a DCJ operation can correspond not only to a translocation but to several structural rearrangements, such as an inversion,
a fusion or a fission.

\medskip

\item {\bf Indel operations modify the content of a genome:}
We can modify the content of a genome with {\em
insertions} and with {\em deletions} of blocks of contiguous
markers, collectively called \emph{indel} operations~\cite{YAN-FRI-2009,BRA-WIL-STO-2011}. As an example, consider the deletion of fragment~$\T{7}\z\T{8}$ from chromosome $\rev{\T{6}}\z\T{1}\z\T{7}\z\T{8}\z\rev{\T{4}}$, resulting in chromosome $\rev{\T{6}}\z\T{1}\z\rev{\T{4}}$.
In the model we consider, we do not allow
that a marker is deleted and reinserted, nor inserted and then deleted.
Furthermore, at most one chromosome can be entirely deleted or inserted at once.

\end{enumerate}

Let $A$ and $B$ be two genomes and let $\ma$ be the set of
markers in genome $A$ and $\mb$ be the set of markers in genome $B$. We consider two distinct settings:

\begin{itemize}
\item In a \emph{family-based setting} markers are grouped into families and each marker from a genome is represented by its family. Therefore, a marker from $\ma$ can occur more than once in $A$, as well as a marker from $\mb$ can occur more than once in $B$. Furthermore, genomes $A$ and $B$ may share a set of \emph{common markers} $\mg = \ma \cap \mb$. We also have sets~$\maa= \ma \setminus \mg$ and~$\mbb= \mb \setminus \mg$ of markers that occur respectively only in $A$ and only in $B$ and are called \emph{exclusive markers}. For example, we could have $A=\{\rev{\T{3}}\z\T{1}\z\T{4}\z\T{3}\z\rev{\T{2}},  \T{3}\z\rev{\T{5}}\z\T{2} \}$ and $B=\{\rev{\T{1}}\z\T{2}\z\rev{\T{1}}\z\T{3}\z\rev{\T{2}}\z\T{6} \}$. In this case we have $\ma = \{\T{1},\T{2},\T{3},\T{4},\T{5}\}$ and $\mb = \{\T{1},\T{2},\T{3},\T{6}\}$. Consequently, $\mg = \{\T{1},\T{2},\T{3}\}$, $\maa = \{\T{4},\T{5}\}$ and $\mbb = \{\T{6}\}$.
\item In a \emph{family-free setting} the markers of $A$ and $B$ are all distinct and unique. In other words, each marker of $\ma$ occurs exactly once in $A$, each marker of $\mb$ occurs exactly once in $B$ and  $\ma \cap \mb=\emptyset$. Consider, for example, genomes  $A=\{\rev{\T{1}}\z\T{3}\z\rev{\T{4}}\z\T{2} \}$ and $B=\{\rev{\T{8}}\z\T{7}\z\rev{\T{5}}, \T{9}\z\rev{\T{6}} \}$.
\end{itemize}
\medskip

\subsection{Relational diagram and distance of family-based singular genomes}
  
Let $A$ and $B$ be two genomes in a family-based setting and assume that both $A$ and $B$ are \emph{singular}, that is, 
each marker from $\mg = \ma \cap \mb$ occurs exactly once in each genome. We will now describe how the DCJ-indel distance can be computed in this case~\cite{BRA-WIL-STO-2011}.


For a given marker $m$, denote its two extremities by~$\tl{m}$ (tail) and~$\hd{m}$ (head).
Given two singular genomes~$A$ and~$B$, the \emph{relational diagram} $R(A,B)$~\cite{BOH-BRA-DOE-STO-2020} has a set of vertices $V = V(A) \cup V(B)$, where $V(A)$ is the set of extremities of markers from $A$ and $V(B)$ is the set of extremities of markers from $B$.
There are three types of edges in $R(A,B)$:
\begin{itemize}
\item \emph{Adjacency edges}:
for each pair of marker extremities $\gamma_1$ and $\gamma_2$ that are adjacent in a chromosome of any of the two genomes, we have the adjacency edge $\gamma_1\gamma_2$. Denote by $\adjedges{A}$ and by $\adjedges{B}$ the adjacency edges in $A$ and in
$B$, respectively.
Marker extremities located at chromosome ends are called \emph{telomeres} and are not connected to any adjacency edge. 
\item \emph{Extremity edges}, whose set is denoted by $E_{\ext}$:
for each common marker~$m\in\mg$,
we have two extremity edges, one connecting the vertex $\hd{m}$ from~$V(A)$ to the vertex $\hd{m}$ from~$V(B)$ and the other connecting the vertex~$\tl{m}$ from $V(A)$ to the vertex $\tl{m}$ from~$V(B)$.
\item \emph{Indel edges}:
for each occurrence of an exclusive marker $m\in\maa \cup \mbb$,
we have the indel edge $\tl{m}\hd{m}$.
Denote by $\selfedges{A}$ and by $\selfedges{B}$ the indel edges in $A$ and in
$B$. 
\end{itemize}

\begin{figure}[ht]
  \begin{center}
    \scriptsize
    \setlength{\unitlength}{.5pt}
    \begin{picture}(500,80)(-70,0)

      \dottedline[$\cdot$]{3}(30,70)(45,70)
      \dottedline[$\cdot$]{3}(75,70)(90,70)
      \dottedline[$\cdot$]{3}(120,70)(135,70)
      \dottedline[$\cdot$]{3}(165,70)(180,70)
      \put(0,0){\color{blue}
        \dottedline[$\cdot$]{3}(255,70)(270,70)
      }
      \put(0,0){\color{red}
        \dottedline[$\cdot$]{3}(300,70)(315,70)
      }
      \dottedline[$\cdot$]{3}(30,10)(45,10)
      \dottedline[$\cdot$]{3}(75,10)(90,10)
      \dottedline[$\cdot$]{3}(120,10)(135,10)
      \put(0,0){\color{red}
        \dottedline[$\cdot$]{3}(165,10)(180,10)
        \dottedline[$\cdot$]{3}(210,10)(225,10)
      }
      \put(0,0){\color{blue}
        \dottedline[$\cdot$]{3}(300,10)(315,10)
      }
      
      \qbezier(45,70)(50,70)(75,70)
      \put(0,0){\color{red}
        \qbezier(180,10)(195,10)(210,10)
      }
      
      \put(0,0){\color{blue}
        \qbezier(0,10)(0,35)(0,70)
      }
      \qbezier(30,10)(30,40)(30,70)
      \qbezier(45,10)(68,40)(90,70)
      \qbezier(75,10)(98,40)(120,70)
      \qbezier(90,10)(128,40)(165,70)
      \qbezier(120,10)(128,40)(135,70)
      \qbezier(135,10)(158,40)(180,70)
      \put(0,0){\color{red}
        \qbezier(165,10)(188,40)(210,70)
        \qbezier(225,10)(225,40)(225,70)
      }
      \put(0,0){\color{blue}
        \qbezier(255,10)(255,40)(255,70)
        \qbezier(300,10)(323,40)(345,70)
        \qbezier(315,10)(293,40)(270,70)
      }          
      \put(0,0){\color{red}
        \qbezier(270,10)(293,40)(315,70)
        \qbezier(345,10)(323,40)(300,70)
      }

      \put(  0,70){\circle*{5}}
      \put( 30,70){\circle*{5}}
      \put( 45,70){\circle*{5}}
      \put( 75,70){\circle*{5}}
      \put( 90,70){\circle*{5}}
      \put(120,70){\circle*{5}}
      \put(135,70){\circle*{5}}
      \put(165,70){\circle*{5}}
      \put(180,70){\circle*{5}}
      \put(210,70){\circle*{5}}
      \put(225,70){\circle*{5}}
      \put(255,70){\circle*{5}}
      \put(270,70){\circle*{5}}
      \put(300,70){\circle*{5}}
      \put(315,70){\circle*{5}}
      \put(345,70){\circle*{5}}
      \put(  0,10){\circle*{5}}
      \put( 30,10){\circle*{5}}
      \put( 45,10){\circle*{5}}
      \put( 75,10){\circle*{5}}
      \put( 90,10){\circle*{5}}
      \put(120,10){\circle*{5}}
      \put(135,10){\circle*{5}}
      \put(165,10){\circle*{5}}
      \put(180,10){\circle*{5}}
      \put(210,10){\circle*{5}}
      \put(225,10){\circle*{5}}
      \put(255,10){\circle*{5}}
      \put(270,10){\circle*{5}}
      \put(300,10){\circle*{5}}
      \put(315,10){\circle*{5}}
      \put(345,10){\circle*{5}}

       \put(-40,82){\makebox(0,0){$A$}}
       \put(  0,82){\makebox(0,0){$\hd{\T{6}}$}}
       \put( 30,82){\makebox(0,0){$\tl{\T{6}}$}}
       \put( 45,82){\makebox(0,0){$\tl{\T{1}}$}}
       \put( 75,82){\makebox(0,0){$\hd{\T{1}}$}}
       \put( 90,82){\makebox(0,0){$\tl{\T{5}}$}}
       \put(120,82){\makebox(0,0){$\hd{\T{5}}$}}
       \put(135,82){\makebox(0,0){$\tl{\T{3}}$}}
       \put(165,82){\makebox(0,0){$\hd{\T{3}}$}}
       \put(180,82){\makebox(0,0){$\tl{\T{4}}$}}
       \put(210,82){\makebox(0,0){$\hd{\T{4}}$}}
       \put(225,82){\makebox(0,0){$\tl{\T{2}}$}}
       \put(255,82){\makebox(0,0){$\hd{\T{2}}$}}
       \put(270,82){\makebox(0,0){$\tl{\T{8}}$}}
       \put(300,82){\makebox(0,0){$\hd{\T{8}}$}}
       \put(315,82){\makebox(0,0){$\tl{\T{9}}$}}
       \put(345,82){\makebox(0,0){$\hd{\T{9}}$}}

       \put(-40,-2){\makebox(0,0){$B$}}
       \put(  0,-2){\makebox(0,0){$\hd{\T{6}}$}}
       \put( 30,-2){\makebox(0,0){$\tl{\T{6}}$}}
       \put( 45,-2){\makebox(0,0){$\tl{\T{5}}$}}
       \put( 75,-2){\makebox(0,0){$\hd{\T{5}}$}}
       \put( 90,-2){\makebox(0,0){$\hd{\T{3}}$}}
       \put(120,-2){\makebox(0,0){$\tl{\T{3}}$}}
       \put(135,-2){\makebox(0,0){$\tl{\T{4}}$}}
       \put(165,-2){\makebox(0,0){$\hd{\T{4}}$}}
       \put(180,-2){\makebox(0,0){$\hd{\T{7}}$}}
       \put(210,-2){\makebox(0,0){$\tl{\T{7}}$}}
       \put(225,-2){\makebox(0,0){$\tl{\T{2}}$}}
       \put(255,-2){\makebox(0,0){$\hd{\T{2}}$}}
       \put(270,-2){\makebox(0,0){$\tl{\T{9}}$}}
       \put(300,-2){\makebox(0,0){$\hd{\T{9}}$}}
       \put(315,-2){\makebox(0,0){$\tl{\T{8}}$}}
       \put(345,-2){\makebox(0,0){$\hd{\T{8}}$}}

    \end{picture}
   \end{center}
   \caption{
   For genomes $A=\{\rev{\T{6}}\z\T{1}\z\T{5}\z\T{3}\z\T{4}, \T{2}\z\T{8}\z\T{9} \}$ and $B=\{\rev{\T{6}}\z\T{5}\z\rev{\T{3}}\z\T{4}\z\rev{\T{7}}\z\T{2}, \T{9}\z\T{8}\}$, the relational diagram contains two cycles, two $\AB$-paths (represented in blue), one $\AA$-path and one $\BB$-path (both represented in red). Short dotted horizontal edges are adjacency edges, long horizontal edges are indel edges, top-down edges are extremity edges.}
   \label{fig:bp-adj-graph}
 \end{figure}
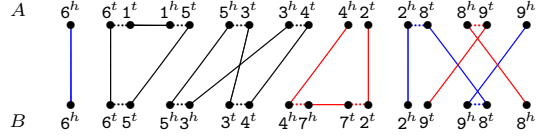

Each vertex has degree one or two: it is connected either to an extremity edge or to an indel edge, and to at most one adjacency edge, therefore~$R(A, B)$ is a simple collection of cycles and paths.
A path that has one endpoint 
in genome~$A$ and the other in genome~$B$ is called an {\em $\AB$-path}. In the same 
way, both endpoints of an {\em $\AA$-path} are in~$A$ and both 
endpoints of
a {\em $\BB$-path} are in~$B$. A cycle contains either zero or an even number of extremity edges. When a cycle has at least two extremity edges, it is called an {\em $\AB$-cycle}.
Moreover, a path (respectively cycle) of~$R(A, B)$ composed exclusively of indel and adjacency edges in one of the two genomes corresponds to a whole linear (respectively circular) chromosome and is called a {\em linear} (respectively {\em circular}) {\em
singleton} in that genome. Actually, linear singletons are particular cases of~$\AA$- or~$\BB$-paths. 
The numbers of telomeres and of $\AB$-paths in $R(A, B)$ are even. 
An example of a relational diagram is given in
Figure~\ref{fig:bp-adj-graph}.


\paragraph*{DCJ distance of canonical genomes:} When singular genomes $A$ and $B$ have no exclusive markers, that is,
$\maa = \mbb =\emptyset$, they are said to be \emph{canonical}. In this case $A$ can be sorted into $B$ with DCJ operations only and their DCJ distance $\ddcj$ can be computed as follows~\cite{BER-MIX-STO-2006}:
\[
\ddcj(A,B) \;=\; |\mg| - c -
\frac{i}{2}\,,
\]
where $c$ is the number of $\AB$-cycles and $i$ is the number of $\AB$-paths in $R(A,B)$.

\paragraph*{Runs and indel-potential:}
When singular genomes $A$ and $B$ have exclusive markers, it is possible to optimally select DCJ operations that group exclusive markers together for minimizing indels~\cite{BRA-WIL-STO-2011}, as follows.

Given two genomes~$A$ and~$B$ and a component~$C$ of~$R(A,B)$, a \emph{run}~\cite{BRA-WIL-STO-2011} is a maximal subpath of~$C$, in which the first and the last edges are indel edges, and all indel edges belong to the same genome. 
It can be an $\ma$-run when its indel edges are in genome~$A$, or a $\mb$-run when its indel edges are in genome~$B$. 
We denote by $\Lambda(C)$ the number of runs in component~$C$. If $\Lambda(C)\geq 1$ the component $C$ is said to be \emph{indel-enclosing}, otherwise $\Lambda(C)=0$ and $C$ is said to be \emph{indel-free}.
The \emph{indel-potential} of a component~$C$, denoted by~$\lambda(C)$, is
the optimal number of indels obtained after ``sorting'' $C$ separately
and can be directly computed from~$\Lambda(C)$~\cite{BRA-WIL-STO-2011}:
\[
\lambda(C) = 
\begin{cases}
~~~~~0\:, & \mbox{ if $\Lambda(C)=0$~~~($C$ is indel-free)}; \\[1mm]
\left\lceil \frac{\Lambda(C)+1}{2}\right\rceil\:, & \mbox{ if $\Lambda(C) \geq 1$~~~($C$ is indel-enclosing)}.
\end{cases}
\] 

Figure~\ref{fig:run} (Appendix~\ref{app:figures}) shows a $\BB$-path with 4 runs, and how its indel-potential can be achieved.
With the indel-potential, an upper bound for the DCJ-indel distance was established~\cite{BRA-WIL-STO-2011}:
\begin{equation}\label{eq:distance-upper}
\ddcjid(A,B) \;\leq\; |\mg| - c -
\frac{i}{2}~+ \!\!\!\! \sum_{C \in R(A,B)} \!\!\!\!\!\! \lambda(C)\\
\end{equation}
\paragraph*{DCJ-indel distance of singular circular genomes:}
For singular circular genomes, the graph $R(A,B)$ is composed of cycles only. 
In this case the upper bound given by Equation~(\ref{eq:distance-upper}) is tight and leads to a simplified formula~\cite{BRA-WIL-STO-2011}:
\[
\ddcjid(A,B) \;=\; |\mg| - c~+ \!\!\!\! \sum_{C \in R(A,B)} \!\!\!\!\!\! \lambda(C)\,.
\]

\paragraph*{DCJ-indel distance of singular linear genomes:}
For singular linear genomes,
the upper bound given by Equation~(\ref{eq:distance-upper}) is achieved
when the components of $R(A,B)$ are sorted separately. However, it can be decreased by {\em recombinations}, that are DCJ operations
that act on two distinct paths of $R(A,B)$.
Such path recombinations are said to be \emph{deducting}.
The total number of types of deducting recombinations is relatively small. By
exhaustively exploring the space of recombination types, it is possible to
identify groups of chained recombinations~\cite{BRA-WIL-STO-2011}, so that the
sources of each group are the original paths of the graph. In other words, a
path that is a resultant of a group is never a source of another group. This
results in a greedy approach (detailed in~\cite{BRA-WIL-STO-2011}) that
optimally finds the value $\delta \geq 0$ to be deducted. We then have the following exact formula~\cite{BRA-WIL-STO-2011}:
\[
\ddcjid(A,B) \;=\; |\mg| - c -
\frac{i}{2}~+ \!\!\!\! \sum_{C \in R(A,B)} \!\!\!\!\!\! \lambda(C) ~~- \delta\,.
\]

\section{The family-free setting}\label{sec:ff}

As already stated, in the family-free setting, each marker in each genome is represented by
a distinct symbol, thus $\ma \cap \mb = \emptyset$. Observe that the cardinalities
$|\ma|$ and $|\mb|$ may be distinct.

\subsection{Marker similarity graph for the family-free setting}

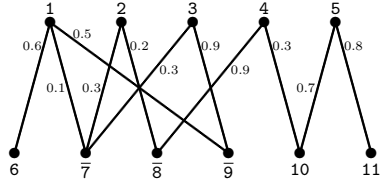
\begin{figure}[hbt]
  \centering
  \setlength{\unitlength}{.45pt}
    \begin{picture}(300,140)
      \scriptsize

      \thicklines
      \qbezier(30,120)(15,65)(0,10)
      \put(15,100){\makebox(0,0){\scalebox{.7}{0.6}}}
      \qbezier(30,120)(45,65)(60,10)
      \put(35,65){\makebox(0,0){\scalebox{.7}{0.1}}}
      \qbezier(30,120)(105,65)(180,10)
      \put(57,110){\makebox(0,0){\scalebox{.7}{0.5}}}
      \qbezier(90,120)(75,65)(60,10)
      \put(65,65){\makebox(0,0){\scalebox{.7}{0.3}}}
      \qbezier(90,120)(105,65)(120,10)
      \put(105,100){\makebox(0,0){\scalebox{.7}{0.2}}}
      \qbezier(150,120)(105,65)(60,10)
      \put(130,80){\makebox(0,0){\scalebox{.7}{0.3}}}
      \qbezier(150,120)(165,65)(180,10)
      \put(165,100){\makebox(0,0){\scalebox{.7}{0.9}}}
      \qbezier(210,120)(165,65)(120,10)
      \put(190,80){\makebox(0,0){\scalebox{.7}{0.9}}}
      \qbezier(210,120)(225,65)(240,10)
      \put(225,100){\makebox(0,0){\scalebox{.7}{0.3}}}
      \qbezier(270,120)(255,65)(240,10)
      \put(245,65){\makebox(0,0){\scalebox{.7}{0.7}}}
      \qbezier(270,120)(285,65)(300,10)
      \put(285,100){\makebox(0,0){\scalebox{.7}{0.8}}}
      
      \put( 30,120){\circle*{8}}
      \put( 90,120){\circle*{8}}
      \put(150,120){\circle*{8}}
      \put(210,120){\circle*{8}}
      \put(270,120){\circle*{8}}

      \put(  0,10){\circle*{8}}
      \put( 60,10){\circle*{8}}
      \put(120,10){\circle*{8}}
      \put(180,10){\circle*{8}}
      \put(240,10){\circle*{8}}
      \put(300,10){\circle*{8}}

      \put( 30,132){\makebox(0,0){$\T{1}$}}
      \put( 90,132){\makebox(0,0){$\T{2}$}}
      \put(150,132){\makebox(0,0){$\T{3}$}}
      \put(210,132){\makebox(0,0){$\T{4}$}}
      \put(270,132){\makebox(0,0){$\T{5}$}}
      
      \put(  0,-4){\makebox(0,0){$\T{6}$}}
      \put( 60,-4){\makebox(0,0){$\rev{\T{7}}$}}
      \put(120,-4){\makebox(0,0){$\rev{\T{8}}$}}
      \put(180,-4){\makebox(0,0){$\rev{\T{9}}$}}
      \put(240,-4){\makebox(0,0){$\T{10}$}}
      \put(300,-4){\makebox(0,0){$\T{11}$}}
      
\end{picture}
  \caption{Graph $\GSx{0.1}(A,B)$ for the two 
  genomes 
  $A=\{\T{1}\z\z\T{2}\z\z\T{3}\z\z\T{4}\z\z\T{5}\}$
  and $B=\{\T{6}\z\z\rev{\T{7}}\z\z\rev{\T{8}}\z\z\rev{\T{9}}\z\z\T{10}\z\z\T{11}\}$.}  
  \label{fig:example-gsg}
\end{figure}

Given a threshold $0 \leq x \leq 1$, we can represent the similarities between the markers of genome $A$ and
the markers of genome $B$ in the so called
\emph{marker similarity graph}~\cite{BCDJSTW13}, denoted by $\GSx{x}(A,
B)$. This is a weighted bipartite graph whose partitions $\A$ and $\B$
are the sets of markers in genomes $A$ and $B$,
respectively. 
Furthermore, for each pair of markers $a\in\ma$ and $b\in\mb$, denote by $\sigma(a, b)$ their \emph{normalized similarity}, a value that ranges in the interval $[0, 1]$. If $\sigma(a, b) \geq x$ there is an edge $e$
connecting $a$ and $b$ in $\GSx{x}(A,B)$ whose weight is $\sigma(e) :=
\sigma(a, b)$.
An example is given in Figure~\ref{fig:example-gsg}. 

\paragraph*{Mapped genomes:}
Let $A$ and $B$ be two genomes with marker
similarity graph $\GSx{x}(A,B)$ and let $M =
\{e_1, e_2, \ldots, e_n\}$ be a matching in $\GSx{x}(A,B)$. Since the endpoints of each edge $e_i = (a, b)$ in
$M$ are not saturated by any other edge of $M$, we can unambiguously
define the function $s(a,M) = s(b,M) = i$. 
We then define the set of \emph{$M$-saturated mapped markers} $\G(M) = \{ s(g,M) \colon g \mbox{ is $M$-saturated }\}=\{1,2,\ldots,n\}$.

Let $\tilde{n}_A$ be the number of unsaturated markers in $\ma$ and $\tilde{n}_B$ be the
number of unsaturated markers in $\mb$. We extend the function $s$, so that it maps each unsaturated marker $a'\in \ma$ to one value in $\{n+1, n+2, \ldots, n+\tilde{n}_A\}$ and each unsaturated marker $b' \in \mb$ to one value in $\{n+\tilde{n}_A+1, n+\tilde{n}_A+2, \ldots, n+\tilde{n}_A+\tilde{n}_B\}$. The sets of \emph{$M$-unsaturated mapped markers} are:
\begin{itemize}
\item $\maa(M) = \{ s(a',M) \colon a' \in \ma \mbox{ is $M$-unsaturated }
\}=\{n+1,n+2,\ldots,n+\tilde{n}_A\}$ and
\item $\mbb(M) = \{ s(b',M) \colon b' \in \mb \mbox{ is $M$-unsaturated }\}=\{n+\tilde{n}_A+1,n+\tilde{n}_A+2,\ldots,n+\tilde{n}_A+\tilde{n}_B\}$.
\end{itemize}

The \emph{mapped genomes} $A^M$ and $B^M$ are then obtained by renaming each marker~$a \in \ma$ to~$s(a,M)$ and each marker $b\in \mb$ to $s(b,M)$, preserving all orientations. 

\paragraph*{Established distances of mapped genomes:}
Let the relational graph $R(A^M,B^M)$ have~$c_M$ $\AB$-cycles and $i_M$ $\AB$-paths. By simply ignoring the exclusive markers of $\maa(M)$ and $\mbb(M)$, we can compute the DCJ distance:
\[
\ddcj(A^M, B^M) = |M| -c_M -\frac{i_M}{2}\,.
\]

Taking into consideration the weight of the matching $M$ defined as $w(M) = \sum_{e \in M} \sigma(e)$, we can also compute the weighted DCJ distance $\wcost(A^M, B^M)$~\cite{MAR-FEI-BRA-STO-15}:
\[
\wcost(A^M, B^M) = \ddcj(A^M, B^M) + |M| - w(M)\,.
\]
Observe that, when all edges of $M$ have the maximum weight 1, we have $w(M)=|M|$ and $\wcost(A^M, B^M) = \ddcj(A^M, B^M)$.

Finally, taking into consideration the exclusive markers of $\maa(M)$ and $\mbb(M)$, but not the weight $w(M)$, we can compute the DCJ-indel distance of mapped genomes $A^M$ and $B^M$:
\begin{align*}
\ddcjid(A^M,B^M)
&= |M| -c_M -\frac{i_M}{2} + \!\!\!\! \sum_{C \in R(A^M,B^M)} \!\!\!\!\!\!\!\! \lambda(C) ~~- \delta_M\,,
\end{align*}
where 
$\delta_M$ is the deduction given by path recombinations in $R(A^M,B^M)$.

\subsection{The family-free DCJ-indel distance}\label{sec:ff-dcj-indel-dist}

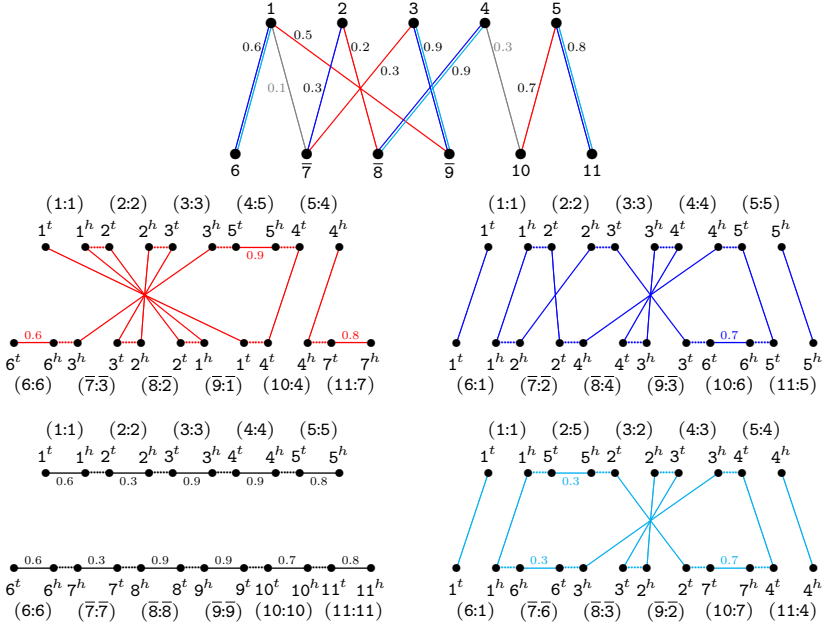
\begin{figure}[hbt]
  \centering
  
\setlength{\unitlength}{.45pt}
    \begin{picture}(300,140)
 \scriptsize
     
      \thinlines
      \put(0,0){\color{blue}
      \qbezier(29,120)(14,65)(-1,10)
      }
      \put(14,100){\makebox(0,0){\scalebox{.7}{0.6}}}
      \put(0,0){\color{cyan}
      \qbezier(31,120)(16,65)(1,10)
      }
      \put(0,0){\color{gray}
      \qbezier(30,120)(45,65)(60,10)
      \put(35,65){\makebox(0,0){\scalebox{.7}{0.1}}}
      }
      \put(0,0){\color{red}
      \qbezier(30,120)(105,65)(180,10)
      }
      \put(57,110){\makebox(0,0){\scalebox{.7}{0.5}}}
      \put(0,0){\color{blue}
      \qbezier(90,120)(75,65)(60,10)
      }
      \put(65,65){\makebox(0,0){\scalebox{.7}{0.3}}}
      \put(0,0){\color{red}
      \qbezier(90,120)(105,65)(120,10)
      \qbezier(150,120)(105,65)(60,10)
      }
      \put(105,100){\makebox(0,0){\scalebox{.7}{0.2}}}
      \put(130,80){\makebox(0,0){\scalebox{.7}{0.3}}}
      \put(0,0){\color{blue}
      \qbezier(149,120)(164,65)(179,10)
      \qbezier(208,120)(163,65)(118,10)
      }
      \put(166,100){\makebox(0,0){\scalebox{.7}{0.9}}}
      \put(190,80){\makebox(0,0){\scalebox{.7}{0.9}}}
      \put(0,0){\color{cyan}
      \qbezier(151,120)(166,65)(181,10)
      \qbezier(211,120)(166,65)(121,10)
      }
      \put(0,0){\color{gray}
      \qbezier(210,120)(225,65)(240,10)
      \put(225,100){\makebox(0,0){\scalebox{.7}{0.3}}}
      }
      \put(0,0){\color{red}
      \qbezier(270,120)(255,65)(240,10)
      }
      \put(245,65){\makebox(0,0){\scalebox{.7}{0.7}}}
      \put(0,0){\color{blue}
      \qbezier(269,120)(284,65)(299,10)
      }
      \put(287,100){\makebox(0,0){\scalebox{.7}{0.8}}}
      \put(0,0){\color{cyan}
      \qbezier(271,120)(286,65)(301,10)
      }
      
      \put( 30,120){\circle*{8}}
      \put( 90,120){\circle*{8}}
      \put(150,120){\circle*{8}}
      \put(210,120){\circle*{8}}
      \put(270,120){\circle*{8}}

      \put(  0,10){\circle*{8}}
      \put( 60,10){\circle*{8}}
      \put(120,10){\circle*{8}}
      \put(180,10){\circle*{8}}
      \put(240,10){\circle*{8}}
      \put(300,10){\circle*{8}}

      \put( 30,132){\makebox(0,0){$\T{1}$}}
      \put( 90,132){\makebox(0,0){$\T{2}$}}
      \put(150,132){\makebox(0,0){$\T{3}$}}
      \put(210,132){\makebox(0,0){$\T{4}$}}
      \put(270,132){\makebox(0,0){$\T{5}$}}
      
      \put(  0,-4){\makebox(0,0){$\T{6}$}}
      \put( 60,-4){\makebox(0,0){$\rev{\T{7}}$}}
      \put(120,-4){\makebox(0,0){$\rev{\T{8}}$}}
      \put(180,-4){\makebox(0,0){$\rev{\T{9}}$}}
      \put(240,-4){\makebox(0,0){$\T{10}$}}
      \put(300,-4){\makebox(0,0){$\T{11}$}}
      
\end{picture}

  \bigskip
  

    \scriptsize
    \setlength{\unitlength}{.6pt}
    \begin{tabular}{c@{~~~~~~~~~~~~}c}    
    \begin{picture}(225,100)

      \put(0,0){\color{red}
        \dottedline[$\cdot$]{2}(25,10)(40,10)
        \dottedline[$\cdot$]{2}(65,10)(80,10)
        \dottedline[$\cdot$]{2}(85,70)(100,70)
        \dottedline[$\cdot$]{2}(145,10)(160,10)
        
        \qbezier(20,70)(82,40)(145,10)
        \qbezier(45,70)(82,40)(120,10)
        \qbezier(60,70)(82,40)(105,10)
        \qbezier(85,70)(82,40)(80,10)
        \qbezier(100,70)(82,40)(65,10)
        \qbezier(125,70)(82,40)(40,10)
        \qbezier(180,70)(170,40)(160,10)
        \qbezier(205,70)(195,40)(185,10)

        \drawline(140,70)(165,70)
        \put(152,65){\makebox(0,0){\scalebox{.65}{0.9}}}
        \drawline(0,10)(25,10)
        \put(12,15){\makebox(0,0){\scalebox{.65}{0.6}}}
        \drawline(200,10)(225,10)
        \put(212,15){\makebox(0,0){\scalebox{.65}{0.8}}}
        
      \dottedline[$\cdot$]{2}(45,70)(60,70)
      \dottedline[$\cdot$]{2}(125,70)(140,70)
      \dottedline[$\cdot$]{2}(165,70)(180,70)
      
      \dottedline[$\cdot$]{2}(105,10)(120,10)
      \dottedline[$\cdot$]{2}(185,10)(200,10)
      
      }

      \put( 20,70){\circle*{5}}
      \put( 45,70){\circle*{5}}
      \put( 60,70){\circle*{5}}
      \put( 85,70){\circle*{5}}
      \put(100,70){\circle*{5}}
      \put(125,70){\circle*{5}}
      \put(140,70){\circle*{5}}
      \put(165,70){\circle*{5}}
      \put(180,70){\circle*{5}}
      \put(205,70){\circle*{5}}
      
      \put( 20,82){\makebox(0,0){$\tl{\T{1}}$}}
      \put( 45,82){\makebox(0,0){$\hd{\T{1}}$}}
      \put( 60,82){\makebox(0,0){$\tl{\T{2}}$}}
      \put( 85,82){\makebox(0,0){$\hd{\T{2}}$}}
      \put(100,82){\makebox(0,0){$\tl{\T{3}}$}}
      \put(125,82){\makebox(0,0){$\hd{\T{3}}$}}
      \put(140,82){\makebox(0,0){$\tl{\T{5}}$}}
      \put(165,82){\makebox(0,0){$\hd{\T{5}}$}}
      \put(180,82){\makebox(0,0){$\tl{\T{4}}$}}
      \put(205,82){\makebox(0,0){$\hd{\T{4}}$}}
      
      \put( 32,97){\makebox(0,0){$(\T{1}\!\!:\!\!\T{1})$}}
      \put( 72,97){\makebox(0,0){$(\T{2}\!\!:\!\!\T{2})$}}
      \put(112,97){\makebox(0,0){$(\T{3}\!\!:\!\!\T{3})$}}
      \put(152,97){\makebox(0,0){$(\T{4}\!\!:\!\!\T{5})$}}
      \put(192,97){\makebox(0,0){$(\T{5}\!\!:\!\!\T{4})$}}

      \put(  0,10){\circle*{5}}
      \put( 25,10){\circle*{5}}
      \put( 40,10){\circle*{5}}
      \put( 65,10){\circle*{5}}
      \put( 80,10){\circle*{5}}
      \put(105,10){\circle*{5}}
      \put(120,10){\circle*{5}}
      \put(145,10){\circle*{5}}
      \put(160,10){\circle*{5}}
      \put(185,10){\circle*{5}}
      \put(200,10){\circle*{5}}
      \put(225,10){\circle*{5}}
      
      \put( 0,-2){\makebox(0,0){$\tl{\T{6}}$}}
      \put( 25,-2){\makebox(0,0){$\hd{\T{6}}$}}
      \put( 40,-2){\makebox(0,0){$\hd{\T{3}}$}}
      \put( 65,-2){\makebox(0,0){$\tl{\T{3}}$}}
      \put( 80,-2){\makebox(0,0){$\hd{\T{2}}$}}
      \put(105,-2){\makebox(0,0){$\tl{\T{2}}$}}
      \put(120,-2){\makebox(0,0){$\hd{\T{1}}$}}
      \put(145,-2){\makebox(0,0){$\tl{\T{1}}$}}
      \put(160,-2){\makebox(0,0){$\tl{\T{4}}$}}
      \put(185,-2){\makebox(0,0){$\hd{\T{4}}$}}
      \put(200,-2){\makebox(0,0){$\tl{\T{7}}$}}
      \put(225,-2){\makebox(0,0){$\hd{\T{7}}$}}
      
      \put( 12,-17){\makebox(0,0){$(\T{6}\!\!:\!\!\T{6})$}}
      \put( 52,-17){\makebox(0,0){$(\rev{\T{7}}\!\!:\!\!\rev{\T{3}})$}}
      \put( 92,-17){\makebox(0,0){$(\rev{\T{8}}\!\!:\!\!\rev{\T{2}})$}}
      \put(132,-17){\makebox(0,0){$(\rev{\T{9}}\!\!:\!\!\rev{\T{1}})$}}
      \put(172,-17){\makebox(0,0){$(\T{10}\!\!:\!\!\T{4})$}}
      \put(212,-17){\makebox(0,0){$(\T{11}\!\!:\!\!\T{7})$}}
      
    \end{picture} &
    \begin{picture}(225,100)

      \put(0,0){\color{blue}
        \dottedline[$\cdot$]{2}(45,70)(60,70)
        \dottedline[$\cdot$]{2}(25,10)(40,10)
        \dottedline[$\cdot$]{2}(65,10)(80,10)
        \dottedline[$\cdot$]{2}(125,70)(140,70)
        \dottedline[$\cdot$]{2}(145,10)(160,10)
        
        \qbezier(20,70)(10,40)(0,10)
        \qbezier(45,70)(35,40)(25,10)
        \qbezier(60,70)(62,40)(65,10)
        \qbezier(85,70)(62,40)(40,10)
        \qbezier(100,70)(122,40)(145,10)
        \qbezier(125,70)(122,40)(120,10)
        \qbezier(140,70)(122,40)(105,10)
        \qbezier(165,70)(122,40)(80,10)
        \qbezier(180,70)(190,40)(200,10)
        \qbezier(205,70)(215,40)(225,10)

        \drawline(160,10)(185,10)
        \put(172,15){\makebox(0,0){\scalebox{.65}{0.7}}}
        
      \dottedline[$\cdot$]{2}(85,70)(100,70)
      \dottedline[$\cdot$]{2}(165,70)(180,70)

      \dottedline[$\cdot$]{2}(105,10)(120,10)
      \dottedline[$\cdot$]{2}(185,10)(200,10)
      
      }

      \put( 20,70){\circle*{5}}
      \put( 45,70){\circle*{5}}
      \put( 60,70){\circle*{5}}
      \put( 85,70){\circle*{5}}
      \put(100,70){\circle*{5}}
      \put(125,70){\circle*{5}}
      \put(140,70){\circle*{5}}
      \put(165,70){\circle*{5}}
      \put(180,70){\circle*{5}}
      \put(205,70){\circle*{5}}
      
      \put( 20,82){\makebox(0,0){$\tl{\T{1}}$}}
      \put( 45,82){\makebox(0,0){$\hd{\T{1}}$}}
      \put( 60,82){\makebox(0,0){$\tl{\T{2}}$}}
      \put( 85,82){\makebox(0,0){$\hd{\T{2}}$}}
      \put(100,82){\makebox(0,0){$\tl{\T{3}}$}}
      \put(125,82){\makebox(0,0){$\hd{\T{3}}$}}
      \put(140,82){\makebox(0,0){$\tl{\T{4}}$}}
      \put(165,82){\makebox(0,0){$\hd{\T{4}}$}}
      \put(180,82){\makebox(0,0){$\tl{\T{5}}$}}
      \put(205,82){\makebox(0,0){$\hd{\T{5}}$}}
      
      \put( 32,97){\makebox(0,0){$(\T{1}\!\!:\!\!\T{1})$}}
      \put( 72,97){\makebox(0,0){$(\T{2}\!\!:\!\!\T{2})$}}
      \put(112,97){\makebox(0,0){$(\T{3}\!\!:\!\!\T{3})$}}
      \put(152,97){\makebox(0,0){$(\T{4}\!\!:\!\!\T{4})$}}
      \put(192,97){\makebox(0,0){$(\T{5}\!\!:\!\!\T{5})$}}
      
      \put(  0,10){\circle*{5}}
      \put( 25,10){\circle*{5}}
      \put( 40,10){\circle*{5}}
      \put( 65,10){\circle*{5}}
      \put( 80,10){\circle*{5}}
      \put(105,10){\circle*{5}}
      \put(120,10){\circle*{5}}
      \put(145,10){\circle*{5}}
      \put(160,10){\circle*{5}}
      \put(185,10){\circle*{5}}
      \put(200,10){\circle*{5}}
      \put(225,10){\circle*{5}}
      
      \put( 0,-2){\makebox(0,0){$\tl{\T{1}}$}}
      \put( 25,-2){\makebox(0,0){$\hd{\T{1}}$}}
      \put( 40,-2){\makebox(0,0){$\hd{\T{2}}$}}
      \put( 65,-2){\makebox(0,0){$\tl{\T{2}}$}}
      \put( 80,-2){\makebox(0,0){$\hd{\T{4}}$}}
      \put(105,-2){\makebox(0,0){$\tl{\T{4}}$}}
      \put(120,-2){\makebox(0,0){$\hd{\T{3}}$}}
      \put(145,-2){\makebox(0,0){$\tl{\T{3}}$}}
      \put(160,-2){\makebox(0,0){$\tl{\T{6}}$}}
      \put(185,-2){\makebox(0,0){$\hd{\T{6}}$}}
      \put(200,-2){\makebox(0,0){$\tl{\T{5}}$}}
      \put(225,-2){\makebox(0,0){$\hd{\T{5}}$}}
      
      \put( 12,-17){\makebox(0,0){$(\T{6}\!\!:\!\!\T{1})$}}
      \put( 52,-17){\makebox(0,0){$(\rev{\T{7}}\!\!:\!\!\rev{\T{2}})$}}
      \put( 92,-17){\makebox(0,0){$(\rev{\T{8}}\!\!:\!\!\rev{\T{4}})$}}
      \put(132,-17){\makebox(0,0){$(\rev{\T{9}}\!\!:\!\!\rev{\T{3}})$}}
      \put(172,-17){\makebox(0,0){$(\T{10}\!\!:\!\!\T{6})$}}
      \put(212,-17){\makebox(0,0){$(\T{11}\!\!:\!\!\T{5})$}}
      
    \end{picture}\\[8mm]
    \begin{picture}(225,100)

      \put(0,0){\color{black}
        \dottedline[$\cdot$]{2}(45,70)(60,70)
        \dottedline[$\cdot$]{2}(25,10)(40,10)
        \dottedline[$\cdot$]{2}(65,10)(80,10)
        \dottedline[$\cdot$]{2}(125,70)(140,70)
        \dottedline[$\cdot$]{2}(145,10)(160,10)
        
        \drawline(20,70)(45,70)
        \put(32,65){\makebox(0,0){\scalebox{.65}{0.6}}}
        \drawline(60,70)(85,70)
        \put(72,65){\makebox(0,0){\scalebox{.65}{0.3}}}
        \drawline(100,70)(125,70)
        \put(112,65){\makebox(0,0){\scalebox{.65}{0.9}}}
        \drawline(140,70)(165,70)
        \put(152,65){\makebox(0,0){\scalebox{.65}{0.9}}}
        \drawline(180,70)(205,70)
        \put(192,65){\makebox(0,0){\scalebox{.65}{0.8}}}
        
        \drawline(0,10)(25,10)
        \put(12,15){\makebox(0,0){\scalebox{.65}{0.6}}}
        \drawline(40,10)(65,10)
        \put(52,15){\makebox(0,0){\scalebox{.65}{0.3}}}
        \drawline(80,10)(105,10)
        \put(92,15){\makebox(0,0){\scalebox{.65}{0.9}}}
        \drawline(120,10)(145,10)
        \put(132,15){\makebox(0,0){\scalebox{.65}{0.9}}}
        \drawline(160,10)(185,10)
        \put(172,15){\makebox(0,0){\scalebox{.65}{0.7}}}
        \drawline(200,10)(225,10)
        \put(212,15){\makebox(0,0){\scalebox{.65}{0.8}}}
        
      \dottedline[$\cdot$]{2}(85,70)(100,70)
      \dottedline[$\cdot$]{2}(165,70)(180,70)

      \dottedline[$\cdot$]{2}(105,10)(120,10)
      \dottedline[$\cdot$]{2}(185,10)(200,10)
      
      }

      \put( 20,70){\circle*{5}}
      \put( 45,70){\circle*{5}}
      \put( 60,70){\circle*{5}}
      \put( 85,70){\circle*{5}}
      \put(100,70){\circle*{5}}
      \put(125,70){\circle*{5}}
      \put(140,70){\circle*{5}}
      \put(165,70){\circle*{5}}
      \put(180,70){\circle*{5}}
      \put(205,70){\circle*{5}}
      
      \put( 20,82){\makebox(0,0){$\tl{\T{1}}$}}
      \put( 45,82){\makebox(0,0){$\hd{\T{1}}$}}
      \put( 60,82){\makebox(0,0){$\tl{\T{2}}$}}
      \put( 85,82){\makebox(0,0){$\hd{\T{2}}$}}
      \put(100,82){\makebox(0,0){$\tl{\T{3}}$}}
      \put(125,82){\makebox(0,0){$\hd{\T{3}}$}}
      \put(140,82){\makebox(0,0){$\tl{\T{4}}$}}
      \put(165,82){\makebox(0,0){$\hd{\T{4}}$}}
      \put(180,82){\makebox(0,0){$\tl{\T{5}}$}}
      \put(205,82){\makebox(0,0){$\hd{\T{5}}$}}
      
      \put( 32,97){\makebox(0,0){$(\T{1}\!\!:\!\!\T{1})$}}
      \put( 72,97){\makebox(0,0){$(\T{2}\!\!:\!\!\T{2})$}}
      \put(112,97){\makebox(0,0){$(\T{3}\!\!:\!\!\T{3})$}}
      \put(152,97){\makebox(0,0){$(\T{4}\!\!:\!\!\T{4})$}}
      \put(192,97){\makebox(0,0){$(\T{5}\!\!:\!\!\T{5})$}}
      
      \put(  0,10){\circle*{5}}
      \put( 25,10){\circle*{5}}
      \put( 40,10){\circle*{5}}
      \put( 65,10){\circle*{5}}
      \put( 80,10){\circle*{5}}
      \put(105,10){\circle*{5}}
      \put(120,10){\circle*{5}}
      \put(145,10){\circle*{5}}
      \put(160,10){\circle*{5}}
      \put(185,10){\circle*{5}}
      \put(200,10){\circle*{5}}
      \put(225,10){\circle*{5}}
      
      \put( 0,-2){\makebox(0,0){$\tl{\T{6}}$}}
      \put( 25,-2){\makebox(0,0){$\hd{\T{6}}$}}
      \put( 40,-2){\makebox(0,0){$\hd{\T{7}}$}}
      \put( 65,-2){\makebox(0,0){$\tl{\T{7}}$}}
      \put( 80,-2){\makebox(0,0){$\hd{\T{8}}$}}
      \put(105,-2){\makebox(0,0){$\tl{\T{8}}$}}
      \put(120,-2){\makebox(0,0){$\hd{\T{9}}$}}
      \put(145,-2){\makebox(0,0){$\tl{\T{9}}$}}
      \put(160,-2){\makebox(0,0){$\tl{\T{10}}$}}
      \put(184,-2){\makebox(0,0){$\hd{\T{10}}$}}
      \put(202,-2){\makebox(0,0){$\tl{\T{11}}$}}
      \put(225,-2){\makebox(0,0){$\hd{\T{11}}$}}
      
      \put( 12,-17){\makebox(0,0){$(\T{6}\!\!:\!\!\T{6})$}}
      \put( 52,-17){\makebox(0,0){$(\rev{\T{7}}\!\!:\!\!\rev{\T{7}})$}}
      \put( 92,-17){\makebox(0,0){$(\rev{\T{8}}\!\!:\!\!\rev{\T{8}})$}}
      \put(132,-17){\makebox(0,0){$(\rev{\T{9}}\!\!:\!\!\rev{\T{9}})$}}
      \put(170,-17){\makebox(0,0){$(\T{10}\!\!:\!\!\T{10})$}}
      \put(214,-17){\makebox(0,0){$(\T{11}\!\!:\!\!\T{11})$}}
      
    \end{picture}
    &
    \begin{picture}(225,100)

      \put(0,0){\color{cyan}
        \dottedline[$\cdot$]{2}(45,70)(60,70)
        \dottedline[$\cdot$]{2}(25,10)(40,10)
        \dottedline[$\cdot$]{2}(65,10)(80,10)
        \dottedline[$\cdot$]{2}(125,70)(140,70)
        \dottedline[$\cdot$]{2}(145,10)(160,10)
        
        \qbezier(20,70)(10,40)(0,10)
        \qbezier(45,70)(35,40)(25,10)
        \qbezier(100,70)(122,40)(145,10)
        \qbezier(125,70)(122,40)(120,10)
        \qbezier(140,70)(122,40)(105,10)
        \qbezier(165,70)(122,40)(80,10)
        \qbezier(180,70)(190,40)(200,10)
        \qbezier(205,70)(215,40)(225,10)
        
        \drawline(60,70)(85,70)
        \put(72,65){\makebox(0,0){\scalebox{.65}{0.3}}}
        \drawline(40,10)(65,10)
        \put(52,15){\makebox(0,0){\scalebox{.65}{0.3}}}
        \drawline(160,10)(185,10)
        \put(172,15){\makebox(0,0){\scalebox{.65}{0.7}}}
        
      \dottedline[$\cdot$]{2}(85,70)(100,70)
      \dottedline[$\cdot$]{2}(165,70)(180,70)
      
      \dottedline[$\cdot$]{2}(105,10)(120,10)
      \dottedline[$\cdot$]{2}(185,10)(200,10)
      }

      \put( 20,70){\circle*{5}}
      \put( 45,70){\circle*{5}}
      \put( 60,70){\circle*{5}}
      \put( 85,70){\circle*{5}}
      \put(100,70){\circle*{5}}
      \put(125,70){\circle*{5}}
      \put(140,70){\circle*{5}}
      \put(165,70){\circle*{5}}
      \put(180,70){\circle*{5}}
      \put(205,70){\circle*{5}}
      
      \put( 20,82){\makebox(0,0){$\tl{\T{1}}$}}
      \put( 45,82){\makebox(0,0){$\hd{\T{1}}$}}
      \put( 60,82){\makebox(0,0){$\tl{\T{5}}$}}
      \put( 85,82){\makebox(0,0){$\hd{\T{5}}$}}
      \put(100,82){\makebox(0,0){$\tl{\T{2}}$}}
      \put(125,82){\makebox(0,0){$\hd{\T{2}}$}}
      \put(140,82){\makebox(0,0){$\tl{\T{3}}$}}
      \put(165,82){\makebox(0,0){$\hd{\T{3}}$}}
      \put(180,82){\makebox(0,0){$\tl{\T{4}}$}}
      \put(205,82){\makebox(0,0){$\hd{\T{4}}$}}
      
      \put( 32,97){\makebox(0,0){$(\T{1}\!\!:\!\!\T{1})$}}
      \put( 72,97){\makebox(0,0){$(\T{2}\!\!:\!\!\T{5})$}}
      \put(112,97){\makebox(0,0){$(\T{3}\!\!:\!\!\T{2})$}}
      \put(152,97){\makebox(0,0){$(\T{4}\!\!:\!\!\T{3})$}}
      \put(192,97){\makebox(0,0){$(\T{5}\!\!:\!\!\T{4})$}}
      
      \put(  0,10){\circle*{5}}
      \put( 25,10){\circle*{5}}
      \put( 40,10){\circle*{5}}
      \put( 65,10){\circle*{5}}
      \put( 80,10){\circle*{5}}
      \put(105,10){\circle*{5}}
      \put(120,10){\circle*{5}}
      \put(145,10){\circle*{5}}
      \put(160,10){\circle*{5}}
      \put(185,10){\circle*{5}}
      \put(200,10){\circle*{5}}
      \put(225,10){\circle*{5}}
      
      \put( 0,-2){\makebox(0,0){$\tl{\T{1}}$}}
      \put( 25,-2){\makebox(0,0){$\hd{\T{1}}$}}
      \put( 40,-2){\makebox(0,0){$\hd{\T{6}}$}}
      \put( 65,-2){\makebox(0,0){$\tl{\T{6}}$}}
      \put( 80,-2){\makebox(0,0){$\hd{\T{3}}$}}
      \put(105,-2){\makebox(0,0){$\tl{\T{3}}$}}
      \put(120,-2){\makebox(0,0){$\hd{\T{2}}$}}
      \put(145,-2){\makebox(0,0){$\tl{\T{2}}$}}
      \put(160,-2){\makebox(0,0){$\tl{\T{7}}$}}
      \put(185,-2){\makebox(0,0){$\hd{\T{7}}$}}
      \put(200,-2){\makebox(0,0){$\tl{\T{4}}$}}
      \put(225,-2){\makebox(0,0){$\hd{\T{4}}$}}
      
      \put( 12,-17){\makebox(0,0){$(\T{6}\!\!:\!\!\T{1})$}}
      \put( 52,-17){\makebox(0,0){$(\rev{\T{7}}\!\!:\!\!\rev{\T{6}})$}}
      \put( 92,-17){\makebox(0,0){$(\rev{\T{8}}\!\!:\!\!\rev{\T{3}})$}}
      \put(132,-17){\makebox(0,0){$(\rev{\T{9}}\!\!:\!\!\rev{\T{2}})$}}
      \put(172,-17){\makebox(0,0){$(\T{10}\!\!:\!\!\T{7})$}}
      \put(212,-17){\makebox(0,0){$(\T{11}\!\!:\!\!\T{4})$}}
      
    \end{picture}
    \end{tabular}
    
    \bigskip
    
    
  \caption{
  Considering the same genomes 
  $A=\{\T{1}\z\z\T{2}\z\z\T{3}\z\z\T{4}\z\z\T{5}\}$
  and $B=\{\T{6}\z\z\rev{\T{7}}\z\z\rev{\T{8}}\z\z\rev{\T{9}}\z\z\T{10}\z\z\T{11}\}$
  as in
  Figure~\ref{fig:example-gsg}, let $M_1$ (red) and $M_2$ (blue) be two
  distinct maximal matchings in $\GSx{0.1}(A,B)$. We also represent the non-maximal matching $M_3$ (cyan) that is a subset of $M_2$.
  In the middle part we show diagrams $\WR(A^{M_1},B^{M_1})$ and
  $\WR(A^{M_2},B^{M_2})$, both with two $\AB$-paths and two $\AB$-cycles.
  In the lower part we show diagrams $\WR(A^{M_\emptyset},B^{M_\emptyset})$, corresponding   to the trivial empty matching $M_\emptyset$ and with two linear singletons (one $\AA$-path and one $\BB$-path), and $\WR(A^{M_3},B^{M_3})$, with two $\AB$-paths
    and two $\AB$-cycles. The labeling $(\T{X\!\!:\!\!Y})$ indicates that $\T{Y}=s(\T{X},M_i)$.}
  \label{fig:matchings}
  \vspace{-2mm}
\end{figure}

Let $A^M$ and $B^M$ be the mapped genomes for a given matching $M$ of
 $\GSx{x}(A,B)$. 
 The \emph{weighted relational diagram} of $A^M$ and $B^M$, denoted by
 $\WR(A^M,B^M)$, is obtained by constructing 
 the relational diagram of $A^M$ and $B^M$ 
 and adding weights to the indel edges as follows.
 For each mapped $M$-unsaturated marker $m \in \maa(M) \cup \mbb(M)$, the indel edge 
 $\hd{m}\tl{m}$
 receives a weight $w(\hd{m}\tl{m})=\max\{ \sigma(uv) | uv \in \GSx{x}(A,B) \text{ and } u\!\!=\!\!s^{-1}(m,M)\}$, that is the maximum similarity among the edges incident to the marker $u=s^{-1}(m,M)$ in $\GSx{x}(A,B)$. We denote by $\CM = \selfedges{A} \cup \selfedges{B}$ the set of indel edges, here also called the \emph{complement} of $M$. 
 The weight of~$\CM$ is $w(\CM) = \sum_{e \in \CM} w(e)$.
Examples of 
diagrams of mapped genomes are shown in
Figure~\ref{fig:matchings}.

In the computation of the weighted DCJ-indel distance of mapped genomes~$A^M$ and~$B^M$, denoted by $\wcostid(A^M, B^M)$, we 
should take into consideration the exclusive markers of~$\maa(M)$ and $\mbb(M)$, and the weights $w(M)$ and $w(\CM)$. An important condition is that $\wcostid(A^M, B^M)$ must be equal to $\ddcjid(A^M, B^M)$ if $w(M)=|M|$  and $w(\CM)=0$.
We can achieve this by extending the formula for computing $\wcost(A^M, B^M)$ as follows:

\begin{align*}
\wcostid(A^M,B^M) & = \wcost(A^M, B^M) + \!\!\!\!\!\!\sum_{C\in
\WR(A^M,B^M)} \!\!\!\!\!\!\!\!\!\!\!\!\!\!\lambda(C) ~~~-\delta_M+ w(\CM) \nonumber \\
& = \ddcj(A^M, B^M) + |M| - w(M) + \!\!\!\!\!\!\sum_{C\in
\WR(A^M,B^M)} \!\!\!\!\!\!\!\!\!\!\!\!\!\!\lambda(C) ~~-\delta_M+  w(\CM) \nonumber \\
& = \ddcjid(A^M, B^M) + |M| - w(M) +  w(\CM)\:.
\end{align*}

Let us now examine the behaviour of the formula above for the examples given in Figure~\ref{fig:matchings}.
Matching $M_1$ is maximal and gives the distance
$\wcostid(A^{M_1},B^{M_1})=8.6$. 
Matching $M_2$ is also maximal 
and gives the distance
$\wcostid(A^{M_2},B^{M_2})=5.2$. 
The empty matching $M_\emptyset$ gives the 
distance
$\wcostid(A^{M_\emptyset},B^{M_\emptyset})=9.7$,
that is the biggest.
And the non-maximal matching $M_3 \subset M_2$ gives the distance
$\wcostid(A^{M_3},B^{M_3})=5.1$,
that is the smallest.

Given that $\mathbb{M}$ is the set of all distinct matchings in~$\GSx{x}(A, B)$, the family-free DCJ-indel distance is defined as follows:
\[
\dffdcjid(A, B, \GSx{x}) = \min_{M \in \mathbb{M}}\{ \wcostid(A^M,B^M) \} \:.
\]

\paragraph*{Complexity:}
If two family-based genomes contain the same number of occurrences of each marker, they are said to be \emph{balanced}. 
The problem of computing the DCJ distance of balanced genomes (BG-DCJ) is NP-hard~\cite{SHA-LIN-MOR-2015}. 
Since the computation of $\dffdcjid$ can be used to solve BG-DCJ, it is also NP-hard. 
See Appendix~\ref{app:complex} for details of the reduction.


\section{Family-free relational diagram}~\label{sec:ff-diagram}

\vspace{-3mm} 

\noindent An efficient way to solve the family-free DCJ-indel distance is to develop an ILP that searches for its solution in a general graph, that represents all possible 
diagrams corresponding to all candidate matchings, in a similar way as the approaches given in~\cite{SHA-LIN-MOR-2015,MAR-FEI-BRA-STO-15,BOH-BRA-DOE-STO-2020}.
Given two genomes $A$ and $B$ and their marker similarity graph $\GSx{x}(A,B)$, the structure that integrates the properties of all 
diagrams of mapped genomes is the \emph{family-free relational diagram} $\FFR(A,B,\GSx{x})$, that
has a set $V(A)$ with a vertex for each of the two extremities of each marker of genome $A$ and a set $V(B)$ with a vertex for each of the
two extremities of each marker of genome~$B$. 

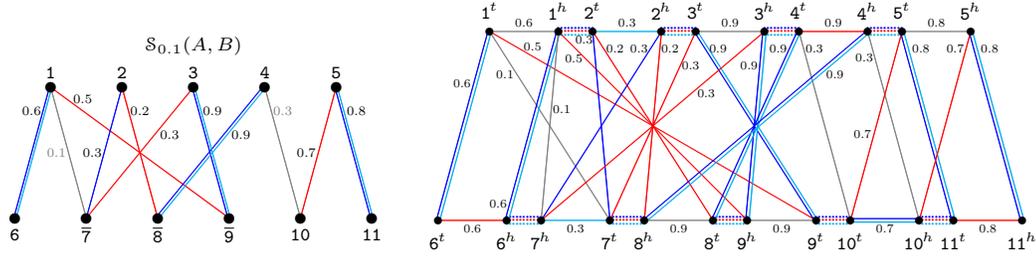
\begin{figure}[ht]
\begin{center}
  \begin{tabular}{m{4.8cm}@{~~~}m{7.3cm}}
\setlength{\unitlength}{.45pt}
    \hspace{-5mm}\begin{picture}(300,165)
     \scriptsize
      \put(150,155){\makebox(0,0){$\GSx{0.1}(A,B)$}}

      \thinlines
      \put(0,0){\color{blue}
      \qbezier(29,120)(14,65)(-1,10)
      }
      \put(14,100){\makebox(0,0){\scalebox{.7}{0.6}}}
      \put(0,0){\color{cyan}
      \qbezier(31,120)(16,65)(1,10)
      }
      \put(0,0){\color{gray}
      \qbezier(30,120)(45,65)(60,10)
      \put(35,65){\makebox(0,0){\scalebox{.7}{0.1}}}
      }
      \put(0,0){\color{red}
      \qbezier(30,120)(105,65)(180,10)
      }
      \put(57,110){\makebox(0,0){\scalebox{.7}{0.5}}}
      \put(0,0){\color{blue}
      \qbezier(90,120)(75,65)(60,10)
      }
      \put(65,65){\makebox(0,0){\scalebox{.7}{0.3}}}
      \put(0,0){\color{red}
      \qbezier(90,120)(105,65)(120,10)
      \qbezier(150,120)(105,65)(60,10)
      }
      \put(105,100){\makebox(0,0){\scalebox{.7}{0.2}}}
      \put(130,80){\makebox(0,0){\scalebox{.7}{0.3}}}
      \put(0,0){\color{blue}
      \qbezier(149,120)(164,65)(179,10)
      \qbezier(208,120)(163,65)(118,10)
      }
      \put(166,100){\makebox(0,0){\scalebox{.7}{0.9}}}
      \put(190,80){\makebox(0,0){\scalebox{.7}{0.9}}}
      \put(0,0){\color{cyan}
      \qbezier(151,120)(166,65)(181,10)
      \qbezier(211,120)(166,65)(121,10)
      }
      \put(0,0){\color{gray}
      \qbezier(210,120)(225,65)(240,10)
      \put(225,100){\makebox(0,0){\scalebox{.7}{0.3}}}
      }
      \put(0,0){\color{red}
      \qbezier(270,120)(255,65)(240,10)
      }
      \put(245,65){\makebox(0,0){\scalebox{.7}{0.7}}}
      \put(0,0){\color{blue}
      \qbezier(269,120)(284,65)(299,10)
      }
      \put(287,100){\makebox(0,0){\scalebox{.7}{0.8}}}
      \put(0,0){\color{cyan}
      \qbezier(271,120)(286,65)(301,10)
      }
      
      \put( 30,120){\circle*{8}}
      \put( 90,120){\circle*{8}}
      \put(150,120){\circle*{8}}
      \put(210,120){\circle*{8}}
      \put(270,120){\circle*{8}}

      \put(  0,10){\circle*{8}}
      \put( 60,10){\circle*{8}}
      \put(120,10){\circle*{8}}
      \put(180,10){\circle*{8}}
      \put(240,10){\circle*{8}}
      \put(300,10){\circle*{8}}

      \put( 30,132){\makebox(0,0){$\T{1}$}}
      \put( 90,132){\makebox(0,0){$\T{2}$}}
      \put(150,132){\makebox(0,0){$\T{3}$}}
      \put(210,132){\makebox(0,0){$\T{4}$}}
      \put(270,132){\makebox(0,0){$\T{5}$}}
      
      \put(  0,-4){\makebox(0,0){$\T{6}$}}
      \put( 60,-4){\makebox(0,0){$\rev{\T{7}}$}}
      \put(120,-4){\makebox(0,0){$\rev{\T{8}}$}}
      \put(180,-4){\makebox(0,0){$\rev{\T{9}}$}}
      \put(240,-4){\makebox(0,0){$\T{10}$}}
      \put(300,-4){\makebox(0,0){$\T{11}$}}
      
\end{picture} &
    \setlength{\unitlength}{.65pt}
\begin{picture}(320,140)
       \scriptsize
        \put(0,0){\color{blue}
        \dottedline[$\cdot$]{2}(40,12)(60,12)
        }
        \put(0,0){\color{red}
        \dottedline[$\cdot$]{2}(40,10)(60,10)
        }
        \put(0,0){\color{cyan}
        \dottedline[$\cdot$]{2}(40,8)(60,8)
        }
        
        \put(0,0){\color{blue}
        \dottedline[$\cdot$]{2}(100,12)(120,12)
        }
        \put(0,0){\color{red}
        \dottedline[$\cdot$]{2}(100,10)(120,10)
        }
        \put(0,0){\color{cyan}
        \dottedline[$\cdot$]{2}(100,8)(120,8)
        }
        
        \put(0,0){\color{blue}
        \dottedline[$\cdot$]{2}(160,12)(180,12)
        }
        \put(0,0){\color{red}
        \dottedline[$\cdot$]{2}(160,10)(180,10)
        }
        \put(0,0){\color{cyan}
        \dottedline[$\cdot$]{2}(160,8)(180,8)
        }
        
        \put(0,0){\color{blue}
        \dottedline[$\cdot$]{2}(220,12)(240,12)
        }
        \put(0,0){\color{red}
        \dottedline[$\cdot$]{2}(220,10)(240,10)
        }
        \put(0,0){\color{cyan}
        \dottedline[$\cdot$]{2}(220,8)(240,8)
        }
        
        \put(0,0){\color{blue}
        \dottedline[$\cdot$]{2}(280,12)(300,12)
        }
        \put(0,0){\color{red}
        \dottedline[$\cdot$]{2}(280,10)(300,10)
        }
        \put(0,0){\color{cyan}
        \dottedline[$\cdot$]{2}(280,8)(300,8)
        }
  
        \put(0,0){\color{blue}
        \dottedline[$\cdot$]{2}(70,122)(90,122)
        }
        \put(0,0){\color{red}
        \dottedline[$\cdot$]{2}(70,120)(90,120)
        }
        \put(0,0){\color{cyan}
        \dottedline[$\cdot$]{2}(70,118)(90,118)
        }
        
        \put(0,0){\color{blue}
        \dottedline[$\cdot$]{2}(130,122)(150,122)
        }
        \put(0,0){\color{red}
        \dottedline[$\cdot$]{2}(130,120)(150,120)
        }
        \put(0,0){\color{cyan}
        \dottedline[$\cdot$]{2}(130,118)(150,118)
        }
        
        \put(0,0){\color{blue}
        \dottedline[$\cdot$]{2}(190,122)(210,122)
        }
        \put(0,0){\color{red}
        \dottedline[$\cdot$]{2}(190,120)(210,120)
        }
        \put(0,0){\color{cyan}
        \dottedline[$\cdot$]{2}(190,118)(210,118)
        }
        
        \put(0,0){\color{blue}
        \dottedline[$\cdot$]{2}(250,122)(270,122)
        }
        \put(0,0){\color{red}
        \dottedline[$\cdot$]{2}(250,120)(270,120)
        }
        \put(0,0){\color{cyan}
        \dottedline[$\cdot$]{2}(250,118)(270,118)
        }
      
        \put(0,0){\color{blue}
        \qbezier(29,120)(14,65)(-1,10)
        }
        \put(0,0){\color{cyan}
        \qbezier(31,120)(16,65)(1,10)
        }
        \put(14,90){\makebox(0,0){\scalebox{.65}{0.6}}}
        
        \put(0,0){\color{gray}
        \qbezier(30,120)(65,65)(100,10)
        }
        \put(39,95){\makebox(0,0){\scalebox{.65}{0.1}}}
        
        \put(0,0){\color{red}
        \qbezier(30,120)(125,65)(220,10)
        }
        \put(55,111){\makebox(0,0){\scalebox{.65}{0.5}}}
        
        \put(0,0){\color{blue}
        \qbezier(69,120)(54,65)(39,10)
        }
        \put(0,0){\color{cyan}
        \qbezier(71,120)(56,65)(41,10)
        }
        \put(35,20){\makebox(0,0){\scalebox{.65}{0.6}}}
        
        \put(0,0){\color{gray}
        \qbezier(70,120)(65,65)(60,10)
        }
        \put(72,75){\makebox(0,0){\scalebox{.65}{0.1}}}
        
        \put(0,0){\color{red}
        \qbezier(70,120)(125,65)(180,10)
        }
        \put(79,104){\makebox(0,0){\scalebox{.65}{0.5}}}
        
        \put(0,0){\color{blue}
        \qbezier(90,120)(95,65)(100,10)
        }
        \put(85,115){\makebox(0,0){\scalebox{.65}{0.3}}}
        
        \put(0,0){\color{red}
        \qbezier(90,120)(125,65)(160,10)
        }
        \put(103,110){\makebox(0,0){\scalebox{.65}{0.2}}}
        
        \put(0,0){\color{blue}
        \qbezier(130,120)(95,65)(60,10)
        }
        \put(117,110){\makebox(0,0){\scalebox{.65}{0.3}}}
        
        \put(0,0){\color{red}
        \qbezier(130,120)(125,65)(120,10)
        }
        \put(135,110){\makebox(0,0){\scalebox{.65}{0.2}}}
        
        \put(0,0){\color{red}
        \qbezier(150,120)(125,65)(100,10)
        }
        \put(147,100){\makebox(0,0){\scalebox{.65}{0.3}}}
        
        \put(0,0){\color{blue}
        \qbezier(149,120)(184,65)(219,10)
        }
        \put(0,0){\color{cyan}
        \qbezier(151,120)(186,65)(221,10)
        }
        \put(163,110){\makebox(0,0){\scalebox{.65}{0.9}}}
        
        \put(0,0){\color{red}
        \qbezier(190,120)(125,65)(60,10)
        }
        \put(156,84){\makebox(0,0){\scalebox{.65}{0.3}}}
        
        \put(0,0){\color{blue}
        \qbezier(189,120)(184,65)(179,10)
        }
        \put(0,0){\color{cyan}
        \qbezier(191,120)(186,65)(181,10)
        }
        \put(181,100){\makebox(0,0){\scalebox{.65}{0.9}}}
        
        \put(0,0){\color{blue}
        \qbezier(209,120)(184,65)(159,10)
        }
        \put(0,0){\color{cyan}
        \qbezier(211,120)(186,65)(161,10)
        }
        \put(198,110){\makebox(0,0){\scalebox{.65}{0.9}}}
        
        \put(0,0){\color{gray}
        \qbezier(210,120)(225,65)(240,10)
        }
        \put(219,110){\makebox(0,0){\scalebox{.65}{0.3}}}
        
        \put(0,0){\color{blue}
        \qbezier(249,120)(184,65)(119,10)
        }
        \put(0,0){\color{cyan}
        \qbezier(252,120)(187,65)(122,10)
        }
        \put(231,95){\makebox(0,0){\scalebox{.65}{0.9}}}
        
        \put(0,0){\color{gray}
        \qbezier(250,120)(265,65)(280,10)
        }
        \put(248,105){\makebox(0,0){\scalebox{.65}{0.3}}}
        
        \put(0,0){\color{red}
        \qbezier(270,120)(255,65)(240,10)
        }
        \put(247,60){\makebox(0,0){\scalebox{.65}{0.7}}}
        
        \put(0,0){\color{blue}
        \qbezier(269,120)(284,65)(299,10)
        }
        \put(0,0){\color{cyan}
        \qbezier(271,120)(286,65)(301,10)
        }
        \put(281,110){\makebox(0,0){\scalebox{.65}{0.8}}}
        
        \put(0,0){\color{red}
        \qbezier(310,120)(295,65)(280,10)
        }
        \put(301,110){\makebox(0,0){\scalebox{.65}{0.7}}}
        
        \put(0,0){\color{blue}
        \qbezier(309,120)(324,65)(339,10)
        }
        \put(0,0){\color{cyan}
        \qbezier(311,120)(326,65)(341,10)
        }
        \put(321,110){\makebox(0,0){\scalebox{.65}{0.8}}}
      
        \put(0,0){\color{red}
        \drawline(0,10)(40,10)
        }
        \put(20,5){\makebox(0,0){\scalebox{.65}{0.6}}}
        \put(0,0){\color{cyan}
        \drawline(60,10)(100,10)
        }
        \put(80,5){\makebox(0,0){\scalebox{.65}{0.3}}}
        \put(0,0){\color{gray}
        \drawline(120,10)(160,10)
        }
        \put(140,5){\makebox(0,0){\scalebox{.65}{0.9}}}
        \put(0,0){\color{gray}
        \drawline(180,10)(220,10)
        }
        \put(200,5){\makebox(0,0){\scalebox{.65}{0.9}}}
        \put(0,0){\color{blue}
        \drawline(240,11)(280,11)
        }
        \put(0,0){\color{cyan}
        \drawline(240,9)(280,9)
        }
        \put(260,5){\makebox(0,0){\scalebox{.65}{0.7}}}
        \put(0,0){\color{red}
        \drawline(300,10)(340,10)
        }
        \put(320,5){\makebox(0,0){\scalebox{.65}{0.8}}}
        \put(0,0){\color{gray}
        \drawline(30,120)(70,120)
        }
        \put(50,125){\makebox(0,0){\scalebox{.65}{0.6}}}
        \put(0,0){\color{cyan}
        \drawline(90,120)(130,120)
        }
        \put(110,125){\makebox(0,0){\scalebox{.65}{0.3}}}
        \put(0,0){\color{gray}
        \drawline(150,120)(190,120)
        }
        \put(170,125){\makebox(0,0){\scalebox{.65}{0.9}}}
        \put(0,0){\color{red}
        \drawline(210,120)(250,120)
        }
        \put(230,125){\makebox(0,0){\scalebox{.65}{0.9}}}
        \put(0,0){\color{gray}
        \drawline(270,120)(310,120)
        }
        \put(290,125){\makebox(0,0){\scalebox{.65}{0.8}}}
      
      \put( 30,120){\circle*{5}}
      \put( 70,120){\circle*{5}}
      \put( 90,120){\circle*{5}}
      \put(130,120){\circle*{5}}
      \put(150,120){\circle*{5}}
      \put(190,120){\circle*{5}}
      \put(210,120){\circle*{5}}
      \put(250,120){\circle*{5}}
      \put(270,120){\circle*{5}}
      \put(310,120){\circle*{5}}
      
      \put( 30,132){\makebox(0,0){$\tl{\T{1}}$}}
      \put( 70,132){\makebox(0,0){$\hd{\T{1}}$}}
      \put( 90,132){\makebox(0,0){$\tl{\T{2}}$}}
      \put(130,132){\makebox(0,0){$\hd{\T{2}}$}}
      \put(150,132){\makebox(0,0){$\tl{\T{3}}$}}
      \put(190,132){\makebox(0,0){$\hd{\T{3}}$}}
      \put(210,132){\makebox(0,0){$\tl{\T{4}}$}}
      \put(250,132){\makebox(0,0){$\hd{\T{4}}$}}
      \put(270,132){\makebox(0,0){$\tl{\T{5}}$}}
      \put(310,132){\makebox(0,0){$\hd{\T{5}}$}}

      \put(  0,10){\circle*{5}}
      \put( 40,10){\circle*{5}}
      \put( 60,10){\circle*{5}}
      \put(100,10){\circle*{5}}
      \put(120,10){\circle*{5}}
      \put(160,10){\circle*{5}}
      \put(180,10){\circle*{5}}
      \put(220,10){\circle*{5}}
      \put(240,10){\circle*{5}}
      \put(280,10){\circle*{5}}
      \put(300,10){\circle*{5}}
      \put(340,10){\circle*{5}}
      
      \put( 0,-2){\makebox(0,0){$\tl{\T{6}}$}}
      \put( 40,-2){\makebox(0,0){$\hd{\T{6}}$}}
      \put( 60,-2){\makebox(0,0){$\hd{\T{7}}$}}
      \put(100,-2){\makebox(0,0){$\tl{\T{7}}$}}
      \put(120,-2){\makebox(0,0){$\hd{\T{8}}$}}
      \put(160,-2){\makebox(0,0){$\tl{\T{8}}$}}
      \put(180,-2){\makebox(0,0){$\hd{\T{9}}$}}
      \put(220,-2){\makebox(0,0){$\tl{\T{9}}$}}
      \put(240,-2){\makebox(0,0){$\tl{\T{10}}$}}
      \put(280,-2){\makebox(0,0){$\hd{\T{10}}$}}
      \put(300,-2){\makebox(0,0){$\tl{\T{11}}$}}
      \put(340,-2){\makebox(0,0){$\hd{\T{11}}$}}
      
    \end{picture}
    
    \end{tabular}
    
    
    \end{center}
  \caption{Given genomes $A=\{\T{1}\z\z\T{2}\z\z\T{3}\z\z\T{4}\z\z\T{5}\}$
  and $B=\{\T{6}\z\z\rev{\T{7}}\z\z\rev{\T{8}}\z\z\rev{\T{9}}\z\z\T{10}\z\z\T{11}\}$, in the left part we represent the marker similarity graph $\GSx{0.1}(A,B)$ and in the right part
  the family-free relational diagram $\FFR(A,B,\GSx{0.1})$. We represent in multiple colors the edges that correspond to multiple matchings.
  }
  \label{fig:multi}
  \vspace{-2mm}
\end{figure}

Again, sets $\adjedges{A}$ and $\adjedges{B}$ contain adjacency edges
 connecting adjacent extremities of markers in $A$ and in $B$.
But here the set $E_{\ext}$ contains, for each edge $ab \in\GSx{x}(A,B)$, an extremity edge connecting $\tl{a}$ to $\tl{b}$, 
and an extremity edge connecting $\hd{a}$ to $\hd{b}$. To both edges $\tl{a}\tl{b}$ and $\hd{a}\hd{b}$, that are called \emph{siblings}, we assign the same weight, that corresponds to the similarity of the edge $ab$ in $\GSx{x}(A,B)$: $w(\tl{a}\tl{b})=w(\hd{a}\hd{b})=\sigma(ab)$.
Furthermore, for each marker $m$ there is an indel edge connecting the vertices $\hd{m}$ and $\tl{m}$. 
The indel edge
 $\hd{m}\tl{m}$ receives a weight $w(\hd{m}\tl{m})=\max\{ \sigma(mv) | mv \in \GSx{x}(A,B) \}$, that is, it is the maximum similarity among the edges incident to the marker $m$ in $\GSx{x}(A,B)$. We denote by $\selfedges{A}$ the set of indel edges of markers in genome $A$ and by $\selfedges{B}$ the set of indel edges of markers in genome $B$.
An example of a family-free relational diagram is given in Figure~\ref{fig:multi}.

\subsection{Consistent decompositions}

The diagram $\FFR(A,B,\GSx{x})$  may contain vertices of degree larger than two. A \emph{decomposition} of $\FFR(A,B,\GSx{x})$ is a collection of vertex-disjoint \emph{components}, that can be cycles and/or paths, covering all vertices of $\FFR(A,B,\GSx{x})$. 
 There can be multiple ways of selecting a decomposition,  
 and we need to find one that 
 allows to identify a matching of $\GSx{x}(A,B)$. 
A set $S \subseteq E_{\ext}$ is a \emph{sibling-set} if it is exclusively composed of pairs of siblings and does not contain any pair of incident edges.
 Thus, a sibling-set $S$ of $\FFR(A,B,\GSx{x})$ corresponds to a
 matching of $\GSx{x}(A,B)$. In other words, there is a clear bijection between matchings of $\GSx{x}(A,B)$ and sibling-sets of $\FFR(A,B,\GSx{x})$ and we denote by $M_S$ the matching corresponding to the sibling-set $S$.

The set of edges~$D[S]$ \emph{induced} by a sibling-set $S$ is said to be a \emph{consistent decomposition} of $\FFR(A,B,\GSx{x})$ and can be obtained as follows. 
 In the beginning, $D[S]$ is the union of $S$ with the sets of adjacency edges $\adjedges{A}$ and $\adjedges{B}$. We then need to determine the \emph{complement} of the sibling-set $S$, denoted by $\widetilde{S}$, that is composed of the indel-edges of $\FFR(A,B,\GSx{x})$ that must be added to $D[S]$: 
  for each indel edge $e$, if its two endpoints have degree one or zero in $D[S]$, then~$e$ is added to both $\widetilde{S}$ and $D[S]$. (Note that $\widetilde{S}=\CM_S$, while $|S| = 2|M_S|$ and $w(S)=2w(M_S)$.) The consistent decomposition $D[S]$ covers all vertices of $\FFR(A,B,\GSx{x})$ and is composed of cycles and paths, allowing us to
 compute the values
 \[
 \ddcjid(D[S]) = \frac{|S|}{2} - c_{D} - \frac{i_{D}}{2} +\sum_{C \in D[S]}\!\! \lambda(C) -\delta_{D} \, \mbox{\ \ \ and}
 \]
 \[
 \wcostid(D[S]) = \ddcjid(D[S]) + \frac{|S|}{2}- \frac{w(S)}{2} +  w(\widetilde{S}) \,,
 \] 
 where $c_{D}$ and
 $i_{D}$ are the numbers of $\AB$-cycles and $\AB$-paths in $D[S]$,
 respectively, and $\delta_{D}$ is the optimal deduction of recombinations of paths from $D[S]$.

Given that $\mathbb{S}$ is the sets of all sibling-sets of $\FFR(A,B,\GSx{x})$, we compute the family-free DCJ-indel distance of $A$ and $B$ with the following equation:
 \[ 
 \dffdcjid(A, B, \GSx{x}) = \min_{S \in \mathbb{S}}\{\wcostid(D[S]) \}\,.
 \]

\subsection{Capping}

Telomeres produce some difficulties for the decomposition of $\FFR(A,B,\GSx{x})$, and a known technique to overcome this problem is called \emph{capping}~\cite{HAN-PEV-1995}. It consists of
modifying the diagram by adding \emph{artificial} markers, also called \emph{caps}, whose extremities should be properly connected to the telomeres of the linear chromosomes of $A$ and $B$. Therefore, usually the capping depends on the numbers $\kappa_{\!A}$ and $\kappa_{\!B}$, that are, respectively, the total numbers of linear chromosomes in genomes $A$ and $B$.

\paragraph*{Family-based singular genomes:}
First we recall the capping of family-based singular genomes.
Here the caps must circularize all linear chromosomes, so that their relational diagram is composed of cycles only, but, if the capping is optimal, the DCJ-indel distance is preserved.

An optimal capping that transforms singular linear genomes $A$ and $B$ into singular circular genomes 
can be obtained after identifying the recombination groups~\cite{BRA-WIL-STO-2011}. The DCJ-indel distance is preserved by properly linking the components of each identified recombination group into a single cycle~\cite{BOH-BRA-DOE-STO-2020}. Such a capping may require some artificial adjacencies between caps. The following result is very useful.

\begin{theorem}[from~\cite{BOH-BRA-DOE-STO-2020}]\label{thm:capping-singular}
We can obtain an optimal capping of singular genomes $A$ and $B$ with exactly $p_* = \max\{\kappa_{\!A},\kappa_{\!B}\}$ caps and $|\kappa_{\!A}-\kappa_{\!B}|$ artificial adjacencies between caps.
\end{theorem}

\paragraph*{Capped family-free relational diagram:}
We transform $\FFR(A,B,\GSx{x})$ into the \emph{capped family-free relational diagram} $\FFRC(A,B,\GSx{x})$ as follows. Again, let $p_* = \max\{\kappa_{\!A},\kappa_{\!B}\}$.
The diagram $\FFRC(A,B,\GSx{x})$ is obtained by adding to $\FFR(A,B,\GSx{x})$ $4p_*$ new vertices, named $\cext{A}{1},\cext{A}{2},\ldots,\cext{A}{2p_*}$ and $\cext{B}{1}, \cext{B}{2},\ldots,\cext{B}{2p_*}$, each one representing a \emph{cap extremity}. 
Each of the $2\kappa_{\!A}$ telomeres of $A$ is connected by an adjacency edge to a distinct cap extremity among $\cext{A}{1},\cext{A}{2},\ldots,\cext{A}{2\kappa_{\!A}}$.
Similarly, each of the $2\kappa_{\!B}$ telomeres of $B$ is connected by an adjacency edge to a distinct cap extremity among $\cext{B}{1},\cext{B}{2},\ldots,\cext{B}{2\kappa_{\!B}}$.
Moreover, if $\kappa_{\!A} < \kappa_{\!B}$, for $i = 2\kappa_{\!A}+1,2\kappa_{\!A}+3,\ldots,2\kappa_{\!B}-1$, connect $\cext{A}{i}$ to $\cext{A}{i+1}$ by an \emph{artificial adjacency edge}. Otherwise, if $\kappa_{\!B} < \kappa_{\!A}$, for $j = 2\kappa_{\!B}+1,2\kappa_{\!B}+3,\ldots,2\kappa_{\!A}-1$, connect $\cext{B}{j}$ to $\cext{B}{j+1}$ by an artificial adjacency edge. 
All these new adjacency edges and artificial adjacency edges are added to $\adjedges{A}$ and $\adjedges{B}$, respectively.
 We also connect each $\cext{A}{i}$, $1\leq i\leq 2p_*$, by a \emph{cap extremity edge} to each $\cext{B}{j}$, $1\leq j\leq 2p_*$, and denote by $E_{\cp}$ the set of cap extremity edges. 

A set $P \subseteq E_{\cp}$ is a \emph{capping-set} if it does not contain any pair of incident edges and is maximal. 
Since each cap extremity of $A$ is connected to each cap extremity of $B$, the size of any (maximal) capping-set is $2p_*$. 
 A consistent decomposition $Q[S,P]$ of $\FFRC(A,B,\GSx{x})$ is induced by a sibling-set $S \subseteq E_\ext$ and a (maximal) capping-set $P \subseteq E_\cp$ and is composed of vertex disjoint cycles that cover all vertices of $\FFRC(A,B,\GSx{x})$.
 An example of a capped family-free relational diagram is given in  Figure~\ref{fig:capped-multi} (Appendix~\ref{app:figures}).

\begin{theorem}
 Let $\mathbb{P}_\textsc{max}$ 
 be the set of all distinct (maximal) capping-sets from 
 $\FFRC(A,B,\GSx{x})$.
 For each sibling-set $S$ of 
 $\FFR(A,B,\GSx{x})$ and $\FFRC(A,B,\GSx{x})$, we have
 \[
 \ddcjid(D[S]) = \min_{P \in \mathbb{P}_\textsc{max}}\{ \ddcjid(Q[S,P])\}\,, \mbox{\ and}
 \]
 \[
 \wcostid(D[S]) = \min_{P \in \mathbb{P}_\textsc{max}}\{ \wcostid(Q[S,P])\}\,.
 \]
 \end{theorem}
 \proof
 Each capping-set corresponds to exactly $p_*$ caps. In addition, all adjacencies, including the $|\kappa_A - \kappa_B|$ artificial adjacencies between cap extremities, are part of each consistent decomposition.
 Recall that each sibling-set $S$ of $\FFRC(A,B,\GSx{x})$ corresponds to a matching $M_S$ of $\GSx{x}(A,B)$.
 The set of consistent decompositions include all possible 
 distinct consistent decompositions induced by $S$ together with one distinct element of $\mathbb{P}_\textsc{max}$.
Theorem~\ref{thm:capping-singular}~states that 
the pair of matched genomes $A^{M_S}$ and $B^{M_S}$
 can be optimally capped with $p_*$ caps and $|\kappa_A - \kappa_B|$ artificial adjacencies. 
 Therefore, it is clear that $\ddcjid(D[S]) = \min_{P \in \mathbb{P}_\textsc{max}}\{ \ddcjid(Q[S,P])\}$. Since the capping does not change the sizes of the sibling-sets and their weights and complements, it is also clear that $\wcostid(D[S]) = \min_{P \in \mathbb{P}_\textsc{max}}\{ \wcostid(Q[S,P])\}$.
\qed

\paragraph*{Alternative formula for computing the indel-potential of cycles:}
The consistent decompositions of the diagram $\FFRC(A,B,\GSx{x})$ are composed exclusively of cycles, and 
the number of runs $\Lambda(C)$ of a cycle $C$ is always in $\{0,1,2,4,6,\ldots\}$. Therefore, the formula to compute the indel-potential of a cycle $C$ can be simplified to
\[
\lambda(C) = 
\begin{cases}
\quad\;\, \Lambda(C)\,, & \mbox{ if } \Lambda(C)\in \{0,1\} \\[1mm]
1+\frac{\Lambda(C)}{2}\,, & \mbox{ if } \Lambda(C) \in \{2,4,6,\ldots\}
\end{cases}
\] 
that can still be redesigned to a form that can be easier implemented in the ILP~\cite{BOH-BRA-DOE-STO-2020}.
First, let a \emph{transition} in a cycle $C$ be an indel-free segment of $C$ that is between a run in one genome and a run in the other genome and denote by $\aleph(C)$ the number of transitions in $C$. 
Observe that, if $C$ is indel-free, then obviously $\aleph(C) = 0$. If $C$ has a single run, then we also have $\aleph(C) = 0$. On the other hand, if $C$ has at least 2 runs, then $\aleph(C)=\Lambda(C)$. 
The new formula is split into two parts. The first part is the function $r(C)$, defined as $r(C)=1$ if $\Lambda(C)\geq1$, otherwise $r(C)=0$, that simply tests whether $C$ is indel-enclosing or indel-free. The second part depends on the number of transitions $\aleph(C)$, and the complete formula stands as follows~\cite{BOH-BRA-DOE-STO-2020}:
\[
\lambda(C)=r(C) + \frac{\aleph(C)}{2}\,.
\]

\paragraph*{Distance formula:}
 Note that the number of indel-enclosing components is $\sum_{C \in Q[S,P]}\! r(C) = c^{r}_{Q} + s_{Q}$, 
where $c^{r}_{Q}$ and $s_{Q}$ are the number of indel-enclosing $\AB$-cycles and
the number of circular singletons in $Q[S,P]$, respectively. Furthermore, the number of indel-free $\AB$-cycles of $Q[S,P]$ is $c^{\tilde{r}}_{Q}=c_Q-c^{r}_{Q}$. We can now compute the values

  \begin{align*}
  \ddcjid(Q[S,P]) & = p_* + \frac{|S|}{2} - c_{Q} +\sum_{C \in Q[S,P]}\!\! \lambda(C)\\
   & = p_* + \frac{|S|}{2} - c_{Q} +\sum_{C \in Q[S,P]}\!\! \left( r(C) + \frac{\aleph(C)}{2}\right)\\
   & = p_* + \frac{|S|}{2} - c^{\tilde{r}}_{Q} + s_{Q} + \sum_{C \in Q[S,P]}\!\! \frac{\aleph(C)}{2}\,, \mbox{\ \ \ and}
  \end{align*}
  \begin{align}
  \wcostid(Q[S,P]) & = \ddcjid(Q[S,P]) + \frac{|S|}{2} - \frac{w(S)}{2} +  w(\widetilde{S}) \notag \\
  & = p_* + |S| - c^{\tilde{r}}_{Q} + s_{Q} + \sum_{C \in Q[S,P]}\!\! \frac{\aleph(C)}{2} - \frac{w(S)}{2}+  w(\widetilde{S})\,. \label{eq:wdcjid}
  \end{align}

Given that $\mathbb{S}$ and $\mathbb{P}_\textsc{max}$ are, respectively, the sets of all sibling-sets and all maximal capping-sets of $\FFRC(A,B,\GSx{x})$, the final version of our optimization problem is
  \[ 
 \dffdcjid(A, B,\GSx{x}) = \min_{S \in \mathbb{S}, P \in \mathbb{P}_\textsc{max}}\big\{\wcostid(Q[S,P])\big\}\,.
 \]


\section{ILP formulation to compute the family-free DCJ-indel distance}\label{sec:ilp}

 Our formulation is an adaptation of the ILP for computing the DCJ-indel distance of family-based natural genomes, by~Bohnenk\"amper {\it et al.}~\cite{BOH-BRA-DOE-STO-2020}, that is itself an extension of the ILP for computing the DCJ distance of family-based balanced genomes, by Shao {\it et al}.~\cite{SHA-LIN-MOR-2015}. The main differences between our approach and the approach from~\cite{BOH-BRA-DOE-STO-2020} are the underlying graphs and the objective functions. The general idea is searching for a sibling-set, that, together with a maximal capping-set, gives an optimal consistent cycle decomposition of the capped diagram $\FFRC(A,B,\GSx{x}) = (V,E)$, where the set of edges comprises all disjoint sets of distinct types: $E=E_\gamma \cup E_\circ \cup \adjedges{A} \cup \adjedges{B} \cup \selfedges{A} \cup \selfedges{B}$. While in the ILP from~\cite{BOH-BRA-DOE-STO-2020} the search space is restricted to maximal sibling-sets, in the family-free DCJ-indel distance the search space includes all sibling-sets, of any size.
 
 In Algorithm~\ref{alg:ilp-w} we give the formulation for computing 
 $\dffdcjid(A,B,\GSx{x})$, distributed in three main parts.
  Counting indel-free cycles in the decomposition makes up the first part, depicted in constraints (C.01)--(C.06), variables and domains (D.01)--(D.03).
 The second part is for counting transitions, described in constraints (C.07)--(C.10), variables and domains (D.04)--(D.05). 
 The last part describes how to count the number of circular singletons, with constraint (C.11), variable and domain (D.06).
The objective function of our ILP minimizes 
the size of the sibling-set, with sum over variables $x_e$, the number of circular singletons, calculated by the sum over variables $s_k$, half the overall number of transitions in indel-enclosing $\AB$-cycles, calculated by the sum over variables $t_e$, and the weight of all indel edges in the decomposition, given by the sum over their weights $w_ex_e$ for all $e \in \selfedges{}$, while maximizing both the number of indel-free cycles, counted by the sum over variables $z_i$, and half of the weights of the edges in the decomposition, given by the sum over their weights $w_ex_e$ for all edges $e \in E_\ext$. 
The minimization is not affected by constant $p_*$, that is  included in the objective function
to keep the correspondence to Equation~\eqref{eq:wdcjid}. 

\begin{algorithm}[ht]
	\caption{\label{alg:ilp-w}ILP for computing the family-free DCJ-indel
	distance}
    
    \footnotesize
    \vspace{-2mm}

    \[
    {\everymath={\displaystyle}
    \arraycolsep=3pt\def\arraystretch{1.2}
    \begin{array}{lrcllr} 
    \min & \multicolumn{5}{c}{p_* + \sum_{\mathclap{e \in E_\ext}} x_e - \sum_{\mathclap{1 \leq i \leq |V|}} z_i + \sum_{\mathclap{k \in K}} s_k + \frac{1}{2} \sum_{\mathclap{e \in E}} t_e - \frac{1}{2} \sum_{\mathclap{e \in E_\ext}} w_e x_e + \sum_{\mathclap{e \in \selfedges{}}} w_e x_e} \\[5ex]
    \text{s.~t.} & x_e & = & 1 & \forall ~ e \in \adjedges{A} \cup \adjedges{B} & \text{(C.01)} \\ \rowcolor[HTML]{FAFAFA}
    & \sum_{\mathclap{uv \in E}} x_{uv} & = & 2 & \forall ~ u \in V & \text{(C.02)} \\
    & x_e & = & x_d & \forall ~ e, d \in E_\ext, \, e,d \text{ are siblings} & \text{(C.03)} \\ \rowcolor[HTML]{FAFAFA}
    & \begin{array}{l}  y_i \\ y_j \end{array} \!\!\! & \begin{array}{l}\leq \\ \leq \end{array} & \!\!\!  \left. \begin{array}{l} y_j + i (1 - x_{v_iv_j}) \\ y_i + j (1 - x_{v_iv_j}) \end{array} \right\} & \forall ~ v_iv_j \in E & \text{(C.04)} \\
    & \begin{array}{l} y_i \\ y_j \end{array} \!\!\! & \begin{array}{l} \leq \\ \leq \end{array} & \!\!\! \left. \begin{array}{l} i (1 - x_{v_iv_j}) \\ j (1 - x_{v_iv_j}) \end{array} \right\} & \forall ~ v_iv_j \in \selfedges{A} \cup \selfedges{B} & \text{(C.05)} \\ \rowcolor[HTML]{FAFAFA}
    & iz_i & \leq & y_i & \forall ~ 1 \leq i \leq |V| & \text{(C.06)} \\
    & \begin{array}{l} r_v \\ r_{v^\prime} \end{array} \!\!\! & \begin{array}{l} \leq \\ \geq \end{array} & \!\!\! \left. \begin{array}{l} 1 - x_{uv} \\ x_{u^\prime v^\prime} \end{array} \right\} & \!\! \begin{array}{l} \forall ~ uv \in \selfedges{A} \\ \forall ~ u^\prime v^\prime \in \selfedges{B} \end{array} & \text{(C.07)} \\ \rowcolor[HTML]{FAFAFA}
    & \begin{array}{l} t_{uv} \\ t_{uv} \end{array} \!\!\! & \begin{array}{l} \geq \\ \geq \end{array} & \!\!\! \left. \begin{array}{l} r_v - r_u - (1 - x_{uv}) \\ r_u - r_v - (1 - x_{uv}) \end{array} \right\} & \forall ~ uv \in E & \text{(C.08)} \\
    & \sum_{\mathclap{\substack{d \in \selfedges{A}\>,\; d \cap e \neq \emptyset}}} x_d - t_e & \geq & 0 & \forall ~ e \in \adjedges{A} & \text{(C.09)} \\ \rowcolor[HTML]{FAFAFA}
    & t_e & = & 0 & \forall ~ e \in E \setminus \adjedges{A} & \text{(C.10)} \\
    & \sum_{\mathclap{e \in \selfedges{k}}} x_e - |k| & \leq & s_k & \forall ~ k \in K & \text{(C.11)} \\ \rowcolor[HTML]{FAFAFA}
    \text{and} & x_e & \in & \{0, 1\} & \forall ~ e \in E & \text{(D.01)} \\
    & 0 & \leq & y_i \;\; \leq \;\; i & \forall ~ 1 \leq i \leq |V| & \text{(D.02)} \\ \rowcolor[HTML]{FAFAFA}
    & z_i & \in & \{0, 1\} & \forall ~ 1 \leq i \leq |V| & \text{(D.03)} \\
    & r_v & \in & \{0, 1\} & \forall ~ v \in V & \text{(D.04)} \\ \rowcolor[HTML]{FAFAFA}
    & t_e & \in & \{0, 1\} & \forall ~ e \in E & \text{(D.05)} \\
    & s_k & \in & \{0, 1\} & \forall ~ k \in K & \text{(D.06)} \\ \rowcolor[HTML]{FAFAFA}
    & p_* & = & \max\{\kappa_A, \kappa_B\} & & \text{(D.07)} \\
    \end{array}
    }
    \]
    
    \vspace{-2mm}
\end{algorithm}

\paragraph*{Comparison to related models:}
Since the pre-requisites of a family-free setting differ substantially from those of a family-based setting, we could not  compare our approach to the one from~\cite{BOH-BRA-DOE-STO-2020}. We intend to perform such a comparison in a future work, for example by using pairwise similarities to cluster the genes into families. Comparing our approach to the original family-free DCJ distance was also not possible, because the ILP provided in~\cite{MAR-FEI-BRA-STO-15} is only suitable for unichromosomal genomes. Again, we intend to perform such a comparison in a future work, after we implement an ILP that is able to compute the family-free DCJ distance of multichromosomal genomes. 

\paragraph*{Unweighted version:}
 In the present work, for comparison purposes, we also implemented a simpler version of the family-free DCJ-indel distance, that simply ignores all weights. This version is called \emph{unweighted} family-free DCJ-indel distance,
and consists of finding a sibling-set in $\FFRC(A,B,\GSx{x})$ that minimizes~$\ddcjid(D[S,P])$. 
But here it is important to observe that smaller sibling-sets, that simply discard blocks of contiguous markers, tend to give the smaller distances.
Considering the similarity graph $\GSx{0.1}(A,B)$ of Figure~\ref{fig:matchings}, the trivial empty matching gives the distance $\ddcjid(A^{M_\emptyset},B^{M_\emptyset})=2$ (deletion of the chromosome of $A$ followed by the insertion of the chromosome of $B$). For the other matchings we have $\ddcjid(A^{M_1},B^{M_1})=4$ and $\ddcjid(A^{M_2},B^{M_2})=\ddcjid(A^{M_3},B^{M_3})=3$. We then restrict the search space to \emph{maximal} sibling-sets only, avoiding that blocks of markers are discarded. 
However, this could also enforce weak connections. In the example shown in Figure~\ref{fig:matchings}, both maximal matchings $M_1$ and $M_2$ have weak edges with weights $0.2$ and $0.3$. Matching $M_3$ has only edges with weight at least $0.6$, but it would be ignored for being non-maximal. 
Enforcing weak connections can be prevented by removing them form the similarity graph, that is, by assigning a higher value to the cutting threshold~$x$.
(see an example in Figure~\ref{fig:matchings2} of Appendix~\ref{app:figures}).
Given that $\mathbb{S}_\textsc{max}$ and $\mathbb{P}_\textsc{max}$ are, respectively, the sets of all maximal sibling-sets and all maximal capping-sets of $\FFRC(A,B,\GSx{x})$, the unweighted version of the problem is then:
 \[ 
 \dunwffdcjid(A, B,\GSx{x}) = \min_{S \in \mathbb{S}_\textsc{max}, P \in \mathbb{P}_\textsc{max}}\big\{\ddcjid(Q[S,P])\big\}\,.
 \]  

For computing the unweighted 
$\dunwffdcjid(A,B,\GSx{x})$ we need to slightly modify the ILP described in Algorithm~\ref{alg:ilp-w}. 
The details are given in Appendix~\ref{app:unweighted}.

\paragraph*{Implementation:}
The ILPs for computing both the family-free DCJ-indel distance and its unweighted version were implemented and can be downloaded from 
\url{https://gitlab.ub.uni-bielefeld.de/gi/gen-diff}.

\paragraph*{Data analysis:} 
For all pairwise comparisons, we obtained gene similarities using the FFGC pipeline\footnote{\url{https://bibiserv.cebitec.uni-bielefeld.de/ffgc}}~\cite{DFS18}, with 
the following parameters: (i) $1$ for the minimum number of genomes for which each gene must share some similarity in, (ii) $0.1$ for the stringency threshold, (iii) $1$ for the BLAST e-value, and (iv) default values for the remaining parameters. As an ILP solver, for all experiments we ran CPLEX with 8 $2.67$GHz cores.

\paragraph*{Cutting threshold:} Differently from the unweighted version, that requires a cutting
thresh\-old of about $x\!=\!0.5$ to give accurate results, the weighted $\dffdcjid$ was designed to be computed with all given
pairwise similarities, i.e., with the cutting threshold $x\!=\!0$, that
leads to a ``complete'' family-free relational diagram. Such a diagram
would be too large to be handled in practice, therefore, if $x\!=\!0$,
we consider only the similarities that are strictly greater than~$0$.
Nevertheless, for bigger instances the diagram with similarities close
to~$0$ might still be too large to be solved in reasonable time. Hence,
for some instances it may be necessary to do a small increase of the
cutting threshold. Our experiments in real data (described in
Section~\ref{sec:real}) show that small similarities have a minor
impact on the computed distance, therefore, by adopting a small cutting
threshold $x$ up to $0.3$, it is possible to reduce the diagram and
solve bigger instances, still with good accuracy.


\subsection{\label{sec:eva}Performance evaluation}

We generated simulated genomes using Artificial Life Simulator (ALF)~\cite{DAGD12} in order to benchmark our algorithm for computing the family-free DCJ-indel distance. We simulated and compared 190 pairs of genomes with different duplication rates, keeping all other parameters fixed (e.g.\ rearrangement, indel and mutation rates). The extant genomes have around $10{,}000$ genes. We obtained gene similarities between simulated genomes using FFGC. For each genome pair, a threshold of $x=0.1$ resulted in up to $8{,}400$ genes with multiple homology relations (i.e.\ vertices with degree $> 1$ in $\GSx{0.1}(A,B)$) and from $2$ to $2.8$ relations on average for those genes. In addition, each pair is about $3{,}000$ rearrangement events away from each other. 
The complete parameter sets used for running ALF, together with additional information on simulated genomes, can be found in Appendix~\ref{app:simulated}.

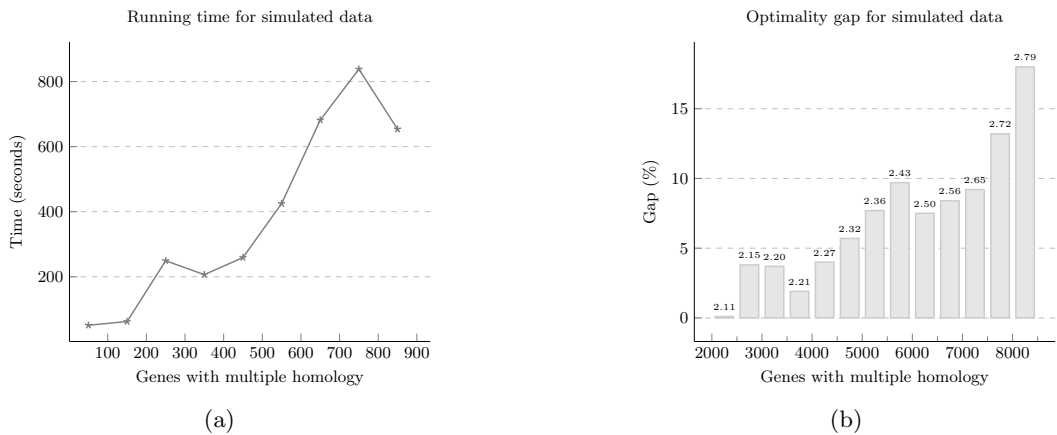
\begin{figure}[ht]\captionsetup[subfigure]{justification=centering}
\centering
\begin{subfigure}{.43\textwidth}
  \centering
  \begin{tikzpicture}[scale=0.70]
  \begin{axis}[ 
    axis x line*=bottom, 
    axis y line*=left,
    align=center,
    title={Running time for simulated data},
    xlabel={Genes with multiple homology},
    ylabel={Time (seconds)},
    ymin=1,
    xmin=1,
    ymajorgrids=true,
    grid style=dashed,
    xtick distance=100,
    every axis plot/.append style={thick},
    ]
    \addplot[gray,mark=star]
    table[x=v_degree_gt1,y=Time(s)] {simulated-time.dat};
  \end{axis}
  \end{tikzpicture}
  \caption{}
  \label{fig:simulated-time}
\end{subfigure}\quad%
\hspace{8mm}
\begin{subfigure}{.43\textwidth}
  \centering
  \begin{tikzpicture}[scale=0.70]
  \begin{axis}[ 
    axis x line*=bottom, 
    axis y line*=left,
    align=center,
    title={Optimality gap for simulated data},
    xlabel={Genes with multiple homology},
    ylabel={Gap (\%)},
    ymajorgrids=true,
    grid style=dashed,
    xtick distance=1000,
    x tick label style={/pgf/number format/1000 sep=},
    minor x tick num=1,
    every axis plot/.append style={thick},
    ]
    \addplot[ybar,gray!40,fill=gray!20,scatter,
    nodes near coords,
    point meta=explicit symbolic,
    mark=none,
    every node near coord/.append style={black,font=\tiny}]
    table[x=v_degree_gt1,y=Gap,meta=Avg.degree>1] {simulated-gap.dat};
  \end{axis}
  \end{tikzpicture}
  \caption{}
  \label{fig:simulated-gap}
\end{subfigure}
\caption{Results of the weighted family-free DCJ-indel distance given by the solver, (a) shows the average running time for instances grouped by the number of vertices with degree $> 1$ in $\GSx{0.1}(A,B)$ (in intervals of 100, those greater than 900 are not shown), and (b) for groups of instances that did not finish within the time limit of 1 hour, the average optimality gap and the average number of homology relations for those genes with multiple homologies (in intervals of 500).}
\label{fig:simulated}

\end{figure}

For computing the family-free DCJ-indel distances, we ran CPLEX with maximum CPU time of 1 hour. Results were grouped depending on the number of genes with multiple homology relations in the respective genome pairs. Figure~\ref{fig:simulated} summarizes the performance of our weighted family-free DCJ-indel distance formulation. The running times escalate quickly as the number of genes with multiple homologies increase (Figure~\ref{fig:simulated-time}, grouped in intervals of 100), reaching the time limit after $2{,}000$ of them (Figure~\ref{fig:simulated-gap}, grouped in intervals of 500). The optimality gap is the relative gap between the best solution found and the upper bound found by the solver, calculated by $(\frac{\text{upper bound}}{\text{best solution}} - 1) \times 100$, and appears to grow, for our simulated data, linearly in the number of genes with multiple homologies (Figure~\ref{fig:simulated-gap}).

The solution time and the optimality gap of our algorithm clearly depends less on genome sizes and more on the multiplicity of homology relations. In our experiments, we were able to find in 1 hour optimal or near-optimal solutions for genomes with $10{,}000$ genes and up to $4{,}000$ genes with $2.2$ homology relations on average. Our formulation should be able to handle, for instance, the complete genomes of bacteria, fungi and insects, or even sets of chromosomes of mammal and plant genomes.

\subsection{Real data analysis}\label{sec:real}

We evaluated the potential of our approach by comparing genomes of fruit flies from the genus \emph{Drosophila}, including the following species:
\emph{D.\,busckii}, \emph{D.\,melanogaster}, \emph{D.\,pseudoobscura}, \emph{D.\,sechellia}, \emph{D.\,simulans} and \emph{D.\,yakuba}~\cite{melanogaster-2000,obscura-2005,drosophila-consortium-2007,busckii-2015}. A reference phylogenetic tree of these species is shown in~Figure~\ref{fig:drosophila}, in Appendix~\ref{app:real}, where we also give the sources of the DNA sequences for each analyzed genome, and additional information on the experiments. Each genome has approximately 150Mb, with about $13{,}000$ genes distributed in 5--6 chromosomes. We obtained gene similarities using FFGC and performed two separate experiments, whose computed distances were used to build phylogenetic trees using Neighbor-Joining~\cite{SN87,KSLKT18}.

\begin{figure}\captionsetup[subfigure]{justification=centering}
\centering
\begin{subfigure}{.46\textwidth}
  \centering
  \begin{minipage}{.42\linewidth}
  \scriptsize
  
  \setlength\arraycolsep{2pt}
  $\begin{matrix}
  \hline
  \hfill\\[-2mm]
  \textsc{ilp} & \textsc{set} & ~\textit{x} \\[1mm]
  \hline
  \hfill\\[-2mm]
  \dffdcjid & \textsc{Gen} &0.3\\[1mm]
  \hline
  \hfill\\[-2mm]
  \dffdcjid & \textsc{Xchr} &0.0\\
  &&0.1\\
  &&0.2\\
  &&0.3\\
  \hline
  \hfill\\[-2mm]
  \dunwffdcjid&\textsc{Xchr} & 0.5\\[1mm]
  \hline
  \end{matrix}$
  \end{minipage}
  \begin{minipage}{.5\linewidth}
  \scriptsize
  
  \vspace{3mm}
  
  \includegraphics[width=\linewidth]{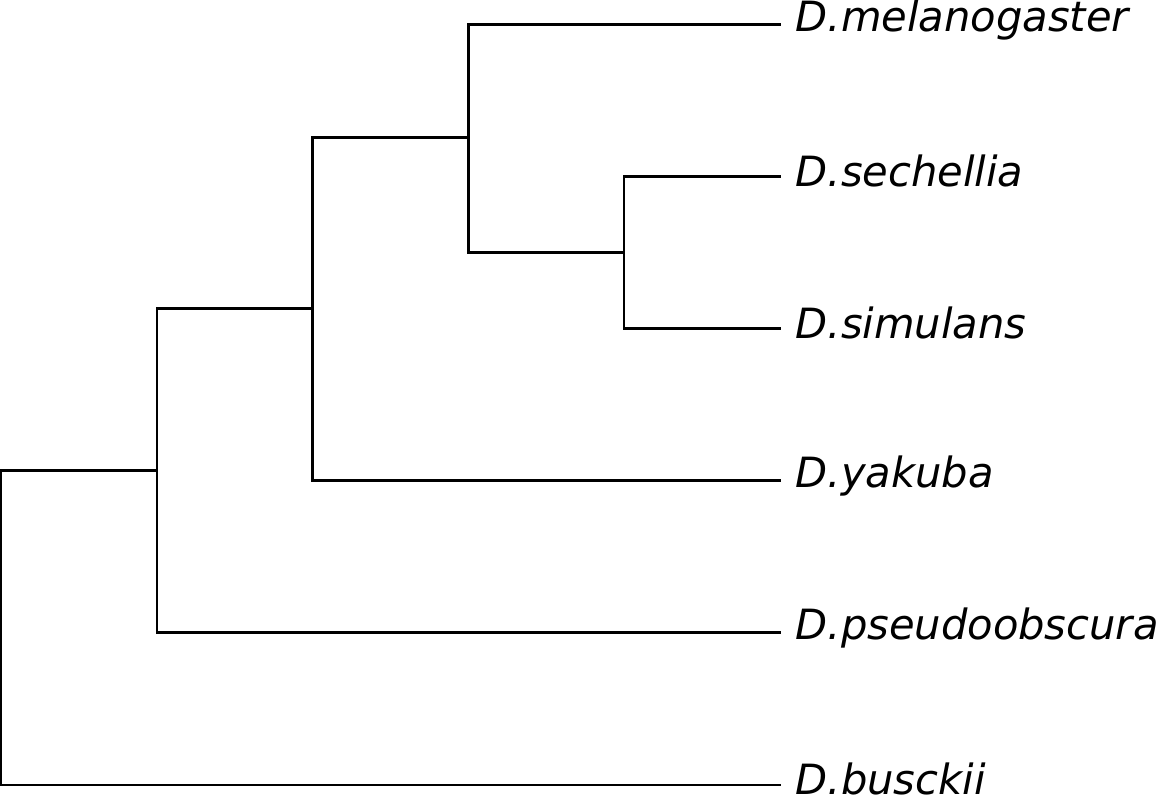}
  \vspace{1mm}
  \end{minipage}
  \caption{}
  \label{fig:drosophilaX_correct}
\end{subfigure}\quad%
\hspace{4mm}
\begin{subfigure}{.46\textwidth}
  \centering
  \begin{minipage}{.42\linewidth}
  \scriptsize
  
  \smallskip
  \setlength\arraycolsep{2pt}
  $\begin{matrix}
  \hline
  \hfill\\[-2mm]
  \textsc{ilp} & \textsc{set} & ~\textit{x}\\[1mm]
  \hline
  \hfill\\[-2mm]
  \dunwffdcjid&\textsc{Gen}&0.3\\[1mm]
  \hline
  \hfill\\[-2mm]
  \dunwffdcjid &\textsc{Xchr}&0.0\\
  &&0.3\\
  \hline
  \end{matrix}$

  \end{minipage}
  \begin{minipage}{.5\linewidth}
  \scriptsize
  
  \vspace{3mm}
  \includegraphics[width=\linewidth]{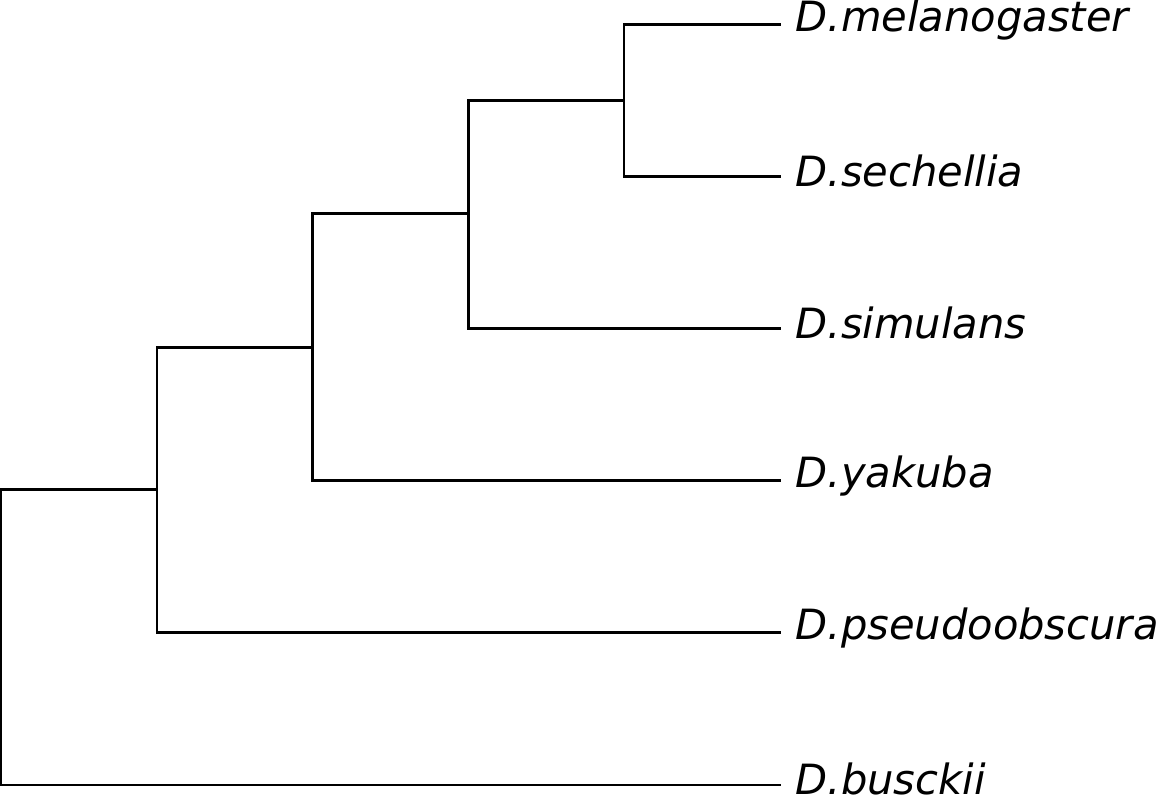}
  \vspace{1mm}
  \end{minipage}
  \caption{}
  \label{fig:drosophilaX_wrong}
\end{subfigure}\quad%

\vspace{-2mm}
\caption{Based on distance matrices calculated by our ILPs for the pairwise comparisons of complete genomes (\textsc{Gen}) or only X chromosomes (\textsc{Xchr}) of \emph{Drosophila}, we built phylogenetic trees computed by the Neighbor-Joining method~\cite{SN87,KSLKT18}. The output of this algorithm is an unrooted tree, and we assumed the most distant species \emph{D. busckii} as the outgroup for rooting the trees. All comparisons converged to exactly two trees, and next to each tree we give a list of comparisons that produced that tree.
The tree in (a) agrees with the reference shown in Figure~\ref{fig:drosophila} (Appendix~\ref{app:real}), while the tree in (b) diverges from the reference in a single branch.}
\label{fig:drosophilaX}
\end{figure}

\paragraph*{Pairwise comparison of complete genomes:} 
In this experiment, genomes in each comparison comprise together $\sim{}13{,}000$ genes with multiple homologies ($11.2$ on average), some of them having about $90$ relations considering similarities that are strictly greater than $x=0$. Since these instances were too large, we set the threshold to $x=0.3$. We then ran CPLEX with maximum CPU time of 3 hours. All $\dffdcjid$ computations finished within the time limit, most of them in less than 10 minutes, whereas the unweighted $\dunwffdcjid$ computations, in spite of having a search space of maximal sibling-sets, that is much smaller, surprisingly took from 1 to 3 hours.
We conjecture that this is due to a large number of co-optimal solutions in the unweighted version, while in $\dffdcjid$ the co-optimality is considerably minimized by weights, which helps the solver to converge faster.
While the tree given by $\dffdcjid$, shown in Figure~\ref{fig:drosophilaX_correct}, agrees with the reference tree, the tree given by $\dunwffdcjid$, shown in Figure~\ref{fig:drosophilaX_wrong}, diverges from the reference in a single branch. Details of the results are given in Appendix~\ref{app:real-complete}.

\paragraph*{Pairwise comparison of X chromosomes:}
We also did an experiment with smaller instances, composed of pairwise comparisons of X chromosomes only, so that we could evaluate the impact of the cutting threshold on the accuracy of the approach.
In this experiment, considering similarities that are strictly greater than $x=0$, each pair comprises $1{,}000$--$2{,}000$ genes with multiple homologies (5 on average) with as many as $30$ relations. 

We computed $\dffdcjid$ with cutting thresholds $x=0$, $x=0.1$, $x=0.2$ and $x=0.3$,  
always obtaining the accurate phylogenetic tree from Figure~\ref{fig:drosophilaX_correct}. These results suggest that a small cutting threshold allows to reduce the size of the instances, without having a big impact in the accuracy of $\dffdcjid$.

In addition, we computed $\dunwffdcjid$ with cutting thresholds $x=0$ and $x=0.3$, both resulting in the slightly inaccurate tree from Figure~\ref{fig:drosophilaX_wrong}, and $x=0.5$, that also resulted in the accurate tree from Figure~\ref{fig:drosophilaX_correct}. 
As expected, in the unweighted formulation the cutting threshold plays a major role in the accuracy of the calculated distances.

The analyses were done with maximum CPU time of 1 hour. The comparisons finished within a few seconds for most of instances, except for $\dunwffdcjid$ with threshold $x=0$, for which the majority of the pairwise comparisons reached the time limit---with an optimality gap of less than $3.5\%$ though (see Appendix~\ref{app:real-X}). 

\paragraph*{Length of indel segments:} As a generalization of the singular DCJ-indel model~\cite{BRA-WIL-STO-2011}, the basic idea behind our approach is that runs can be merged and accumulated with DCJ operations. This is a more parsimonious alternative to the trivial approach of inserting or deleting exclusive markers individually. However, it raises the question of whether the indels then tend to be very long, and whether this makes biological sense. Considering that it is possible to distribute the runs so that each indel is composed of 1-2 runs, we can say that the lengths of the runs play a major role in defining the length of indel segments. In the particular analysis of {\it Drosophila} complete genomes, we have an average run length of 5.1, while the maximum run length is 121. We conjecture that the long runs are mostly composed of genes that are part of a contiguous segment from the beginning, and are not really accumulated by DCJ operations. 
In a future work we intend to have a closer look into the long runs, so that we can characterize their structures and verify this conjecture. 

\section{Conclusions and discussion}\label{sec:conclusion}

In this work we proposed a new genomic distance, for the first time integrating DCJ and indel operations in a family-free setting. In this setting the whole analysis requires less pre-processing and no classification of the data, since it can be performed based on the pairwise similarities of markers in both genomes. Based on the positions and orientations of markers in both genomes we build the \emph{family-free relational diagram}. We then assign weights to the edges of the diagram, according to the given pairwise similarities. A \emph{sibling-set} of edges corresponds to a set of matched markers in both genomes. Our approach transfers weights from the edges to  matched and unmatched markers, so that, again for the first time, an optimal solution does not necessarily need to maximize the number of matched markers. Instead, the search space of our approach allows solutions composed of any number of matched markers.
The computation of our new family-free DCJ-indel distance is NP-hard and we provide an efficient ILP formulation to solve it.

The experiments on simulated data show that our ILP can handle not only bacterial genomes, but also complete genomes of fungi and insects, or sets of chromosomes of mammals and plants. We performed a comparison study of six fruit fly genomes, using the obtained distances to reconstruct the phylogenetic tree of the six species, obtaining accurate results.
This study was a first validation of the quality of our method and a more rigorous evaluation will be performed in a future work. In particular, we intend to analyze the reasons behind insertions and deletions of long segments and verify the quality of the obtained gene matchings, by comparing them to the annotated orthologies given by public databases.
Furthermore, as already mentioned, we plan to compare our ILP to the one given in~\cite{BOH-BRA-DOE-STO-2020}, once we manage to cluster the genes into families, and also to implement an ILP that is able to compute the family-free DCJ distance described in~\cite{MAR-FEI-BRA-STO-15} for multichromosomal genomes, so that we can compare it to our ILP.

\bibliographystyle{splncs04}
\bibliography{arxiv}

\appendix

\section{Supplementary figures}\label{app:figures}

\begin{figure}[ht]
\vspace{-3mm}
\begin{center}
\tiny
\setlength{\unitlength}{.45pt}
\begin{tabular}{ccc}
{\bf (i)} &&{\bf (ii)}\\
  \begin{picture}(310,90)

    \drawline(10,70)(10,30)
    \dottedline{1}(10,70)(20,70)
    \put(0,0){\color{red}\drawline(20,70)(40,70)}
    \dottedline{1}(40,70)(50,70)
    \drawline(50,70)(50,30)
    \dottedline{1}(50,30)(60,30)
    \drawline(60,70)(60,30)
    \dottedline{1}(60,70)(70,70)
    \drawline(70,70)(70,30)
    \dottedline{1}(70,30)(80,30)
    \put(0,0){\color{red}\drawline(80,30)(100,30)}
    \dottedline{1}(100,30)(110,30)
    \drawline(110,70)(110,30)
    \dottedline{1}(110,70)(120,70)
    \put(0,0){\color{red}\drawline(120,70)(140,70)}
    \dottedline{1}(140,70)(150,70)
    \drawline(150,70)(150,30)
    \dottedline{1}(150,30)(160,30)
    \drawline(160,70)(160,30)
    \dottedline{1}(160,70)(170,70)
    \put(0,0){\color{red}\drawline(170,70)(190,70)}
    \dottedline{1}(190,70)(200,70)
    \drawline(200,70)(200,30)
    \dottedline{1}(200,30)(210,30)
    \put(0,0){\color{red}\drawline(210,30)(230,30)}
    \dottedline{1}(230,30)(240,30)
    \drawline(240,70)(240,30)
    \dottedline{1}(240,70)(250,70)
    \drawline(250,70)(250,30)
    \dottedline{1}(250,30)(260,30)
    \drawline(260,70)(260,30)
    \dottedline{1}(260,70)(270,70)
    \drawline(270,70)(270,30)
    \dottedline{1}(270,30)(280,30)
    \put(0,0){\color{red}\drawline(280,30)(300,30)}
    
    \put( 10,70){\circle*{3}}
    \put( 20,70){\circle*{3}}
    \put( 40,70){\circle*{3}}
    \put( 50,70){\circle*{3}}
    \put( 60,70){\circle*{3}}
    \put( 70,70){\circle*{3}}
    \put(110,70){\circle*{3}}
    \put(120,70){\circle*{3}}
    \put(140,70){\circle*{3}}
    \put(150,70){\circle*{3}}
    \put(160,70){\circle*{3}}
    \put(170,70){\circle*{3}}
    \put(190,70){\circle*{3}}
    \put(200,70){\circle*{3}}
    \put(240,70){\circle*{3}}
    \put(250,70){\circle*{3}}
    \put(260,70){\circle*{3}}
    \put(270,70){\circle*{3}}
    
    \put( 30,82){\makebox(0,0){$e_1$}}
    \put(130,82){\makebox(0,0){$e_3$}}
    \put(115,70){\makebox(0,0){$\backslash$}}
    \put(180,82){\makebox(0,0){$e_4$}}
    \put(195,70){\makebox(0,0){$/$}}
    \put(270,82){\makebox(0,0){}}
    \put(330,82){\makebox(0,0){}}
    
    \put( 10,30){\circle*{3}}
    \put( 50,30){\circle*{3}}
    \put( 60,30){\circle*{3}}
    \put( 70,30){\circle*{3}}
    \put( 80,30){\circle*{3}}
    \put(100,30){\circle*{3}}
    \put(110,30){\circle*{3}}
    \put(150,30){\circle*{3}}
    \put(160,30){\circle*{3}}
    \put(200,30){\circle*{3}}
    \put(210,30){\circle*{3}}
    \put(230,30){\circle*{3}}
    \put(240,30){\circle*{3}}
    \put(250,30){\circle*{3}}
    \put(260,30){\circle*{3}}
    \put(270,30){\circle*{3}}
    \put(280,30){\circle*{3}}
    \put(300,30){\circle*{3}}

    \put(30,-2){\makebox(0,0){$\underbrace{~~~}_{\mathcal
    A\mbox{-run}}$}} 
    \put(90,2){\makebox(0,0){$\underbrace{~e_2~}_{\mathcal
    B\mbox{-run}}$}}
    \put(155,-4){\makebox(0,0){$\underbrace{~~~~~~~~~~~~~~~~}_{\mathcal
    A\mbox{-run}}$}} 
    \put(255,2){\makebox(0,0){$\underbrace{e_5~~~~~~~~~~~e_6}_{\mathcal
    B\mbox{-run}}$}}
    
  \end{picture}  &
    \begin{picture}(100,90)
     \put(50,60){\makebox(0,0){$\rightarrow$}}
     \put(50,40){\makebox(0,0){optimal}}
     \put(50,20){\makebox(0,0){DCJ}}
    \end{picture} &
    \begin{picture}(310,90)
    \drawline(10,70)(10,30)
    \dottedline{1}(10,70)(20,70)
    \put(0,0){\color{red}\drawline(20,70)(40,70)}
    \dottedline{1}(40,70)(50,70)
    \drawline(50,70)(50,30)
    \dottedline{1}(50,30)(60,30)
    \drawline(60,70)(60,30)
    \dottedline{1}(60,70)(70,70)
    \drawline(70,70)(70,30)
    \dottedline{1}(70,30)(80,30)
    \put(0,0){\color{red}\drawline(80,30)(100,30)}
    \dottedline{1}(100,30)(110,30)
    \drawline(110,70)(110,30)
    \dottedline{1}(110,70)(120,70)
    \drawline(120,70)(120,30)
    \dottedline{1}(120,30)(130,30)
    \put(0,0){\color{red}\drawline(130,30)(150,30)}
    \dottedline{1}(150,30)(160,30)
    \drawline(160,70)(160,30)
    \dottedline{1}(160,70)(170,70)
    \drawline(170,70)(170,30)
    \dottedline{1}(170,30)(180,30)
    \drawline(180,70)(180,30)
    \dottedline{1}(180,70)(190,70)
    \drawline(190,70)(190,30)
    \dottedline{1}(190,30)(200,30)
    \put(0,0){\color{red}\drawline(200,30)(220,30)}
    
    \put( 10,70){\circle*{3}}
    \put( 20,70){\circle*{3}}
    \put( 40,70){\circle*{3}}
    \put( 50,70){\circle*{3}}
    \put( 60,70){\circle*{3}}
    \put( 70,70){\circle*{3}}
    \put(110,70){\circle*{3}}
    \put(120,70){\circle*{3}}
    \put(160,70){\circle*{3}}
    \put(170,70){\circle*{3}}
    \put(180,70){\circle*{3}}
    \put(190,70){\circle*{3}}
    
    \put( 30,82){\makebox(0,0){$e_1$}}
    \put(150,82){\makebox(0,0){}}
    \put(210,82){\makebox(0,0){}}
    \put(270,82){\makebox(0,0){}}
    
    \put( 10,30){\circle*{3}}
    \put( 50,30){\circle*{3}}
    \put( 60,30){\circle*{3}}
    \put( 70,30){\circle*{3}}
    \put( 80,30){\circle*{3}}
    \put(100,30){\circle*{3}}
    \put(110,30){\circle*{3}}
    \put(115,70){\circle{7}}
    \put(120,30){\circle*{3}}
    \put(130,30){\circle*{3}}
    \put(150,30){\circle*{3}}
    \put(160,30){\circle*{3}}
    \put(170,30){\circle*{3}}
    \put(180,30){\circle*{3}}
    \put(190,30){\circle*{3}}
    \put(200,30){\circle*{3}}
    \put(220,30){\circle*{3}}

    \put(30,-2){\makebox(0,0){$\underbrace{~~~}_{\mathcal
    A\mbox{-run}}$}} 
 
    \put(150,2){\makebox(0,0){$\underbrace{e_2~~\;~~~~e_5~~~\;~~~~~~~e_6}_{\mathcal
    B\mbox{-run}}$}}
    
    \put(0,0){\color{red}\drawline(250,70)(270,70)}
    \put(0,0){\color{red}\drawline(280,70)(300,70)}
    \qbezier(240,70)(255,50)(270,30)
    \qbezier(310,70)(295,50)(280,30)
    \dottedline{1}(240,70)(250,70)
    \dottedline{1}(270,70)(280,70)
    \dottedline{1}(300,70)(310,70)
    \dottedline{1}(270,30)(280,30)
    
    \put(240,70){\circle*{3}}
    \put(250,70){\circle*{3}}
    \put(270,70){\circle*{3}}
    \put(275,70){\circle{7}}
    \put(280,70){\circle*{3}}
    \put(300,70){\circle*{3}}
    \put(310,70){\circle*{3}}
    \put(270,30){\circle*{3}}
    \put(280,30){\circle*{3}}
    \put(260,85){\makebox(0,0){$e_4$}}
    \put(290,85){\makebox(0,0){$e_3$}}

    \put(275,-4){\makebox(0,0){$\underbrace{~~~~~~~~~~~}_{\ma\mbox{-run}}$}}
    \end{picture}
  \end{tabular}
\end{center}
\vspace{-1mm}
\caption{ (i) A $\BB$-path with 4 runs. 
(ii) After an optimal DCJ that creates a new cycle, one $\ma$-run is accumulated (between edges $e_4$ and $e_3$ there is only an adjacency edge) and two $\mb$-runs are
merged ($e_2$ is in the same run with $e_5$ and $e_6$).
Indeed the indel-potential of the original $\BB$-path is three.}
\label{fig:run}
\vspace{-3mm}
\end{figure}
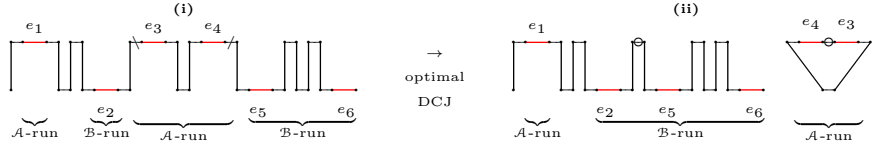

Figure~\ref{fig:run} shows a $\BB$-path with 4 runs, and how its indel-potential can be achieved.


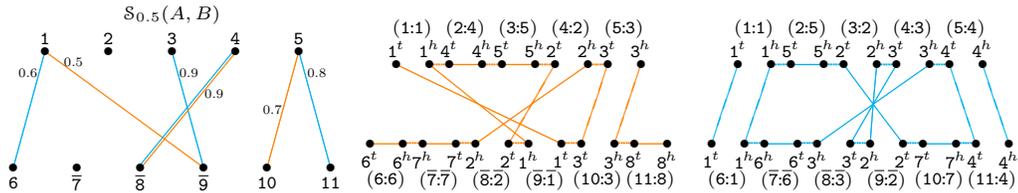
\begin{figure}[ht]
\vspace{-2mm}
  \centering

\setlength{\unitlength}{.4pt}

  \begin{tabular}{m{4.2cm}@{~~~~~}m{4cm}@{~~~~~}m{4cm}}
    \begin{picture}(300,165)
      \scriptsize
      \put(150,155){\makebox(0,0){$\GSx{0.5}(A,B)$}}
      \thinlines
      \put(0,0){\color{cyan}
      \qbezier(30,120)(15,65)(0,10)
      }
      \put(14,100){\makebox(0,0){\scalebox{.7}{0.6}}}

      \put(0,0){\color{orange}
      \qbezier(30,120)(105,65)(180,10)
      }
      \put(57,110){\makebox(0,0){\scalebox{.7}{0.5}}}
      \put(0,0){\color{cyan}
      \qbezier(150,120)(165,65)(180,10)
      \qbezier(208,120)(163,65)(118,10)
      }
      \put(166,100){\makebox(0,0){\scalebox{.7}{0.9}}}
      \put(190,80){\makebox(0,0){\scalebox{.7}{0.9}}}
      
      \put(0,0){\color{orange}
      \qbezier(211,120)(166,65)(121,10)
      \qbezier(270,120)(255,65)(240,10)
      }
      \put(245,65){\makebox(0,0){\scalebox{.7}{0.7}}}
      \put(0,0){\color{cyan}
      \qbezier(270,120)(285,65)(300,10)
      }
      \put(287,100){\makebox(0,0){\scalebox{.7}{0.8}}}

      \put( 30,120){\circle*{8}}
      \put( 90,120){\circle*{8}}
      \put(150,120){\circle*{8}}
      \put(210,120){\circle*{8}}
      \put(270,120){\circle*{8}}

      \put(  0,10){\circle*{8}}
      \put( 60,10){\circle*{8}}
      \put(120,10){\circle*{8}}
      \put(180,10){\circle*{8}}
      \put(240,10){\circle*{8}}
      \put(300,10){\circle*{8}}

      \put( 30,132){\makebox(0,0){$\T{1}$}}
      \put( 90,132){\makebox(0,0){$\T{2}$}}
      \put(150,132){\makebox(0,0){$\T{3}$}}
      \put(210,132){\makebox(0,0){$\T{4}$}}
      \put(270,132){\makebox(0,0){$\T{5}$}}
      
      \put(  0,-4){\makebox(0,0){$\T{6}$}}
      \put( 60,-4){\makebox(0,0){$\rev{\T{7}}$}}
      \put(120,-4){\makebox(0,0){$\rev{\T{8}}$}}
      \put(180,-4){\makebox(0,0){$\rev{\T{9}}$}}
      \put(240,-4){\makebox(0,0){$\T{10}$}}
      \put(300,-4){\makebox(0,0){$\T{11}$}}
\end{picture} &
    \setlength{\unitlength}{.5pt}
    \begin{picture}(225,100)
      \scriptsize
      \put(0,0){\color{orange}
        \dottedline[$\cdot$]{2}(25,10)(40,10)
        \dottedline[$\cdot$]{2}(65,10)(80,10)
        \dottedline[$\cdot$]{2}(85,70)(100,70)
        \dottedline[$\cdot$]{2}(145,10)(160,10)
        
        \qbezier(20,70)(82,40)(145,10)
        \qbezier(45,70)(82,40)(120,10)
        \drawline(140,70)(105,10)
        \qbezier(165,70)(122,40)(80,10)
        \drawline(180,70)(160,10)
        \drawline(205,70)(185,10)
      
        \drawline(60,70)(85,70)
        \drawline(100,70)(125,70)
        \drawline(0,10)(25,10)
        \drawline(40,10)(65,10)
        \drawline(200,10)(225,10)
        
      \dottedline[$\cdot$]{2}(45,70)(60,70)
      \dottedline[$\cdot$]{2}(125,70)(140,70)
      \dottedline[$\cdot$]{2}(165,70)(180,70)
      
      \dottedline[$\cdot$]{2}(105,10)(120,10)
      \dottedline[$\cdot$]{2}(185,10)(200,10)

      }

      \put( 20,70){\circle*{5}}
      \put( 45,70){\circle*{5}}
      \put( 60,70){\circle*{5}}
      \put( 85,70){\circle*{5}}
      \put(100,70){\circle*{5}}
      \put(125,70){\circle*{5}}
      \put(140,70){\circle*{5}}
      \put(165,70){\circle*{5}}
      \put(180,70){\circle*{5}}
      \put(205,70){\circle*{5}}

      \put( 20,82){\makebox(0,0){$\tl{\T{1}}$}}
      \put( 45,82){\makebox(0,0){$\hd{\T{1}}$}}
      \put( 60,82){\makebox(0,0){$\tl{\T{4}}$}}
      \put( 85,82){\makebox(0,0){$\hd{\T{4}}$}}
      \put(100,82){\makebox(0,0){$\tl{\T{5}}$}}
      \put(125,82){\makebox(0,0){$\hd{\T{5}}$}}
      \put(140,82){\makebox(0,0){$\tl{\T{2}}$}}
      \put(165,82){\makebox(0,0){$\hd{\T{2}}$}}
      \put(180,82){\makebox(0,0){$\tl{\T{3}}$}}
      \put(205,82){\makebox(0,0){$\hd{\T{3}}$}}
      
      \put( 32,97){\makebox(0,0){$(\T{1}\!\!:\!\!\T{1})$}}
      \put( 72,97){\makebox(0,0){$(\T{2}\!\!:\!\!\T{4})$}}
      \put(112,97){\makebox(0,0){$(\T{3}\!\!:\!\!\T{5})$}}
      \put(152,97){\makebox(0,0){$(\T{4}\!\!:\!\!\T{2})$}}
      \put(192,97){\makebox(0,0){$(\T{5}\!\!:\!\!\T{3})$}}

      \put(  0,10){\circle*{5}}
      \put( 25,10){\circle*{5}}
      \put( 40,10){\circle*{5}}
      \put( 65,10){\circle*{5}}
      \put( 80,10){\circle*{5}}
      \put(105,10){\circle*{5}}
      \put(120,10){\circle*{5}}
      \put(145,10){\circle*{5}}
      \put(160,10){\circle*{5}}
      \put(185,10){\circle*{5}}
      \put(200,10){\circle*{5}}
      \put(225,10){\circle*{5}}
      
      \put( 0,-2){\makebox(0,0){$\tl{\T{6}}$}}
      \put( 25,-2){\makebox(0,0){$\hd{\T{6}}$}}
      \put( 40,-2){\makebox(0,0){$\hd{\T{7}}$}}
      \put( 65,-2){\makebox(0,0){$\tl{\T{7}}$}}
      \put( 80,-2){\makebox(0,0){$\hd{\T{2}}$}}
      \put(105,-2){\makebox(0,0){$\tl{\T{2}}$}}
      \put(120,-2){\makebox(0,0){$\hd{\T{1}}$}}
      \put(145,-2){\makebox(0,0){$\tl{\T{1}}$}}
      \put(160,-2){\makebox(0,0){$\tl{\T{3}}$}}
      \put(185,-2){\makebox(0,0){$\hd{\T{3}}$}}
      \put(200,-2){\makebox(0,0){$\tl{\T{8}}$}}
      \put(225,-2){\makebox(0,0){$\hd{\T{8}}$}}
      
      \put( 12,-17){\makebox(0,0){$(\T{6}\!\!:\!\!\T{6})$}}
      \put( 52,-17){\makebox(0,0){$(\rev{\T{7}}\!\!:\!\!\rev{\T{7}})$}}
      \put( 92,-17){\makebox(0,0){$(\rev{\T{8}}\!\!:\!\!\rev{\T{2}})$}}
      \put(132,-17){\makebox(0,0){$(\rev{\T{9}}\!\!:\!\!\rev{\T{1}})$}}
      \put(172,-17){\makebox(0,0){$(\T{10}\!\!:\!\!\T{3})$}}
      \put(212,-17){\makebox(0,0){$(\T{11}\!\!:\!\!\T{8})$}}

    \end{picture} &
    \setlength{\unitlength}{.5pt}
    \begin{picture}(225,100)
      \scriptsize
      \put(0,0){\color{cyan}
        \dottedline[$\cdot$]{2}(45,70)(60,70)
        \dottedline[$\cdot$]{2}(25,10)(40,10)
        \dottedline[$\cdot$]{2}(65,10)(80,10)
        \dottedline[$\cdot$]{2}(125,70)(140,70)
        \dottedline[$\cdot$]{2}(145,10)(160,10)
        
        \drawline(20,70)(0,10)
        \drawline(45,70)(25,10)
        \qbezier(100,70)(122,40)(145,10)
        \qbezier(125,70)(122,40)(120,10)
        \qbezier(140,70)(122,40)(105,10)
        \qbezier(165,70)(122,40)(80,10)
        \drawline(180,70)(200,10)
        \drawline(205,70)(225,10)

        \drawline(60,70)(85,70)
        \drawline(40,10)(65,10)
        \drawline(160,10)(185,10)
        
      \dottedline[$\cdot$]{2}(85,70)(100,70)
      \dottedline[$\cdot$]{2}(165,70)(180,70)

      \dottedline[$\cdot$]{2}(105,10)(120,10)
      \dottedline[$\cdot$]{2}(185,10)(200,10)

      }

      \put( 20,70){\circle*{5}}
      \put( 45,70){\circle*{5}}
      \put( 60,70){\circle*{5}}
      \put( 85,70){\circle*{5}}
      \put(100,70){\circle*{5}}
      \put(125,70){\circle*{5}}
      \put(140,70){\circle*{5}}
      \put(165,70){\circle*{5}}
      \put(180,70){\circle*{5}}
      \put(205,70){\circle*{5}}

      \put( 20,82){\makebox(0,0){$\tl{\T{1}}$}}
      \put( 45,82){\makebox(0,0){$\hd{\T{1}}$}}
      \put( 60,82){\makebox(0,0){$\tl{\T{5}}$}}
      \put( 85,82){\makebox(0,0){$\hd{\T{5}}$}}
      \put(100,82){\makebox(0,0){$\tl{\T{2}}$}}
      \put(125,82){\makebox(0,0){$\hd{\T{2}}$}}
      \put(140,82){\makebox(0,0){$\tl{\T{3}}$}}
      \put(165,82){\makebox(0,0){$\hd{\T{3}}$}}
      \put(180,82){\makebox(0,0){$\tl{\T{4}}$}}
      \put(205,82){\makebox(0,0){$\hd{\T{4}}$}}
      
      \put( 32,97){\makebox(0,0){$(\T{1}\!\!:\!\!\T{1})$}}
      \put( 72,97){\makebox(0,0){$(\T{2}\!\!:\!\!\T{5})$}}
      \put(112,97){\makebox(0,0){$(\T{3}\!\!:\!\!\T{2})$}}
      \put(152,97){\makebox(0,0){$(\T{4}\!\!:\!\!\T{3})$}}
      \put(192,97){\makebox(0,0){$(\T{5}\!\!:\!\!\T{4})$}}
      
      \put(  0,10){\circle*{5}}
      \put( 25,10){\circle*{5}}
      \put( 40,10){\circle*{5}}
      \put( 65,10){\circle*{5}}
      \put( 80,10){\circle*{5}}
      \put(105,10){\circle*{5}}
      \put(120,10){\circle*{5}}
      \put(145,10){\circle*{5}}
      \put(160,10){\circle*{5}}
      \put(185,10){\circle*{5}}
      \put(200,10){\circle*{5}}
      \put(225,10){\circle*{5}}
      
      \put( 0,-2){\makebox(0,0){$\tl{\T{1}}$}}
      \put( 25,-2){\makebox(0,0){$\hd{\T{1}}$}}
      \put( 40,-2){\makebox(0,0){$\hd{\T{6}}$}}
      \put( 65,-2){\makebox(0,0){$\tl{\T{6}}$}}
      \put( 80,-2){\makebox(0,0){$\hd{\T{3}}$}}
      \put(105,-2){\makebox(0,0){$\tl{\T{3}}$}}
      \put(120,-2){\makebox(0,0){$\hd{\T{2}}$}}
      \put(145,-2){\makebox(0,0){$\tl{\T{2}}$}}
      \put(160,-2){\makebox(0,0){$\tl{\T{7}}$}}
      \put(185,-2){\makebox(0,0){$\hd{\T{7}}$}}
      \put(200,-2){\makebox(0,0){$\tl{\T{4}}$}}
      \put(225,-2){\makebox(0,0){$\hd{\T{4}}$}}
      
      \put( 12,-17){\makebox(0,0){$(\T{6}\!\!:\!\!\T{1})$}}
      \put( 52,-17){\makebox(0,0){$(\rev{\T{7}}\!\!:\!\!\rev{\T{6}})$}}
      \put( 92,-17){\makebox(0,0){$(\rev{\T{8}}\!\!:\!\!\rev{\T{3}})$}}
      \put(132,-17){\makebox(0,0){$(\rev{\T{9}}\!\!:\!\!\rev{\T{2}})$}}
      \put(172,-17){\makebox(0,0){$(\T{10}\!\!:\!\!\T{7})$}}
      \put(212,-17){\makebox(0,0){$(\T{11}\!\!:\!\!\T{4})$}}

    \end{picture}
    \end{tabular}
    
    \smallskip
    
    
  \caption{
  Considering the same genomes 
  $A=\{\T{1}\z\z\T{2}\z\z\T{3}\z\z\T{4}\z\z\T{5}\}$
  and $B=\{\T{6}\z\z\rev{\T{7}}\z\z\rev{\T{8}}\z\z\rev{\T{9}}\z\z\T{10}\z\z\T{11}\}$
  as in
  Figure~\ref{fig:example-gsg}, let $M_1$ (orange) and $M_2$ (cyan) be two
  distinct maximal matchings in $\GSx{0.5}(A,B)$.
  In the middle part we show the diagram $R(A^{M_1},B^{M_1})$, that
  has two $\AB$-paths and one $\AB$-cycle, giving $\ddcjid(A^{M_1},B^{M_1})=4$. 
  In the right part we show the diagram 
  $R(A^{M_2},B^{M_2})$, that has two $\AB$-paths
    and two $\AB$-cycles, giving  $\ddcjid(A^{M_2},B^{M_2})=3$.
    }
  \label{fig:matchings2}
\end{figure}

\noindent Figure~\ref{fig:matchings2} shows -- in contrast to the marker similarity graph $\GSx{0.1}(A,B)$ of Figure~\ref{fig:matchings} -- the graph $\GSx{0.5}(A,B)$, two of its matchings and the corresponding weighted relational diagrams. 



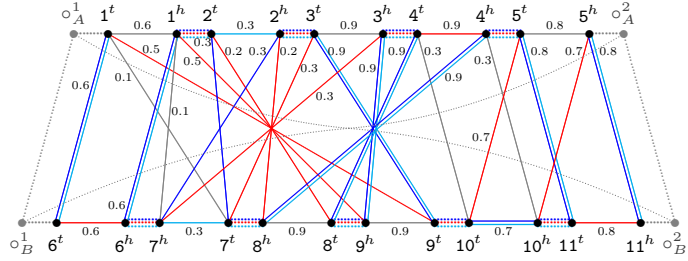
\begin{figure}[ht]
\centering
\setlength{\unitlength}{.65pt}
\begin{picture}(340,140)
       \scriptsize
        \put(0,0){\color{blue}
        \dottedline[$\cdot$]{2}(40,12)(60,12)
        }
        \put(0,0){\color{red}
        \dottedline[$\cdot$]{2}(40,10)(60,10)
        }
        \put(0,0){\color{cyan}
        \dottedline[$\cdot$]{2}(40,8)(60,8)
        }
        
        \put(0,0){\color{blue}
        \dottedline[$\cdot$]{2}(100,12)(120,12)
        }
        \put(0,0){\color{red}
        \dottedline[$\cdot$]{2}(100,10)(120,10)
        }
        \put(0,0){\color{cyan}
        \dottedline[$\cdot$]{2}(100,8)(120,8)
        }
        
        \put(0,0){\color{blue}
        \dottedline[$\cdot$]{2}(160,12)(180,12)
        }
        \put(0,0){\color{red}
        \dottedline[$\cdot$]{2}(160,10)(180,10)
        }
        \put(0,0){\color{cyan}
        \dottedline[$\cdot$]{2}(160,8)(180,8)
        }
        
        \put(0,0){\color{blue}
        \dottedline[$\cdot$]{2}(220,12)(240,12)
        }
        \put(0,0){\color{red}
        \dottedline[$\cdot$]{2}(220,10)(240,10)
        }
        \put(0,0){\color{cyan}
        \dottedline[$\cdot$]{2}(220,8)(240,8)
        }
        
        \put(0,0){\color{blue}
        \dottedline[$\cdot$]{2}(280,12)(300,12)
        }
        \put(0,0){\color{red}
        \dottedline[$\cdot$]{2}(280,10)(300,10)
        }
        \put(0,0){\color{cyan}
        \dottedline[$\cdot$]{2}(280,8)(300,8)
        }
  
        \put(0,0){\color{blue}
        \dottedline[$\cdot$]{2}(70,122)(90,122)
        }
        \put(0,0){\color{red}
        \dottedline[$\cdot$]{2}(70,120)(90,120)
        }
        \put(0,0){\color{cyan}
        \dottedline[$\cdot$]{2}(70,118)(90,118)
        }
        
        \put(0,0){\color{blue}
        \dottedline[$\cdot$]{2}(130,122)(150,122)
        }
        \put(0,0){\color{red}
        \dottedline[$\cdot$]{2}(130,120)(150,120)
        }
        \put(0,0){\color{cyan}
        \dottedline[$\cdot$]{2}(130,118)(150,118)
        }
        
        \put(0,0){\color{blue}
        \dottedline[$\cdot$]{2}(190,122)(210,122)
        }
        \put(0,0){\color{red}
        \dottedline[$\cdot$]{2}(190,120)(210,120)
        }
        \put(0,0){\color{cyan}
        \dottedline[$\cdot$]{2}(190,118)(210,118)
        }
        
        \put(0,0){\color{blue}
        \dottedline[$\cdot$]{2}(250,122)(270,122)
        }
        \put(0,0){\color{red}
        \dottedline[$\cdot$]{2}(250,120)(270,120)
        }
        \put(0,0){\color{cyan}
        \dottedline[$\cdot$]{2}(250,118)(270,118)
        }
      
        \put(0,0){\color{blue}
        \qbezier(29,120)(14,65)(-1,10)
        }
        \put(0,0){\color{cyan}
        \qbezier(31,120)(16,65)(1,10)
        }
        \put(14,90){\makebox(0,0){\scalebox{.65}{0.6}}}
        
        \put(0,0){\color{gray}
        \qbezier(30,120)(65,65)(100,10)
        }
        \put(39,95){\makebox(0,0){\scalebox{.65}{0.1}}}
        
        \put(0,0){\color{red}
        \qbezier(30,120)(125,65)(220,10)
        }
        \put(55,111){\makebox(0,0){\scalebox{.65}{0.5}}}
        
        \put(0,0){\color{blue}
        \qbezier(69,120)(54,65)(39,10)
        }
        \put(0,0){\color{cyan}
        \qbezier(71,120)(56,65)(41,10)
        }
        \put(35,20){\makebox(0,0){\scalebox{.65}{0.6}}}
        
        \put(0,0){\color{gray}
        \qbezier(70,120)(65,65)(60,10)
        }
        \put(72,75){\makebox(0,0){\scalebox{.65}{0.1}}}
        
        \put(0,0){\color{red}
        \qbezier(70,120)(125,65)(180,10)
        }
        \put(79,104){\makebox(0,0){\scalebox{.65}{0.5}}}
        
        \put(0,0){\color{blue}
        \qbezier(90,120)(95,65)(100,10)
        }
        \put(85,115){\makebox(0,0){\scalebox{.65}{0.3}}}
        
        \put(0,0){\color{red}
        \qbezier(90,120)(125,65)(160,10)
        }
        \put(103,110){\makebox(0,0){\scalebox{.65}{0.2}}}
        
        \put(0,0){\color{blue}
        \qbezier(130,120)(95,65)(60,10)
        }
        \put(117,110){\makebox(0,0){\scalebox{.65}{0.3}}}
        
        \put(0,0){\color{red}
        \qbezier(130,120)(125,65)(120,10)
        }
        \put(135,110){\makebox(0,0){\scalebox{.65}{0.2}}}
        
        \put(0,0){\color{red}
        \qbezier(150,120)(125,65)(100,10)
        }
        \put(147,100){\makebox(0,0){\scalebox{.65}{0.3}}}
        
        \put(0,0){\color{blue}
        \qbezier(149,120)(184,65)(219,10)
        }
        \put(0,0){\color{cyan}
        \qbezier(151,120)(186,65)(221,10)
        }
        \put(163,110){\makebox(0,0){\scalebox{.65}{0.9}}}
        
        \put(0,0){\color{red}
        \qbezier(190,120)(125,65)(60,10)
        }
        \put(156,84){\makebox(0,0){\scalebox{.65}{0.3}}}
        
        \put(0,0){\color{blue}
        \qbezier(189,120)(184,65)(179,10)
        }
        \put(0,0){\color{cyan}
        \qbezier(191,120)(186,65)(181,10)
        }
        \put(181,100){\makebox(0,0){\scalebox{.65}{0.9}}}
        
        \put(0,0){\color{blue}
        \qbezier(209,120)(184,65)(159,10)
        }
        \put(0,0){\color{cyan}
        \qbezier(211,120)(186,65)(161,10)
        }
        \put(198,110){\makebox(0,0){\scalebox{.65}{0.9}}}
        
        \put(0,0){\color{gray}
        \qbezier(210,120)(225,65)(240,10)
        }
        \put(219,110){\makebox(0,0){\scalebox{.65}{0.3}}}
        
        \put(0,0){\color{blue}
        \qbezier(249,120)(184,65)(119,10)
        }
        \put(0,0){\color{cyan}
        \qbezier(252,120)(187,65)(122,10)
        }
        \put(231,95){\makebox(0,0){\scalebox{.65}{0.9}}}
        
        \put(0,0){\color{gray}
        \qbezier(250,120)(265,65)(280,10)
        }
        \put(248,105){\makebox(0,0){\scalebox{.65}{0.3}}}
        
        \put(0,0){\color{red}
        \qbezier(270,120)(255,65)(240,10)
        }
        \put(247,60){\makebox(0,0){\scalebox{.65}{0.7}}}
        
        \put(0,0){\color{blue}
        \qbezier(269,120)(284,65)(299,10)
        }
        \put(0,0){\color{cyan}
        \qbezier(271,120)(286,65)(301,10)
        }
        \put(281,110){\makebox(0,0){\scalebox{.65}{0.8}}}
        
        \put(0,0){\color{red}
        \qbezier(310,120)(295,65)(280,10)
        }
        \put(301,110){\makebox(0,0){\scalebox{.65}{0.7}}}
        
        \put(0,0){\color{blue}
        \qbezier(309,120)(324,65)(339,10)
        }
        \put(0,0){\color{cyan}
        \qbezier(311,120)(326,65)(341,10)
        }
        \put(321,110){\makebox(0,0){\scalebox{.65}{0.8}}}
      
        \put(0,0){\color{red}
        \drawline(0,10)(40,10)
        }
        \put(20,5){\makebox(0,0){\scalebox{.65}{0.6}}}
        \put(0,0){\color{cyan}
        \drawline(60,10)(100,10)
        }
        \put(80,5){\makebox(0,0){\scalebox{.65}{0.3}}}
        \put(0,0){\color{gray}
        \drawline(120,10)(160,10)
        }
        \put(140,5){\makebox(0,0){\scalebox{.65}{0.9}}}
        \put(0,0){\color{gray}
        \drawline(180,10)(220,10)
        }
        \put(200,5){\makebox(0,0){\scalebox{.65}{0.9}}}
        \put(0,0){\color{blue}
        \drawline(240,11)(280,11)
        }
        \put(0,0){\color{cyan}
        \drawline(240,9)(280,9)
        }
        \put(260,5){\makebox(0,0){\scalebox{.65}{0.7}}}
        \put(0,0){\color{red}
        \drawline(300,10)(340,10)
        }
        \put(320,5){\makebox(0,0){\scalebox{.65}{0.8}}}
        \put(0,0){\color{gray}
        \drawline(30,120)(70,120)
        }
        \put(50,125){\makebox(0,0){\scalebox{.65}{0.6}}}
        \put(0,0){\color{cyan}
        \drawline(90,120)(130,120)
        }
        \put(110,125){\makebox(0,0){\scalebox{.65}{0.3}}}
        \put(0,0){\color{gray}
        \drawline(150,120)(190,120)
        }
        \put(170,125){\makebox(0,0){\scalebox{.65}{0.9}}}
        \put(0,0){\color{red}
        \drawline(210,120)(250,120)
        }
        \put(230,125){\makebox(0,0){\scalebox{.65}{0.9}}}
        \put(0,0){\color{gray}
        \drawline(270,120)(310,120)
        }
        \put(290,125){\makebox(0,0){\scalebox{.65}{0.8}}}
        
        \put(0,0){\color{gray}
        
        \dottedline[$\cdot$]{2}(10,120)(30,120)
        \dottedline[$\cdot$]{2}(310,120)(330,120)
        \dottedline[$\cdot$]{2}(-20,10)(0,10)
        \dottedline[$\cdot$]{2}(340,10)(360,10)
        
       \qbezier[110](10,120)(90,72)(170,65)
       \qbezier[120](170,65)(265,57)(360,10)
       \dottedline[{\tiny .}]{2}(10,120)(-20,10)
       \qbezier[110](330,120)(250,72)(170,65)
       \qbezier[120](170,65)(75,57)(-20,10)
       \dottedline[{\tiny .}]{2}(330,120)(360,10)
       }
      
      \put( 10,120){\color{gray}\circle*{5}}
      \put( 30,120){\circle*{5}}
      \put( 70,120){\circle*{5}}
      \put( 90,120){\circle*{5}}
      \put(130,120){\circle*{5}}
      \put(150,120){\circle*{5}}
      \put(190,120){\circle*{5}}
      \put(210,120){\circle*{5}}
      \put(250,120){\circle*{5}}
      \put(270,120){\circle*{5}}
      \put(310,120){\circle*{5}}
      \put(330,120){\color{gray}\circle*{5}}
      
      \put( 10,132){\makebox(0,0){\color{darkgray}$\circ_A^1$}}
      \put( 30,132){\makebox(0,0){$\tl{\T{1}}$}}
      \put( 70,132){\makebox(0,0){$\hd{\T{1}}$}}
      \put( 90,132){\makebox(0,0){$\tl{\T{2}}$}}
      \put(130,132){\makebox(0,0){$\hd{\T{2}}$}}
      \put(150,132){\makebox(0,0){$\tl{\T{3}}$}}
      \put(190,132){\makebox(0,0){$\hd{\T{3}}$}}
      \put(210,132){\makebox(0,0){$\tl{\T{4}}$}}
      \put(250,132){\makebox(0,0){$\hd{\T{4}}$}}
      \put(270,132){\makebox(0,0){$\tl{\T{5}}$}}
      \put(310,132){\makebox(0,0){$\hd{\T{5}}$}}
      \put(330,132){\makebox(0,0){\color{darkgray}$\circ_A^2$}}
      
      \put(-20,10){\color{gray}\circle*{5}}
      \put(  0,10){\circle*{5}}
      \put( 40,10){\circle*{5}}
      \put( 60,10){\circle*{5}}
      \put(100,10){\circle*{5}}
      \put(120,10){\circle*{5}}
      \put(160,10){\circle*{5}}
      \put(180,10){\circle*{5}}
      \put(220,10){\circle*{5}}
      \put(240,10){\circle*{5}}
      \put(280,10){\circle*{5}}
      \put(300,10){\circle*{5}}
      \put(340,10){\circle*{5}}
      \put(360,10){\color{gray}\circle*{5}}
      
      \put(-20,-2){\makebox(0,0){\color{darkgray}$\circ_B^1$}}
      \put( 0,-2){\makebox(0,0){$\tl{\T{6}}$}}
      \put( 40,-2){\makebox(0,0){$\hd{\T{6}}$}}
      \put( 60,-2){\makebox(0,0){$\hd{\T{7}}$}}
      \put(100,-2){\makebox(0,0){$\tl{\T{7}}$}}
      \put(120,-2){\makebox(0,0){$\hd{\T{8}}$}}
      \put(160,-2){\makebox(0,0){$\tl{\T{8}}$}}
      \put(180,-2){\makebox(0,0){$\hd{\T{9}}$}}
      \put(220,-2){\makebox(0,0){$\tl{\T{9}}$}}
      \put(240,-2){\makebox(0,0){$\tl{\T{10}}$}}
      \put(280,-2){\makebox(0,0){$\hd{\T{10}}$}}
      \put(300,-2){\makebox(0,0){$\tl{\T{11}}$}}
      \put(340,-2){\makebox(0,0){$\hd{\T{11}}$}}
      \put(360,-2){\makebox(0,0){\color{darkgray}$\circ_B^2$}}
      
    \end{picture}
    \vspace{2mm}
  \caption{The capped version of the family-free relational diagram from Figure~\ref{fig:multi}.
  }
  \label{fig:capped-multi}
\end{figure}

\noindent Figure~\ref{fig:capped-multi} shows an example of a capped family-free relational diagram.


\section{\label{app:complex}Computational complexity of the family-free DCJ-indel distance}

In the family-based setting, 
if two genomes contain the same number of occurrences of each marker, they are said to be \emph{balanced}. Notice that there are no exclusive  markers in this case. The problem of computing the DCJ distance of balanced genomes is NP-hard~\cite{SHA-LIN-MOR-2015}. We use this problem in straightforward reductions to show that computing the family-free DCJ-indel distance, both weighted and unweighted, are NP-hard problems. The first step of the reduction is the creation of a similarity graph and transforming family-based into family-free genomes as follows. Let $A'$ and $B'$ be two balanced genomes. Rename each occurrence of each marker in each of the two genomes $A'$ and $B'$, so that we get two family-free genomes $A$ and $B$; and build the marker similarity graph $\GSx{1}(A, B)$ by connecting markers $a$ in $A$ and $b$ in $B$ if they correspond to occurrences of the same marker in the original family-based genomes 
and setting $\sigma(ab) = 1$. Note that, for any $0 \leq x \leq 1$, we have $\GSx{x}(A,B)=\GSx{1}(A,B)$.

\begin{theorem} \label{thm:NP-hardness-unw}
For given genomes $A$ and $B$ and a marker similarity graph $\GSx{x}(A, B)$ for any $0 \leq x \leq 1$, computing the unweighted family-free DCJ-indel distance $\dunwffdcjid(A, B, \GSx{x})$ is NP-hard.
\end{theorem}
\proof
Given balanced genomes $A'$ and $B'$, we obtain family-free genomes $A$ and $B$ and their marker similarity graph $\GSx{x}(A, B)=\GSx{1}(A,B)$ as described above. Then, a maximal matching in $\GSx{x}(A, B)$ that finds the unweighted family-free DCJ-indel distance $\dunwffdcjid(A, B, \GS{x})$ implies immediately in finding the DCJ distance of balanced genomes $A'$ and $B'$.
\qed

\begin{theorem} \label{thm:NP-hardness-w}
For given genomes $A$ and $B$ and a marker similarity graph $\GSx{x}(A, B)$ for any $0 \leq x \leq 1$, computing the weighted family-free DCJ-indel distance $\dffdcjid(A, B, \GSx{x})$ is NP-hard.
\end{theorem}
\proof
Given balanced genomes $A'$ and $B'$, we obtain family-free genomes $A$ and $B$ and their marker similarity graph $\GSx{x}(A, B)=\GSx{1}(A,B)$ as described above.
We now show that $\dffdcjid(A, B, \GSx{1})$ is given by a matching in $\GSx{1}(A, B)$ of maximal cardinality.

Let $n := |A| = |B|$. Let $M$ be a matching in $\GSx{1}(A, B)$. First, notice that $|M| = w(M)$, for any matching $M$ in $\GSx{1}(A, B)$. Thus, we can 
compute the weighted DCJ-indel distance $\wcostid(A^M,B^M)$ 
as follows:
\[
\wcostid(A^M,B^M) = \ddcj(A^M, B^M) + \sum_{\mathclap{C\in
\WR(A^M,B^M)}} \lambda(C) - \delta_M +  w(\CM) = \ddcjid(A^M, B^M) + w(\CM)\:.
\]

Notice that $w(\CM) = 2(n - |M|)$ for any matching $M$ in $\GSx{1}(A, B)$. Then, if $M$ is maximal, i.e., if $|M| = n$, no indel operation is performed on the genomes and thus 
\[
\wcostid(A^M, B^M) = \ddcj(A^M, B^M) \leq n\,.
\]
On the other hand, if we take the trivial empty matching $M_\emptyset$, no DCJ operation is performed and we have at least 2 indel operations (one per chromosome of $A$ and one per chromosome of $B$). Since $w(\CM_\emptyset) = 2n$, we have 
\[
\wcostid(A^{M_\emptyset}, B^{M_\emptyset}) = \ddcjid(A^{M_\emptyset}, B^{M_\emptyset}) + w(\CM_\emptyset) \geq 2 + 2n\,.
\]
Therefore, the trivial empty matching is definitely not a candidate for giving the optimal solution.


Now let $M_1, M_{2}, \ldots, M_n$ be a sequence of matchings in $\GSx{1}(A, B)$ such that, for any $1 \leq i \leq n$, $|M_i| = i$ and $M_{i+1}=M_{i} \cup \{e\}$, with $e \notin M_{i}$, i.e., $M_{i+1}$ is obtained by adding exactly one edge to the matching $M_i$. We necessarily have
\[w(\CM_{i+1}) = w(\CM_i) - 2\,, \text{\ and}\] 
\[\ddcjid(A^{M_{i+1}}, B^{M_{i+1}}) \leq \ddcjid(A^{M_i}, B^{M_i}) + 2\,,\] 
meaning that, in the worst case, we increase the size of a matching and keep the same distance: $$\wcostid(A^{M_{i+1}},B^{M_{i+1}}) \leq \wcostid(A^{M_{i}},B^{M_{i}})\,.$$ 

However, for the last pair of consecutive matchings $M_{n-1}$ and $M_n$, we know that the number of indels decrease from 2 to 0, while the DCJ part of the formula increases at most +2, that is $\ddcjid(A^{M_{n}}, B^{M_{n}}) \leq \ddcjid(A^{M_{n-1}}, B^{M_{n-1}})$. Since we still have $w(\CM_{n}) = w(\CM_{n-1}) - 2$,  it is clear that
\begin{align*}
\wcostid(A^{M_n}, B^{M_n}) &\leq \wcostid(A^{M_{n-1}}, B^{M_{n-1}}) - 2\,.
\end{align*}
Therefore, the weighted family-free DCJ-indel distance $\dffdcjid(A, B, \GSx{1})$ corresponds to a maximal matching of $\GSx{1}(A,B)$. 
And since all maximum matchings of $\GSx{1}(A, B)$ give mapped genomes without exclusive markers, $\dffdcjid(A, B, \GSx{1})$ is exactly the DCJ distance of original balanced genomes $A'$ and $B'$.
\qed


\section{ILP for computing the unweighted family-free DCJ distance}\label{app:unweighted}

For computing the unweighted 
$\dunwffdcjid(A,B,\GSx{x})$ we need to slightly modify the ILP described in Algorithm~\ref{alg:ilp-w}. Besides all its constraints, variables and domains, to ensure that a matching is of maximal cardinality, we add a new constraint as follows:
\[
{\everymath={\displaystyle}
\arraycolsep=3pt
\qquad 
\begin{array}{lcll} 
x_d + x_{d^\prime} & \leq & 1 \qquad \qquad \left\{ \begin{array}{l} \forall ~ d \in \selfedges{A}, d^\prime \in \selfedges{B}\,, \\ e, e^\prime \in E_\ext \text{ and } e,e^\prime \text{ are siblings, and } \\ d \cap (e \cup e') \neq \emptyset \text{ and } d' \cap (e \cup e') \neq \emptyset \end{array} \right. & \qquad \qquad \text{(C.12)}
\end{array}
}
\]
We also simplify the objective function to
\[
\min \qquad p_* + \frac{1}{2} \sum_{\mathclap{e \in E_\ext}} x_e - \sum_{\mathclap{1 \leq i \leq |V|}} z_i + \sum_{\mathclap{k \in K}} s_k + \frac{1}{2} \sum_{\mathclap{e \in E}} t_e\,.
\]


\section{Generation of simulated data}\label{app:simulated}

Here we describe the process and the parameters used in Artificial Life Simulator (ALF)~\cite{DAGD12} for generating our simulated data. Each one of the 190 instances generated consists of a pair of simulated genomes. We used the default values for parameters not mentioned.
PAM units were used as time scale for simulation, starting with a randomly generated root genome with $10{,}000$ genes, whose lengths where drawn from a Gamma distribution with $k = 2.4019$ and $\theta = 133.8063$ (minimum length $100$). We used a custom evolutionary tree defining an speciation event after 25 time units, resulting in two leaf species, which evolved for additional 25 time units. The WAG substitution model was used together with Zipfian indels in DNA sequences with rate $0.0002$ (maximum length $50$). Such rate varies among sites according to a Gamma distribution with shape $1$ and $10$ classes. In addition, we set the rate of invariable sites to $0.001$. Inversions and translocations of up to $30$ genes were allowed at a rate of $0.0025$. Finally, for generating instances comprising genes with different numbers of homologies, we varied the gene duplication and the gene loss rates between $1 \times 10^{-5}$ and $2 \times 10^{-3}$.

\section{Analysis of \emph{Drosophila} genomes}\label{app:real}

\begin{figure}[ht]
\centering
  \footnotesize
\setlength{\tabcolsep}{3pt}
\begin{tabular}{ll@{~~~~~~~~~~}ll}
\bf Species & \bf NCBI Assembly & \bf Species & \bf NCBI Assembly\\[2mm]
\emph{Drosophila busckii}  & ASM1175060v1 & \emph{Drosophila sechellia}  & ASM438219v1\\[2mm]
\emph{Drosophila melanogaster}  & Release 6 plus ISO1 MT & \emph{Drosophila simulans}  & ASM75419v2\\[2mm]
\emph{Drosophila pseudoobscura}  & UCI\_Dpse\_MV25 & \emph{Drosophila yakuba}  & dyak\_caf1
\end{tabular}

\vspace{5mm}

  \includegraphics[width=12cm]{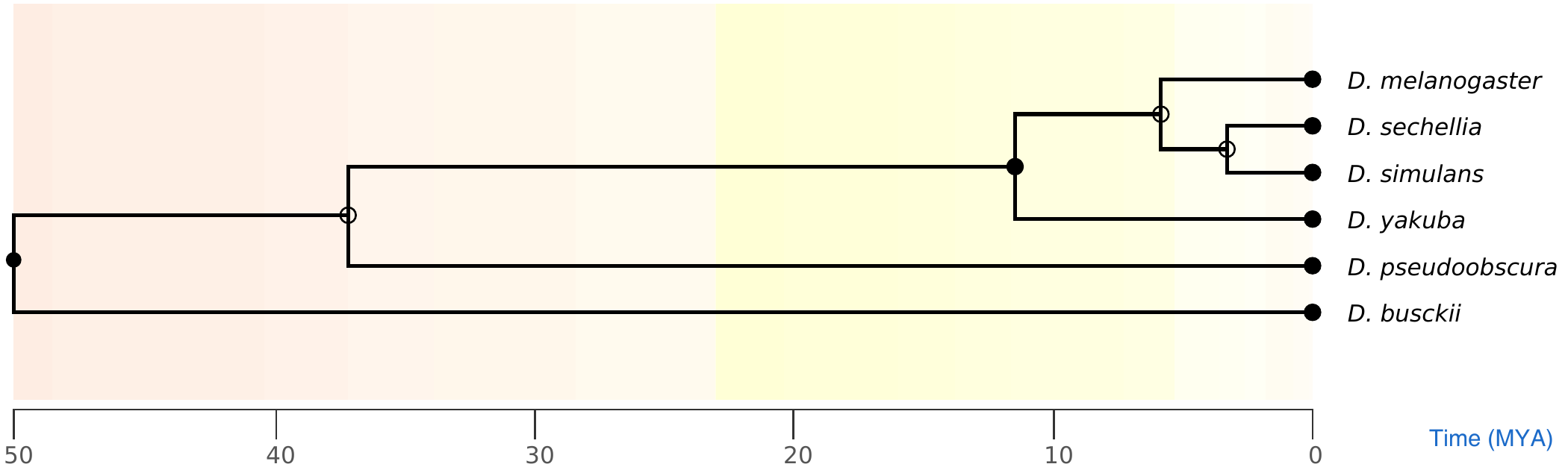}
  \caption{List of genomes used in our experiments and a reference phylogenetic tree of the respective species of \emph{Drosophila} given by TimeTree~\cite{KSSH17}, a public knowledge-base for information on the tree-of-life and its evolutionary timescale.}
  \label{fig:drosophila}
\end{figure}

We downloaded the genomes of 6 species of \emph{Drosophila}~\cite{melanogaster-2000,obscura-2005,drosophila-consortium-2007,busckii-2015}
from NCBI\footnote{\url{https://www.ncbi.nlm.nih.gov}}. In our experiments we used the assemblies listed in Figure~\ref{fig:drosophila}, with their respective gene annotations.

As already mentioned, we obtained pairwise similarities between genes of \emph{Drosophila} genomes using the FFGC pipeline\footnote{\url{https://bibiserv.cebitec.uni-bielefeld.de/ffgc}}~\cite{DFS18}
with the following parameters: (i) $1$ for the minimum number of genomes for which each gene must share some similarity in, (ii) $0.1$ for the stringency threshold, (iii) $1$ for the BLAST e-value, and (iv) default values for the remaining parameters.


In the following subsections, in-depth information is provided on the results for experiments using complete genomes and X chromosomes of the listed \emph{Drosophila} species.

\subsection{Complete genomes}\label{app:real-complete}

The first tables in this section detail the results of the comparison of complete genomes in terms of the BLAST alignment performed for all genes, and the corresponding similarity graphs for each genome pair without cutting threshold. This data was generated using the FFCG pipeline with the parameters described above. Unplaced scaffolds were discarded, 
decreasing the number of genes, from $\sim 15{,}000$ to $\sim 13{,}000$.
Table~\ref{tbl:distr-cmp} outlines the number of gene pairs in each similarity range for each pair of genomes.
Table~\ref{tbl:assoc-cmp} shows the number of genes with no homology relations 
(which induce trivial selections of indel edges in the relational diagram), the number of genes with exactly one homology relation and 
 the number of genes with multiple homologies  
(which pose a significant challenge to the solver). 
The computed distances and elapsed time (or gap in \% when the solver reaches the time limit) in the pairwise comparisons with cutting threshold $0.3$ are shown in Tables~\ref{tbl:ff-cmpl-0.3-exec} and~\ref{tbl:unwff-cmpl-0.3-exec}. 
The solver was set to stop after finding a solution with optimality gap smaller than 0.5\% or after 3 hours.

\vspace{5mm}

\ \

\begin{table}[ht]\setlength{\tabcolsep}{.8pt} 
\caption{\label{tbl:distr-cmp} Distribution of similarities between genes (and percentage) in pairwise comparisons of complete genomes.}
\vspace{3mm}
\scriptsize
\centering
\begin{tabular}{lcccccc}
species & similarity & \emph{pseudoobscura} & \emph{sechellia} & \emph{simulans} & \emph{yakuba} & \emph{busckii} \\ \hline 
 & (0.0-0.2) & 53648 (60.09\pct) & 33409 (48.69\pct) & 34803 (49.15\pct) & 38143 (51.71\pct) & 53733 (65.42\pct) \\
 & [0.2-0.4) & 19034 (21.32\pct) & 17822 (25.97\pct) & 18566 (26.22\pct) & 18748 (25.42\pct) & 16129 (19.64\pct) \\
\emph{melanogaster} & [0.4-0.6) & \, 6036 \, (6.76\pct) & \, 3896 \, (5.68\pct) & \, 4019 \, (5.68\pct) & \, 4195 \, (5.69\pct) & \, 5207 \, (6.34\pct) \\
 & [0.6-0.8) & \, 4993 \, (5.59\pct) & \, 1826 \, (2.66\pct) & \, 1909 \, (2.70\pct) & \, 3010 \, (4.08\pct) & \, 4300 \, (5.23\pct) \\
 & [0.8-1.0] & \, 5570 \, (6.24\pct) & 11666 (17.00\pct) & 11513 (16.26\pct) & \, 9663 (13.10\pct) & \, 2772 \, (3.37\pct) \\
 &  &  \: 89281 \;\, (100\pct) \: & \: 68619 \;\, (100\pct) \: & \: 70810 \;\, (100\pct) \: & \: 73759 \;\, (100\pct) \: & \: 82141 \;\, (100\pct) \: \\ \cline{2-7}
 & (0.0-0.2) &  & 53777 (62.13\pct) & 54221 (61.83\pct) & 54147 (61.96\pct) & 54104 (65.78\pct) \\
 & [0.2-0.4) &  & 18169 (20.99\pct) & 18724 (21.35\pct) & 18645 (21.34\pct) & 15940 (19.38\pct) \\
\emph{pseudoobscura} & [0.4-0.6) &  & \, 5466 \, (6.32\pct) & \, 5601 \, (6.39\pct) & \, 5595 \,(6.40\pct) & \, 5183 \, (6.30\pct) \\
 & [0.6-0.8) &  & \, 4838 \, (5.59\pct) & \, 4895 \, (5.58\pct) & \, 4797 \, (5.49\pct) & \, 4223 \, (5.13\pct) \\
 & [0.8-1.0] &  & \, 4303 \, (4.97\pct) & \, 4255 \, (4.85\pct) & \, 4202 \, (4.81\pct) & \, 2798 \, (3.40\pct) \\
 &  &  & \: 86553 \;\, (100\pct) \: & \: 87696 \;\, (100\pct) \:& \: 87386 \;\, (100\pct) \: & \: 82248 \;\, (100\pct) \: \\ \cline{2-7}
 & (0.0-0.2) &  &  & 34227 (49.87\pct) & 38169 (52.98\pct) & 53105 (66.03\pct) \\
 & [0.2-0.4) &  &  & 17325 (25.25\pct) & 17430 (24.19\pct) & 15521 (19.30\pct) \\
\emph{sechellia} & [0.4-0.6) &  &  & \, 3721 \, (5.42\pct) & \, 4075 \, (5.66\pct) & \, 5003 \, (6.22\pct) \\
 & [0.6-0.8) &  &  & \, 1277 \, (1.86\pct) & \, 2987 \, (4.15\pct) & \, 4175 \, (5.19\pct) \\
 & [0.8-1.0] &  &  & 12077 (17.60\pct) & \, 9379 (13.02\pct) & \, 2626 \, (3.26\pct) \\
 &  &  &  & \: 68627 \;\, (100\pct) \: & \: 72040 \;\, (100\pct) \: & \: 80430 \;\, (100\pct) \: \\ \cline{2-7}
 & (0.0-0.2) &  &  &  & 39218 (52.89\pct) & 54066 (66.32\pct) \\
 & [0.2-0.4) &  &  &  & 18288 (24.66\pct) & 15648 (19.20\pct) \\
\emph{simulans} & [0.4-0.6) &  &  &  & \, 4287 \, (5.78\pct) & \, 5115 \, (6.27\pct) \\
 & [0.6-0.8) &  &  &  & \, 2960 \, (3.99\pct) & \, 4103 \, (5.03\pct) \\
 & [0.8-1.0] &  &  &  & \, 9395 (12.67\pct) & \, 2589 \, (3.18\pct) \\
 &  &  &  &  & \: 74148 \;\, (100\pct) \: & \: 81521 \;\, (100\pct) \: \\ \cline{2-7}
 & (0.0-0.2) &  &  &  &  & 54022 (66.32\pct) \\
 & [0.2-0.4) &  &  &  &  & 15767 (19.36\pct) \\
\emph{yakuba} & [0.4-0.6) &  &  &  &  & \, 5027 \, (6.17\pct) \\
 & [0.6-0.8) &  &  &  &  & \, 4105 \, (5.04\pct) \\
 & [0.8-1.0] &  &  &  &  & \, 2540 \, (3.12\pct) \\
 &  &  &  &  &  & \: 81461 \;\, (100\pct) \: \\ \hline
\end{tabular}
\end{table}


\begin{table}[ht]\setlength{\tabcolsep}{6pt} 
\caption{\label{tbl:assoc-cmp} Association between genes 
in pairwise comparisons of complete genomes, considering pairwise similarities strictly greater than 0. The tables show the number of genes with zero, one and multiple homology relations, respectively. For all of them, the element stored in line $i$ and column $j$ represents the number of genes of the species $i$ in the pairwise comparison of genomes $i$ and $j$.}
\vspace{3mm}
\centering
\scriptsize
\begin{tabular}{llrrrrrr}
\multicolumn{8}{c}{\bf Number of unassociated genes 
}\\[2mm]
\hline
species &  & \emph{melanog} & \emph{pseudoob} & \emph{sechellia} & \emph{simulans} & \emph{yakuba} & \emph{busckii} \\ \hline
& \#genes & 13049 & 13399 & 13037 & 13023 & 12835 & 11371 \\ \hline 
\emph{melanogaster} & 13049 & --- & 570 & 213 & 277 & 352 & 1183 \\
\emph{pseudoobscura} & 13399 & 565 & --- & 583 & 694 & 710 & 1211 \\
\emph{sechellia} & 13037 & 189 & 620 & --- & 263 & 393 & 1189 \\
\emph{simulans} & 13023 & 335 & 779 & 345 & --- & 484 & 1358 \\
\emph{yakuba} & 12835 & 306 & 666 & 323 & 327 & --- & 1225 \\
\emph{busckii} & 11371 & 304 & 354 & 321 & 380 & 400 & ---\\
\hline
\rule{0em}{4ex}\\
\multicolumn{8}{c}{\bf Number of genes uniquely associated}\\[2mm]
\hline
species &  & \emph{melanog} & \emph{pseudoob} & \emph{sechellia} & \emph{simulans} & \emph{yakuba} & \emph{busckii} \\ \hline
& \#genes & 13049 & 13399 & 13037 & 13023 & 12835 & 11371 \\ \hline 
\emph{melanogaster}  & 13049 &  --- & 5439 & 6624 & 6533 & 6361 & 5107 \\
\emph{pseudoobscura} & 13399 & 5775 & --- & 5746 & 5704 & 5707 & 5205 \\
\emph{sechellia}     & 13037 & 6650 & 5487 & --- & 6656 & 6307 & 5099 \\
\emph{simulans}      & 13023 & 6516 & 5394 & 6594 & --- & 6237 & 4985 \\
\emph{yakuba}        & 12835 & 6288 & 5358 & 6242 & 6251 & --- & 4982 \\
\emph{busckii}       & 11371 & 4797 & 4654 & 4749 & 4730 & 4725 &  --- \\
\hline
\rule{0em}{4ex}\\
\multicolumn{8}{c}{\bf Number of genes associated to at least two other genes}\\[2mm]
\hline
species &  & \emph{melanog} & \emph{pseudoob} & \emph{sechellia} & \emph{simulans} & \emph{yakuba} & \emph{busckii} \\ \hline
& \#genes & 13049 & 13399 & 13037 & 13023 & 12835 & 11371 \\ \hline   
\emph{melanogaster} & 13049 & --- & 7040 & 6212 & 6239 & 6336 & 6759 \\
\emph{pseudoobscura} & 13399 & 7059 & --- & 7070 & 7001 & 6982 & 6983 \\
\emph{sechellia} & 13037 & 6198 & 6930 & --- & 6118 & 6337 & 6749 \\
\emph{simulans} & 13023 & 6172 & 6850 & 6084 & --- & 6302 & 6680 \\
\emph{yakuba} & 12835 & 6241 & 6811 & 6270 & 6257 & --- & 6628 \\
\emph{busckii} & 11371 & 6270 & 6363 & 6301 & 6261 & 6246 & ---\\[2mm] 
\hline
\end{tabular}
\end{table}

\begin{table}[ht]\setlength{\tabcolsep}{2pt}
\caption{\label{tbl:ff-cmpl-0.3-exec} Computed $\dffdcjid$ and elapsed time (or gap in \%) in pairwise comparisons of complete genomes, with  
cutting threshold $x=0.3$. The time limit for execution of the ILP solver is 10800s.}
\vspace{3mm}
\scriptsize
\centering
\begin{tabular}{lc@{~~~}c@{~~~}c@{~~~}c@{~~~}c}
species & \emph{pseudoobscura} & \emph{sechellia} & \emph{simulans} & \emph{yakuba} & \emph{busckii} \\ \hline
\emph{melanogaster} & 7373.7 (0.76\pct) & 1925.5 (4431.78s) & 2094.7 (109.60s) & 3193.2 \, (201.49s) & 7764.6 (540.19s) \\
\emph{pseudoobscura} &  & 7326.0 \, (163.12s) & 7355.5 (764.24s) & 7351.2 (5782.73s) & 7784.0 (290.12s) \\
\emph{sechellia} &  &  & 1661.0 (103.33s) & 3259.0 \, (146.88s) & 7710.4 (415.23s) \\
\emph{simulans} &  &  &  & 3306.0 \, (216.77s) & 7699.9 (115.54s) \\
\emph{yakuba} &  &  &  &  & 7667.4 (153.36s)\\
\hline
\end{tabular}
\end{table}

\begin{table}[ht]
\caption{\label{tbl:unwff-cmpl-0.3-exec} Computed $\dunwffdcjid$ and elapsed time (or gap in \%) in pairwise comparisons of complete genomes, with
cutting threshold $x=0.3$. The time limit for execution of the ILP solver is 10800s.}
\vspace{3mm}
\scriptsize
\centering
\begin{tabular}{lccccc}
species & \emph{pseudoobscura} & \emph{sechellia} & \emph{simulans} & \emph{yakuba} & \emph{busckii} \\ \hline
\emph{melanogaster} & 4084 (2.67\pct) & \, 708 (0.96\pct) & 933 (0.62\pct) \;\;\; & 1269 (1.35\pct) & 4791 (0.95\pct) \\
\emph{pseudoobscura} &  & 4088 (1.50\pct) & 4176 (1.47\pct) \;\;\;\;\; & 4142 (1.22\pct) & 4797 (1.20\pct) \\
\emph{sechellia} &  &  & 905 (2812.89s) & 1341 (1.10\pct) & 4817 (0.98\pct) \\
\emph{simulans} &  &  &  & 1478 (1.44\pct) & 4866 (0.84\pct) \\
\emph{yakuba} &  &  &  &  & 4820 (1.00\pct)\\
\hline
\end{tabular}
\end{table}

\clearpage 

\subsection{X chromosomes}\label{app:real-X}

Unplaced scaffolds of X chromosomes were discarded, decreasing the overall number of genes from $\sim 2{,}500$ to $\sim 2{,}000$. 
Similarity values in pairwise comparisons are given in Table~\ref{tbl:distr-X}. 
The number of genes with 0, 1 and multiple homologies are given in Table~\ref{tbl:assoc-X}.
Results for $\dffdcjid$ and for $\dunwffdcjid$ are shown in
Tables~\ref{tbl:perform-ff} and~\ref{tbl:perform-gc}
(CPLEX was set to stop after finding a solution with optimality gap smaller than 0.1\% or after 1 hour). 

\begin{table}[ht]\setlength{\tabcolsep}{4pt}
\caption{\label{tbl:distr-X} Distribution of similarities (and percentage) in pairwise comparisons of X chromosomes.}
\vspace{3mm}
\scriptsize
\centering
\begin{tabular}{lcccccc}
species & similarity & \emph{pseudoobscura} & \emph{sechellia} & \emph{simulans} & \emph{yakuba} & \emph{busckii} \\ \hline 
 & (0.0-0.2) & 4710 (63.70\pct) & \, 829 (22.38\pct) & \, 897 (25.63\pct) & \, 987 (28.81\pct) & 2072 (48.64\pct) \\
 & [0.2-0.4) & \, 980 (13.25\pct) & \, 576 (15.55\pct) & \, 536 (15.31\pct) & \, 528 (15.41\pct) & \, 738 (17.23\pct) \\
\emph{melanogaster} & [0.4-0.6) & \, 541 \, (7.32\pct) & \, 352 \, (9.50\pct) & \, 256 \, (7.31\pct) & \, 242 \, (7.06\pct) & \, 475 (11.15\pct) \\
 & [0.6-0.8) & \, 605 \, (8.18\pct) & \, 271 \, (7.32\pct) & \, 256 \, (7.31\pct) & \, 412 (12.03\pct) & \, 584 (13.71\pct) \\
 & [0.8-1.0] & \, 558 \, (7.55\pct) & 1676 (45.25\pct) & 1555 (44.43\pct) & 1257 (36.69\pct) & \, 391 \, (9.18\pct) \\
 & & 7394 \:\, (100\pct)  & 3704 \;\, (100\pct) & 3500 \;\, (100\pct) & 3426 \;\, (100\pct) & 4260 \;\, (100\pct) \\ \cline{2-7}
 & (0.0-0.2) &  & 4849 (64.60\pct) & 4703 (65.15\pct) & 4588 (64.64\pct) & 5021 (66.59\pct) \\
 & [0.2-0.4) &  & \, 962 (12.82\pct) & \, 907 (12.56\pct) & \, 898 (12.65\pct) & \, 953 (12.64\pct) \\
\emph{pseudoobscura} & [0.4-0.6) &  & \, 539 \, (7.18\pct) & \, 498 \, (6.90\pct) & \, 495 \, (6.97\pct) & \, 563 \, (7.47\pct) \\
 & [0.6-0.8) &  & \, 600 \, (7.99\pct) & \, 585 \, (8.10\pct) & \, 574 \, (8.09\pct) & \, 584 \, (7.75\pct) \\
 & [0.8-1.0] &  & \, 556 \, (7.41\pct) & \, 526 \, (7.29\pct) & \, 543 \, (7.65\pct) & \, 419 \, (5.56\pct) \\
 &  &  & 7506 \;\, (100\pct) & 7219 \;\, (100\pct) & 7098 \;\, (100\pct) & 7540 \;\, (100\pct) \\ \cline{2-7}
 & (0.0-0.2) &  &  & \, 773 (22.62\pct) & \, 961 (28.16\pct) & 2014 (47.90\pct) \\
 & [0.2-0.4) &  &  & \, 521 (15.24\pct) & \, 532 (15.59\pct) & \, 741 (17.62\pct) \\
\emph{sechellia} & [0.4-0.6) &  &  & \, 191 \, (5.59\pct) & \, 266 \, (7.79\pct) & \, 486 (11.56\pct) \\
 & [0.6-0.8) &  &  & \, 139 \, (4.07\pct) & \, 423 (12.39\pct) & \, 574 (13.65\pct) \\
 & [0.8-1.0] &  &  & 1795 (52.50\pct) & 1231 (36.07\pct) & \, 390 \, (9.27\pct) \\
 &  &  &  & 3419 \;\, (100\pct) & 3413 \;\, (100\pct) & 4205 \;\, (100\pct) \\ \cline{2-7}
 & (0.0-0.2) &  &  &  & 1038 (30.77\pct) & 2069 (49.95\pct) \\
 & [0.2-0.4) &  &  &  & \, 506 (15.00\pct) & \, 697 (16.83\pct) \\
\emph{simulans} & [0.4-0.6) &  &  &  & \, 254 \, (7.53\pct) & \, 448 (10.82\pct) \\
 & [0.6-0.8) &  &  &  & \, 403 (11.95\pct) & \, 556 (13.42\pct) \\
 & [0.8-1.0] &  &  &  & 1172 (34.75\pct) & \, 372 \, (8.98\pct) \\
 &  &  &  &  & 3373 \;\, (100\pct) & 4142 \;\, (100\pct) \\ \cline{2-7}
 & (0.0-0.2) &  &  &  &  & 2110 (50.62\pct) \\
 & [0.2-0.4) &  &  &  &  & \, 668 (16.03\pct) \\
\emph{yakuba} & [0.4-0.6) &  &  &  &  & \, 456 (10.94\pct) \\
 & [0.6-0.8) &  &  &  &  & \, 561 (13.46\pct) \\
 & [0.8-1.0] &  &  &  &  & \, 373 \, (8.95\pct) \\
 &  &  &  &  &  & 4168 \;\, (100\pct) \\ \hline
\end{tabular}
\end{table}

\begin{table}[ht]\setlength{\tabcolsep}{6pt}
\caption{\label{tbl:assoc-X} Association between genes 
in pairwise comparisons of the corresponding X chromosomes, considering pairwise similarities strictly greater than 0. For the three tables, the element stored in line $i$ and column $j$ represents the number of genes of the species $i$ in the pairwise comparison of genomes $i$ and $j$. The X chromosome of \emph{D.~pseudoobscura} was fused with another chromosome during evolution~\cite{obscura-2005}, therefore it presents a larger number of unassociated genes when compared to the other species.}
\vspace{3mm}
\centering
\scriptsize
\begin{tabular}{llrrrrrr}
\multicolumn{8}{c}{\bf Number of unassociated genes 
}\\[2mm]
\hline

species & & \emph{melanog} & \emph{pseudoob} & \emph{sechellia} & \emph{simulans} & \emph{yakuba} & \emph{busckii} \\ \hline
 & \#genes & 2043 & 4770 & 2107 & 2007 & 1956 & 1953 \\ \hline
\emph{melanogaster} & 2043 & ---& 152 & 23 & 84 & 100 & 221 \\
\emph{pseudoobscura} & 4770 & 2076 &--- & 2025 & 2127 & 2113 & 2110 \\
\emph{sechellia} & 2107 & 57 & 174 &--- & 102 & 133 & 257 \\
\emph{simulans} & 2007 & 84 & 187 & 74 & ---& 130 & 272 \\
\emph{yakuba} & 1956 & 80 & 153 & 80 & 107 & ---& 227 \\
\emph{busckii} & 1953 & 167 & 124 & 173 & 217 & 201 & --- \\
\hline
\rule{0em}{4ex}\\
\multicolumn{8}{c}{\bf Number of genes uniquely associated}\\[2mm]
\hline
species &  & \emph{melanog} & \emph{pseudoob} & \emph{sechellia} & \emph{simulans} & \emph{yakuba} & \emph{busckii} \\ \hline
 & \#genes & 2043 & 4770 & 2107 & 2007 & 1956 & 1953 \\ \hline
\emph{melanogaster}  & 2043 &  ---   & 1052 & 1440 & 1402 & 1428 & 1191 \\
\emph{pseudoobscura} & 4770 & 1613 &  ---   & 1651 & 1579 & 1646 & 1565 \\
\emph{sechellia}     & 2107 & 1479 & 1084 &  ---   & 1454 & 1449 & 1224 \\
\emph{simulans}      & 2007 & 1382 & 1024 & 1392 &  --- & 1352 & 1126 \\
\emph{yakuba}        & 1956 & 1352 & 1017 & 1353 & 1331 &  ---   & 1123 \\
\emph{busckii}       & 1953 & 1155 &  977 & 1154 & 1125 & 1147 & ---    \\
\hline
\rule{0em}{4ex}\\
\multicolumn{8}{c}{\bf Number of genes associated to at least two other genes}\\[2mm]
\hline
species & & \emph{melanog} & \emph{pseudoob} & \emph{sechellia} & \emph{simulans} & \emph{yakuba} & \emph{busckii} \\ \hline
& \#genes & 2043 & 4770 & 2107 & 2007 & 1956 & 1953 \\ \hline
\emph{melanogaster} & 2043 &--- & 839 & 580 & 557 & 515 & 631 \\
\emph{pseudoobscura} & 4770 & 1081 & ---& 1094 & 1064 & 1011 & 1095 \\ 
\emph{sechellia} & 2107 & 571 & 849 &--- & 551 & 525 & 626 \\
\emph{simulans} & 2007 & 541 & 796 & 541 & ---& 525 & 609 \\
\emph{yakuba} & 1956 & 524 & 786 & 523 & 518 & --- & 606 \\
\emph{busckii} & 1953 & 631 & 852 & 626 & 611 & 605 & --- \\
\hline
\end{tabular}
\end{table}

\begin{table}[hbt]
\caption{\label{tbl:perform-ff} Computed $\dffdcjid$ and elapsed time (or gap in \%) in pairwise comparisons of X chromosomes, 
with cutting thresholds ranging between $x=0.0$ and $x=0.3$. The time limit is 3600s.}
\vspace{3mm}
\scriptsize
\centering
\setlength{\tabcolsep}{1pt}
\begin{tabular}{lc@{~~~~}rr@{~~~~}rr@{~~~~}rr@{~~~~}rr@{~~~~}rr}
species & $x$ & \multicolumn{2}{c}{\emph{pseudoobscura}} & \multicolumn{2}{c}{\emph{sechellia}} & \multicolumn{2}{c}{\emph{simulans}} & \multicolumn{2}{c}{\emph{yakuba}} & \multicolumn{2}{c}{\emph{busckii}} \\ \hline
\multirow{4}{*}{\emph{melanogaster}} & 0.0 & 1390.3 & (255.22s) & 407.3 & (0.32\pct) & 432.4 & (9.60s) & 587.4 & (4.21s)  & 1362.7 & (109.59s) \\
 & 0.1 & 1370.1 & (30.01s) & 408.3 & (0.34\pct) & 433.8 & (10.12s) & 590.0 & (4.08s) & 1363.9 & (11.49s) \\
 & 0.2 & 1326.0 & (6.28s) & 412.7 & (174.06s) & 436.6 & (5.25s) & 601.0 & (2.17s) & 1344.7 & (5.12s) \\
 & 0.3 & 1296.0  &(4.13s) & 416.7 & (24.47s) & 445.4 & (3.55s) & 609.3 & (1.64s) & 1321.8 & (2.87s) \\ \cline{2-12}
\multirow{4}{*}{\emph{pseudoobscura}} & 0.0 & & & 1417.1 & (258.89s) & 1375.6 & (281.95s) & 1361.7 & (94.68s) & 1515.7 & (368.78s) \\
 & 0.1 & & & 1394.1 & (36.51s) & 1355.1 & (45.6s) & 1337.0 & (17.74s) & 1491.7 & (33.27s) \\
 & 0.2 & & & 1344.1 & (3.64s) & 1309.7 & (329.25s) & 1299.5 & (3.34s) & 1433.3 & (5.61s) \\
 & 0.3 & & & 1308.0 & (5.56s) & 1278.0 & (3.73s) & 1262.3 & (3.69s) & 1374.1 & (4.69s) \\ \cline{2-12}
\multirow{4}{*}{\emph{sechellia}} & 0.0 & & & & & 352.5 & (5.90s) &  626.8 & (4.70s) & 1378.2 & (74.01s) \\
 & 0.1 & & & & & 352.8 & (5.83s) & 630.4 & (3.92s) & 1377.1 & (23.36s) \\
 & 0.2 & & & & & 351.9 & (3.56s) & 635.3 & (3.08s) & 1354.2 & (5.38s) \\
 & 0.3 & & & & & 355.0 & (2.55s) & 641.3 & (1.92s) & 1328.3 & (4.18s) \\ \cline{2-12}
\multirow{4}{*}{\emph{simulans}} & 0.0 & & & & & & & 617.8 & (7.78s) & 1344.0 & (80.84s)\\
 & 0.1 & & & & & & & 621.3 & (5.27s) & 1342.7 & (29.58s) \\
 & 0.2 & & & & & & & 626.2 & (2.04s) & 1316.7 & (5.50s) \\
 & 0.3 & & & & & & & 637.8 & (1.99s) & 1295.5 & (3.25s) \\ \cline{2-12}
\multirow{4}{*}{\emph{yakuba}} & 0.0 & & & & & & & & & 1325.5 & (69.40s) \\
 & 0.1 & & & & & & & & & 1323.7 & (24.32s)\\
 & 0.2 & & & & & & & & & 1304.9 & (6.27s)\\
 & 0.3 & & & & & & & & & 1280.8 & (3.73s)\\ \hline
\end{tabular}
\end{table}

\begin{table}[hbt]
\caption{\label{tbl:perform-gc} Computed $\dunwffdcjid$ and elapsed time (or gap in \%) in pairwise comparisons of X chromosomes, 
with cutting thresholds $x=0.0$, $x=0.3$ and $x=0.5$. The time limit is 3600s.}
\vspace{3mm}
\scriptsize
\centering
\setlength{\tabcolsep}{1pt}
\begin{tabular}{lc@{~~~~}rr@{~~~~}rr@{~~~~}rr@{~~~~}rr@{~~~~}rr}
species & $x$ & \multicolumn{2}{c}{\emph{pseudoobscura}} & \multicolumn{2}{c}{\emph{sechellia}} & \multicolumn{2}{c}{\emph{simulans}} & \multicolumn{2}{c}{\emph{yakuba}} & \multicolumn{2}{c}{\emph{busckii}} \\ \hline
\multirow{3}{*}{\emph{melanogaster}} & 0.0 & 720 & (2.80\pct) & 132 & (0.38\pct) & 178 & (30.93s) & 218 & (4.99s) & 832 & (3.49\pct) \\
 & 0.3 & 829 & (1.70s) & 160 & (9.31s) & 218 & (0.98s) & 293 & (0.66s) & 972  & (0.90s)\\
 & 0.5 & 940 & (0.46s) & 228 & (0.52s) & 298 & (0.40s) & 397 & (0.34s) & 1003 & (0.27s) \\ \cline{2-12}
\multirow{3}{*}{\emph{pseudoobscura}} & 0.0 & & & 743 & (3.13\pct) & 743 & (2.14\pct) & 724 & (1.40\pct) & 912 & (3.76\pct)\\
 & 0.3 & & & 836 &(1.06s)& 849 &(1.03s)& 837 &(0.96s) & 980 &  (1.06s)\\
 & 0.5 & & & 929 &(0.44s)& 938 &(0.43s)& 908 &(0.43s)& 1015 & (0.39s) \\ \cline{2-12}
\multirow{3}{*}{\emph{sechellia}} & 0.0 & & & & & 171 & (45.95s) & 236 & (6.41s) & 850 & (2.19\pct)  \\
 & 0.3 & & & & & 194 & (1.06s) & 301 & (0.74s) & 982 & (2.11s) \\
 & 0.5 & & & & & 244 & (0.44s) & 423 & (0.40s)  & 1014 & (0.28s) \\ \cline{2-12}
\multirow{3}{*}{\emph{simulans}} & 0.0 & & & & & & & 265 & (18.14s) & 863 & (2.35\pct)\\
 & 0.3 & & & & & & & 336 & (0.63s) & 994 & (0.87s) \\
 & 0.5 & & & & & & & 453 & (0.33s) & 1005 & (0.25s)\\ \cline{2-12}
\multirow{3}{*}{\emph{yakuba}} & 0.0 & & & & & & & & & 830 & (1.73\pct) \\
 & 0.3 & & & & & & & & & 972 & (0.72s)\\
 & 0.5 & & & & & & & & & 992 &(0.24s)\\ \hline
\end{tabular}
\end{table}

\end{document}